\documentclass[12pt]{article}
\usepackage{arxiv}
\usepackage{amssymb}
\usepackage{nicefrac}
\usepackage[utf8]{inputenc} 
\usepackage[T1]{fontenc}    
\usepackage{lipsum}
\usepackage{parskip}
\usepackage{comment}

\usepackage{titlesec}
\titleclass{\subsubsubsection}{straight}[\subsection]
\newcounter{subsubsubsection}[subsubsection]
\renewcommand\thesubsubsubsection{\thesubsubsection.\arabic{subsubsubsection}}

\titleformat{\subsubsubsection}
  {\normalfont\normalsize\bfseries}{\thesubsubsubsection}{1em}{}
\titlespacing*{\subsubsubsection}
{0pt}{3.25ex plus 1ex minus .2ex}{1.5ex plus .2ex}

\makeatletter
\renewcommand\paragraph{\@startsection{paragraph}{5}{\z@}%
  {3.25ex \@plus1ex \@minus.2ex}%
  {-1em}%
  {\normalfont\normalsize\bfseries}}
\renewcommand\subparagraph{\@startsection{subparagraph}{6}{\parindent}%
  {3.25ex \@plus1ex \@minus .2ex}%
  {-1em}%
  {\normalfont\normalsize\bfseries}}
\def\toclevel@subsubsubsection{4}
\def\toclevel@paragraph{5}
\def\toclevel@paragraph{6}
\def\l@subsubsubsection{\@dottedtocline{4}{7em}{4em}}
\def\l@paragraph{\@dottedtocline{5}{10em}{5em}}
\def\l@subparagraph{\@dottedtocline{6}{14em}{6em}}
\makeatother
\usepackage{stmaryrd}
\usepackage[sort&compress,square,numbers]{natbib}
\usepackage{mathrsfs}
\usepackage[colorlinks]{hyperref}
\usepackage{graphicx}
\usepackage{amsmath}
\usepackage{amssymb}
\usepackage[dvipsnames]{xcolor}
\usepackage{float} 
\usepackage{caption} 
\usepackage{subcaption}
\usepackage{comment}
\usepackage{appendix}
\usepackage{here}
\usepackage{gensymb}
\usepackage{mathrsfs}
\usepackage{ctable}
\usepackage{bm}
\usepackage{multirow}
\usepackage{textcomp}
\newcommand{\bbGamma}{{\mathpalette\makebbGamma\relax}}
\newcommand{\makebbGamma}[2]{%
  \raisebox{\depth}{\scalebox{1}[-1]{$\mathsurround=0pt#1\mathbb{L}$}}%
}
\usepackage{siunitx}[=v2]
\DeclareSIUnit{\wtpercent}{wt\%}
\newcommand{\murm}{%
  \ifmmode
    \mathchoice
        {\hbox{\normalsize\textmu}}
        {\hbox{\normalsize\textmu}}
        {\hbox{\scriptsize\textmu}}
        {\hbox{\tiny\textmu}}%
  \else
    \textmu
  \fi
}
\title{Kinetic equations and level-set approach for simulating solid-state microstructure evolutions at the mesoscopic scale: state of the art, limitations, and prospects}

\renewcommand{\headeright}{}
\renewcommand{\undertitle}{}
\renewcommand{\shorttitle}{Kinetic equations and level-set approach for simulating solid-state microstructure evolutions }

\author{ 
	\href{https://orcid.org/0000-0002-6677-2850}{\includegraphics[scale=0.06]{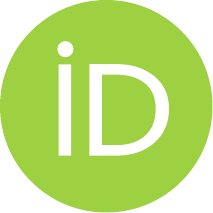}\hspace{1mm}Marc ~Bernacki}\thanks{Corresponding author: marc.bernacki@mines-paristech.fr} $^1$ \\
	\\
	$^1$ CEMEF – Centre de mise en forme des mat\'{e}riaux, CNRS UMR 7635, \\ Mines-ParisTech, PSL-Research University \\
	CS 10207 rue Claude Daunesse, 06904 Sophia Antipolis Cedex, France \\
}

\begin{document}
\maketitle

\begin{abstract}
For over three decades, the front-capturing level-set method has demonstrated its prowess for the simulation, at the mesoscopic scale, of numerous mechanisms in the context of microstructure evolution occurring during complex thermomechanical paths. This review delves into the foundations of this numerical framework, charting its evolution concerning polycrystalline materials, examining its recent advancements, scrutinizing its current shortcomings, and exploring future possibilities. Special attention will be given to the context of hot metal forming processes. In this context, this article also aims to reintroduce, as simply as possible, the kinetic equations related to the grain boundary migration.
\end{abstract}

\keywords{level-set method \and front-capturing approaches \and recrystallization \and grain-growth \and solid-state diffusive phase transformations \and kinetic equations.}


\section{Introduction}\label{Introduction}
The in-use properties of metallic materials are strongly related to their microstructures, which are themselves inherited from the thermomechanical treatments. Hence, understanding, predicting and optimizing microstructure evolution remain open academic questions and are nowadays a key to the competitiveness of major industries, with direct environmental, economic and societal benefits. The prediction of the microstructures take place in the realm of Integrated Computational Materials Engineering (ICME) \citep{allison2013integrated} and the remarkable evolution of computational resources has enabled within the academic world the extensive modeling of solid materials at all scales of interest sometimes with multiscale strategies. In an industrial R\&D context, where computational resources are inevitably more limited and codes need to be robust and realistic, macroscale finite element codes with minimal information about the material itself remain the standard.\\

Indeed, microstructure predictions on multi-pass processes are very challenging due to the strong evolution of the microstructure topology from the beginning to the end of the process. The competition between grain and sub-grain boundary migration, recovery, solid-state phase transformation, recrystallization through nucleation of new grains or rearrangement of substructures lead to complex coupled effects. Minor variations of the process parameters (interpass and/or reheating times, temperature and strain rate values at each pass) may have huge effects on the way the previous cited mechanisms can take place. In this context, macroscopic and homogenized models, i.e. phenomenological models such as those based on the well-known Johnson-Mehl-Avrami-Kolmogorov (JMAK) equations \citep{Avrami1939,Kolmogorov1937,Johnson1939} are widely used in the industry, mainly owing to their low computational cost. If this phenomenological framework is quite convenient, the validity range of these models, associated with a given set of material parameters is often limited to a given process and initial material state. To push these limits, mean field models, based on an implicit description of the microstructure by considering grains or precipitates as spherical entities and statistical evolution related to different characteristics (grain size, precipitate size, dislocation density), have been developed \citep{Hillert1965, Montheillet2009, Cram2009, Bernard2011, Maire2018, perez2008, seret2020}. Mean-field models generally provide acceptable predictions in terms of recrystallization kinetics,  grain size and/or precipitate distribution evolution with very interesting computational cost. However, facing multi-pass processes, they rapidly reach their limits.\\

Then, there is a tremendous demand for predictive models at the mesoscopic scale which explains the development of the hierarchical scale-bridging strategies, where the so-called mesoscopic ”full-field” numerical methods come into play. These approaches, are based on a full description of the polycrystalline microstructure and high-fidelity calculations at the scale of representative volume elements (RVE) where local macroscopic information, such as temperature, strain, and strain rate, determined by macroscopic computations without coupling with the mesoscale characteristics, can serve as boundary conditions. They have demonstrated an exciting potential for an extensive range of microstructure evolutions like the precise modeling of recrystallization \citep{Rollett2017,Miodownik2002,Hallberg2011} in dynamic or post-dynamic conditions, grain growth \citep{Rollett2017}, diffusive solid-state phase transformations \citep{Militzer2011}, spheroidization \citep{poly2017introduction}, and sintering \citep{bruchon20103d}.\\

 The primary numerical frameworks involved include Monte Carlo Potts \citep{rollett1989computer}, cellular automata \citep{Raabe2002,Janssens2010,golab2014}, multi-phase field \citep{Steinbach1996,Moelans2008,KrillIII2002}, front-tracking/vertex \citep{BarralesMora2008,Florez2020b}, and level-set models \citep{Merriman1994,Zhao1996,Bernacki2008,Hallberg2019}. These numerical methods are currently used and developed by many researchers \citep{Rollett2017} and regularly compared for particular metallurgical mechanisms \citep{Jin2015}.\\

Of course, all the mentioned models have their own strengths and weaknesses. Probabilistic voxel-based approaches such as Monte Carlo Potts and some Cellular Automata formulations are very popular. These models consider uniform grids composed of cells to model microstructure and stochastic laws to predict the motion of interfaces. These simulations are efficient in term of computational cost and the scalability is excellent. On the other hand, deterministic approaches, based on the resolution of partial differential equations, are accurate in the description of the involved physical mechanisms although they are numerically more expensive. For instance, front-tracking or vertex approaches are based on an explicit description of interfaces in terms of vertices. Interfaces motion is imposed at each increment by computing the velocity of a set of points. A major difficulty of these approaches is related to the complexity of handling all the possible topological events, such as disappearance and appearance of new grains, which is not straightforward especially in 3D. Other deterministic approaches, also called front-capturing approaches, avoid these topological problems since they are based on an implicit description of the interfaces: the multi-phase field and the level-set (LS) methods.\\

LS simulations in context of regular grids and Fourier transform resolution, with very large number of grains, can be found for grain growth \citep{Elsey2009,Miessen2015} and for static recrystallization \citep{Elsey2011} modeling. When global or local meshing/remeshing operations have to be considered (large deformation, presence of second phase particles…), LS approach in context of unstructured finite element (FE-LS) mesh, and reasonable number of grains can be considered \citep{Bernacki2008, Loge2008, Bernacki2011, Hallberg2013, Scholtes2016, Maire2017}. One of the major advantages of this approach lies in its ability to simultaneously model numerous concomitant mechanisms in the context of large deformations, which explains its used to simulate realistic industrial thermomechanical paths.\\

In the following, the state of the art concerning the use of LS method for the modeling of microstructure evolution will be summarized. First, an introduction to the mathematical description of grain boundaries and relevant kinetics equations linked to their evolution will be recalled. Second, the LS methodology in context of polycrystal description will be introduced. Third, algorithms for modeling recrystallization, grain growth, and other diffusive phenomena will be described. Discussions concerning the anisotropy of grain boundary energy and how to take into account static or evolving second phase particles will be illustrated.  Finally some limits, current developments and perspectives of this approach will be discussed.

\section{Mathematical description of grain boundaries and kinetic equations of microstructure evolution at the mesoscopic scale}\label{sec:gb}
\begin{figure}[ht!]
  \centering
  \includegraphics[scale=0.7]{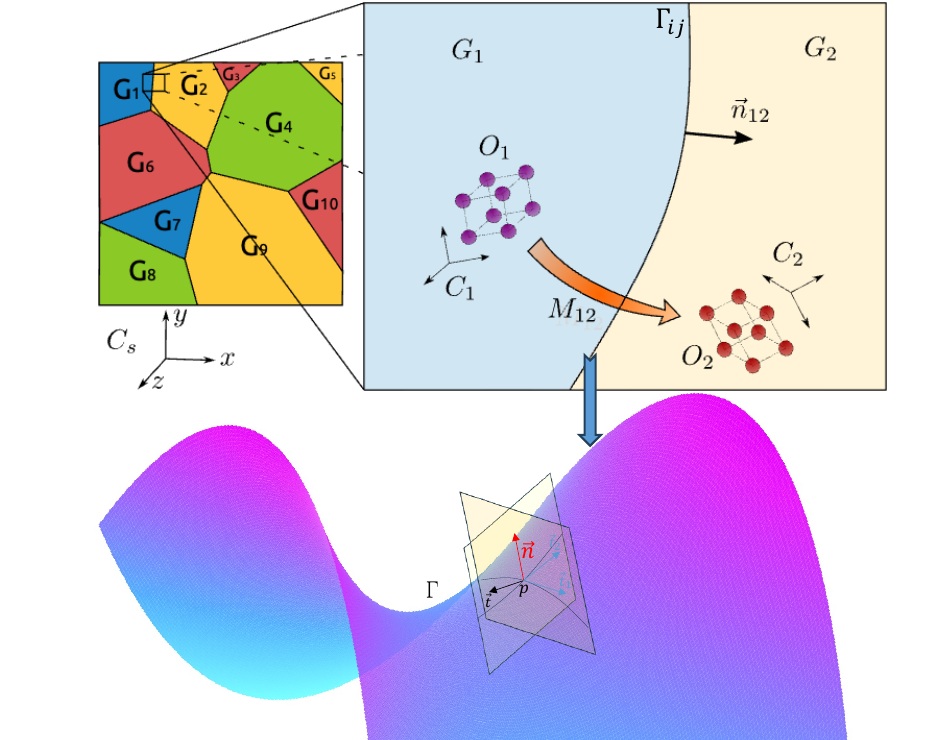}
  \caption{Scheme depicting one GB and its parameters. Inspired from an image available online at Flickr (https://flic.kr/p/2m5JQkz, Uploaded on 15 June 2021) licensed under CC BY 2.0 (https://creativecommons.org/licenses/by/2.0/). Title: 10GGBParam. Author: Brayan Murgas.}\label{fig:tuple}
\end{figure}

Most common metallic parts have defects that change the structure of the material and their arrangement are responsible of the properties of the material. Defects types can be classified into point (0D), linear (1D), planar (2D) and volume (3D) defects. Vacancies, interstitial, substitutional atoms are point defects, dislocations are 1D defects, grain/phase interfaces, and stacking faults are 2D defects, and pores, cracks are 3D defects. Microstucture evolutions during thermomechanical treatments can be explained by the minimization of the stored energies linked to these different defects. 
By depicting typical grain boundary network in a polycrystal as in Fig.\ref{fig:tuple}, two neighboring grains $G_i$ and $G_j$ constitute a grain boundary (GB) $\Gamma_{ij}$. It is characterized by its morphology and its crystallographic properties which may be summarized by a tuple $\mathcal{B}_{ij} =\left(\pmb{\mathbb{M}}_{ij},\mathbf{n}_{ij}\right)$ with two shape parameters describing the interfaces through the unitary-outward normal direction $\mathbf{n}_{ij}$, and three crystallographic parameters describing the orientation relationship between the two adjacent grains known as the misorientation tensor $\pmb{\mathbb{M}}_{ij}$ (see Figure~\ref{fig:tuple} top right). This GB space $\mathcal{B}$ parameterized by the misorientation and the normal direction is illustrated in figure~\ref{fig:tuple}. The misorientation is frequently defined with the axis-angle parameterization, i.e. $\pmb{\mathbb{M}}_{ij}\left(\mathbf{a}_{ij},\theta_{ij}\right)$, where $\mathbf{a}_{ij}$ is the misorientation axis and $\theta_{ij}$ the disorientation \citep{Morawiec2003}. \medbreak
As described in Figure~\ref{fig:tuple}, the GB must be mathematically defined as a 2D closed surface with an interior (the grain itself) in a 3D domain $\Omega$, and many driving pressures discussed thereafter are related to metric properties associated with this surface and intrinsic quantities that exist only on it. Therefore, it is necessary to correctly introduce the surface differential operators to describe the characteristics residing on it. Additionally, different ambiguous notation exist in the literature, which can be confusing. Typically, it is common to find in many writings the indices $\Gamma$, $s$, and $\mathbf{n}$ used, sometimes simultaneously. For the sake of consistency with the conventional notation of the GB stiffness tensor \citep{gurtin1988,gurtin1988toward,gurtin1990}, the '$\mathbf{n}$' index notation referring to the outward unit normal (inclination) will be used in the following to describe all surface-related differential operators on the closed surface $\Gamma$. By defining in $\Omega$ the sign Euclidean distance $d\left(\mathbf{x}\right)$ to $\Gamma$ with positive value inside the closed surface and negative outside (classical in LS framework as discussed in the next section), we have in all points $P$ of $\Gamma$, $\mathbf{n}_P=-\nabla d\left(\mathbf{x}_P\right)$. In the following and for the sake of clarity, the index $P$ will be omitted.
Once $\mathbf{n}$ is clearly introduced, different surface derivative operators and operations for a scalar function $u:\ \Gamma\rightarrow\mathbb{R}$ and vector functions $\mathbf{u}:\ \Gamma\rightarrow\mathbb{R}^3$ can be defined:
\begin{equation}\label{eq:surfgradsca}
\nabla_{\mathbf{n}} u=\nabla u - \left(\nabla u\cdot\mathbf{n}\right)\mathbf{n}=\pmb{\mathbb{P}}_{\mathbf{n}}\nabla u = \left(\nabla u\right)_{\parallel},
\end{equation}
with $\pmb{\mathbb{P}}_{\mathbf{n}}=\pmb{\mathbb{Id}}-\mathbf{n}\mathbf{n}^{T}=\pmb{\mathbb{Id}}-\mathbf{n}\otimes\mathbf{n}$ the orthogonal projection,
\begin{equation}\label{eq:surfgradvec}
\nabla_{\mathbf{n}} \mathbf{u}=\pmb{\mathbb{P}}_{\mathbf{n}}\nabla \mathbf{u}\pmb{\mathbb{P}}_{\mathbf{n}},
\end{equation}
\begin{equation}\label{eq:surfdiv}
\pmb{\nabla}_{\mathbf{n}}\cdot\mathbf{u} = \pmb{\nabla}\cdot\left(\pmb{\mathbb{P}}_{\mathbf{n}} \mathbf{u}\right)=\pmb{\nabla}\cdot \mathbf{u_{\parallel}},
\end{equation}
\begin{equation}\label{eq:surflapl}
\Delta_{\mathbf{n}}u = \nabla_{\mathbf{n}}^{2}u=\pmb{\nabla}_{\mathbf{n}}\cdot\nabla_{\mathbf{n}} u =\pmb{\nabla}\cdot\nabla_{\mathbf{n}}u,
\end{equation}

as such a clear framework to introduce the 3D curvature tensor (also called the Weingarten map). It exists different equivalent ways to introduce this tensor. A 2D one, living in the $\Gamma$ surface, consists to considering the map $P\longrightarrow \pmb{\kappa}$ that assigns each point of $\Gamma$ to the function that measures the directional curvatures \citep{carmo1976} $\pmb{\kappa}\left(\mathbf{t}\right)$ of $\Gamma$ at $P$ in the unitary vector $\mathbf{t}$, tangent to $\Gamma$ at $P$ (see bottom of Fig.~\ref{fig:tuple}). The function $\pmb{\kappa}\left(\cdot\right)$ is a quadratic form :
\begin{equation}\label{eq:tensorcurvasurf}
\pmb{\kappa}\left(\mathbf{t}\right)=\left(\begin{array}{cc}
\alpha_1 & \alpha_2
\end{array}\right)\underbrace{\left(\begin{array}{cc}
\kappa_{11} & \kappa_{12}\\
\kappa_{21} & \kappa_{22}
\end{array}\right)}_{\pmb{\mathbb{K}}_{\mid\Gamma}}\left(\begin{array}{c}
\alpha_1 \\
 \alpha_2
\end{array}\right),
\end{equation}
where $\mathbf{t}=\alpha_1\mathbf{t_1}+\alpha_2\mathbf{t_2}$ with $\left(\mathbf{t_1},\mathbf{t_2}\right)$ an orthonormal basis of the tangent space to $\Gamma$ in $P$, $\kappa_{11}=\pmb{\kappa}\left(\mathbf{t_1}\right),\ \kappa_{22}=\pmb{\kappa}\left(\mathbf{t_2}\right)$ and $\kappa_{12}=\kappa_{21}$. If the vectors $\left(\mathbf{t_1},\mathbf{t_2}\right)$ corresponds to the eigenvalues of the diagonalizable second order tensor $\pmb{\mathbb{K}}_{\mid\Gamma}$ as choosen in Fig.~\ref{fig:tuple}, $\kappa_{12}=\kappa_{21}=0$ and the eigenvalues $\kappa_{11}=\kappa_{1}$ and $\kappa_{22}=\kappa_{2}$, i.e. the corresponding directional curvatures are known as the principal curvatures. The invariants $det\left(\pmb{\mathbb{K}}_{\mid\Gamma}\right)=\kappa_1\kappa_2$ and $Tr\left(\pmb{\mathbb{K}}_{\mid\Gamma}\right)=\kappa_1+ \kappa_2=\kappa$ are well-known as the Gaussian curvature and two times the mean curvature. It must be highlighted that, in material science community, the mean curvature is the classical terminology often used to describe the trace itself.  A 3D extension of $\pmb{\kappa}$ can easily be obtained by considering the $\left(\mathbf{n},\mathbf{t_1},\mathbf{t_2}\right)$ $\mathbb{R}^3-$orthonormal basis and :
\begin{equation}\label{eq:tensorcurvasurf2}
\pmb{\kappa}\left(\mathbf{t}\right)=\left(\begin{array}{ccc}
n_1 & \alpha_1 & \alpha_2
\end{array}\right)\underbrace{\left(\begin{array}{ccc}
0 & 0 & 0\\
0 & \kappa_{1} & 0\\
0 & 0 & \kappa_{2}
\end{array}\right)}_{\pmb{\mathbb{K}}}\left(\begin{array}{c}
n_1 \\
\alpha_1 \\
 \alpha_2
\end{array}\right),
\end{equation}
for any vector $\mathbf{t}=n_1\mathbf{n}+\alpha_1\mathbf{t_1}+\alpha_2\mathbf{t_2}$ and where $\pmb{\mathbb{K}}$ is the curvature tensor, expressed here in the $\left(\mathbf{n},\mathbf{t_1},\mathbf{t_2}\right)$ orthonormal basis. Finally, a classical way to evaluates the curvature tensor in the reference frame consists to use the surface gradient operator defined in Eq.\ref{eq:surfgradvec}. Indeed by using the already precised convention concerning the sign of $d$ and then the sens of $\mathbf{n}$, we have $\pmb{\mathbb{K}}=\nabla_{\mathbf{n}} \mathbf{n}=\pmb{\mathbb{P}}_{\mathbf{n}}\nabla \mathbf{n}\pmb{\mathbb{P}}_{\mathbf{n}}$. Interestingly, if the surface gradient operator is quasi-systematically used in the literature to introduced the curvature tensor, it is in fact completely superfluous here. Indeed, by considering firstly that the normal is unitary, and so $\mathbf{n}\cdot\partial_i\left(\mathbf{n}\right)=0\ \forall i$, we have $\nabla \mathbf{n}\pmb{\mathbb{P}}_{\mathbf{n}}=\nabla \mathbf{n}$. Moreover, by considering now that $\Gamma$ is a smooth closed surface, i.e. that $\mathbf{n}$ derived from $d$ and so $\partial_i n_j=\partial_j n_i\ \forall i,j$, we have $\pmb{\mathbb{P}}_{\mathbf{n}}\nabla \mathbf{n}=\nabla \mathbf{n}$. Thus $\pmb{\mathbb{K}}$ and $\kappa$ can be defined without surface derivative operator as:
\begin{equation}\label{eq:eqdtokappa}
\pmb{\mathbb{K}}=\nabla \mathbf{n}=-\nabla\nabla d=-Hess\left(d\right)\text{, and }\kappa=Tr\left(\pmb{\mathbb{K}}\right)=\nabla\cdot\mathbf{n}=-\Delta d.
\end{equation}

Several metallurgical phenomena have been cited. Therefore, it is important to draw up here an assessment of the kinetics equations existing in the state of the art, at the mesoscopic scale, and their limits and to come back to the relevant physical mechanisms.  When recrystallization and related phenomena or diffusive solid-state phase transformations are considered, the classical framework consists in defining the interface kinetics as the cross product between a mobility, $\mu$ [\SI{}{\raiseto{4}\metre\per\second\per\joule}], generally viewed as an intrinsic scalar property of the considered interface and function of temperature, and the considered driving pressures, $P$ [\SI{}{\joule\per\metre\cubed}]:
\begin{equation}\label{eq:global kinetic equation}
\mathbf{v}=\mu P\mathbf{n},
\end{equation}

The first driving pressure of interest, is related to the minimization of the energy correlated to surface defects in materials (as grain and phase interfaces). This energy field [\SI{}{\joule\per\square\metre}] is classically denoted $\gamma$ for GB and  $\sigma$ for PB. Concerning GB, the corresponding driving pressure can be defined by several way depending on the anisotropy degree of $\gamma$. Indeed, as largely exhibited experimentally \citep{Rollett2017}, $\gamma$ can be seen as a 5-parameter functions $\gamma\left(\pmb{\mathbb{M}},\mathbf{n}\right)$. Following the chain rules perfectly depicted by Herring in \citep{herring1999}, the capillarity driving pressure $P_c$ linked to a GB interface, moving from the $S_0$ surface to $S=S_0+dS$ surface with a $\delta v$ resulting volume variation, can be described as:

\begin{equation}\label{eq:herring1}
P_c\delta v= -\delta\left(\int\gamma d\,S\right)=-\int\delta\gamma d\,S_0 -\int\gamma \delta\left(d\,S\right),
\end{equation}
with,

\begin{equation}\label{eq:herring2}
\int\gamma \delta\left(d\,S\right)\approx \gamma\delta v\pmb{\mathbb{Id}}:\pmb{\mathbb{K}}=\gamma\delta v Tr\left(\pmb{\mathbb{K}}\right),
\end{equation}
and,
\begin{equation}\label{eq:herring3}
\int\delta\gamma d\,S_0\approx -\nabla_{\mathbf{n}}\gamma\cdot\mathbf{n}\delta v + \nabla_{\mathbf{n}}\nabla_{\mathbf{n}}\gamma:\pmb{\mathbb{K}}\delta v.
\end{equation}
By combining Eqs.\ref{eq:herring1}, \ref{eq:herring2} and \ref{eq:herring3}, we obtain at the first order of the small displacement performed:
\begin{equation}\label{eq:herring4}
P_c \approx -\left(\underbrace{\gamma \pmb{\mathbb{Id}} + \nabla_{\mathbf{n}}\nabla_{\mathbf{n}}\gamma}_{\pmb{\bbGamma}\left(\mathbf{n}\right)}\right):\pmb{\mathbb{K}} + \underbrace{\nabla_{\mathbf{n}}\gamma\cdot\mathbf{n}}_{0}.
\end{equation}
This equation invites various comments. Firstly, as it will be illustrated in section \ref{sec:LS}, if the last term is equal, by construction, to 0 for a field existing strictly on a smooth closed $\Gamma$ interface, this term can become ambiguous if front-capturing type methods are used to describe the interface with an extension of the $\gamma$ field over a certain thickness around the $\Gamma$ interface but also near multiple junctions. The term $\pmb{\bbGamma}\left(\mathbf{n}\right)$ is well known as the GB stiffness tensor \citep{gurtin1988,gurtin1988toward,gurtin1990}. Generally, $\pmb{\bbGamma}\left(\mathbf{n}\right)$ and $\pmb{\mathbb{K}}$ are not diagonalizable in the same basis of orthonormal eigenvectors and the calculation of the tensor double product contracted is not straightforward and must be carefully evaluated in the right basis by also taken into account the dependence of $\gamma$ to $\pmb{\mathbb{M}}$ and crystallographic symmetries \citep{Abdeljawad2018,du2007}. 
Finally, when $\gamma$ is assumed not dependent of the inclination $\mathbf{n}$, Eq.\ref{eq:herring4} simplifies in the well-known mesoscopic kinetic equation for curvature flow:
\begin{equation}\label{eq:herring6}
P_c \approx -\gamma\left(\kappa_1 + \kappa_2\right)=-\gamma\kappa.
\end{equation}
Equations \ref{eq:herring4} and \ref{eq:herring6} are interesting, in my opinion, to discuss the current criticism concerning the real link between grain growth and curvature thanks to reverse engineering of  recent 3D in-situ data or 3D full-field high-fidelity simulations \citep{Chen2020,Bhattacharya2021,Xu2023,bizana_kinetics_2023}. Typically, in view of the different proposed equations, the sometimes recently used statement that "capillarity pressure is not correlated to curvature" is inherently ambiguous. Are we speaking of the trace of the curvature tensor (Eq.\ref{eq:herring6}) or the curvature tensor itself (Eq.\ref{eq:herring4})? In the state of the art, it is often the equation (Eq.\ref{eq:herring6}) that is questioned without taken into account the torque terms or additional neglected driving pressures in experimental or numerical analysis. Recent results proposed in \citep{Florez2022,bizana_kinetics_2023} seem to reinforce this discussion. The underlying question then being, can one reasonably use equation Eq.\ref{eq:herring6} in place of Eq.\ref{eq:herring4} when 3D in-situ data at interface scale are considered ? Is the Eq.\ref{eq:herring6} acceptable only when an averaged approach at the polycrystalline scale is considered? These questions, as such as the straightness of the 5-parameter manifold used to describe $\gamma$, will likely be widely addressed in the coming years via the development of new algorithms for characterizing the GB stiffness tensor in 3D in-situ data.

In the context of recrystallization and related phenomena \citep{Rollett2017}, $P$ is classically defined as:
\begin{equation}\label{eq:kinetic equation ReX}
P=P_e + P_c=\tau\llbracket\rho\rrbracket + P_c,
\end{equation} 
where $\tau$ [\SI{}{\joule\per\metre}] is the dislocation line energy and $\llbracket\rho\rrbracket$ [\SI{}{\raiseto{2}\per\metre}] is the homogenized dislocation density jump across interfaces estimated from the modeling of plastic deformation.  At the mesoscopic scale,  it is often assumed that dislocation density of each grain under subsequent deformation ($\epsilon$) evolves according the following type equation :

\begin{eqnarray}
\partial_{\epsilon}\rho =  K_1 \rho^{\xi} - K_2 \rho,
\label{eq:YLJKM}       
\end{eqnarray}

where $K_1$ and $K_2$ are two material constants which respectively describe the strain hardening and the recovery and $\xi$ a model dependent parameter. The well-known Yoshie-Laasraoui-Jonas (YLJ) model \citep{Laasraoui1991} corresponds to $\xi=0$ and Kocks-Mecking one to $\xi=1/2$ \citep{kocks_laws_1976,mecking_kinetics_1981}. Of course, more precise numerical framework involving crystal plasticity simulations, and many intermediate models in terms of complexity, can also be considered. In context of hot metal forming and large plastic deformation, crystal plasticity finite element method (CPFEM) is often used. In CPFEM, the stress-strain response of each finite element is defined by a single crystal model following an elastoviscoplastic formulation \citep{Cuitino1993,Marin1998,Asaro1985} and a Lagrangian framework is often used to update the mesh nodes positions. In this context, the local evolution of the dislocation density is defined through the hardening rule depending of the slip rates $\dot{\gamma}_{\alpha}$, over all the slip systems $\alpha$, themselves defined through the adopted flow rule. Typically, for FCC materials, the equivalent crystal plasticity YLJ law than Eq.\ref{eq:YLJKM} to described density dislocation evolution is obtained with the following equation:
\begin{equation}\label{eq:YLJCP}
\partial_t\rho=\frac{1}{M}\left(K_1 - K_2 \rho \right)\sum_{\alpha=1}^{n}\mid\partial_t\gamma_{\alpha}\mid,
\end{equation}
with $M$ the Taylor factor.\\

In the context of diffusive solid-state phase transformations, $P$ is generally defined through the well-known Gibbs-Thomson equation applying onto interphase boundaries :
\begin{equation}\label{eq:kinetic equation sspt}
 P=\llbracket G\rrbracket + P_e + P_c,
 \end{equation}
 where $\llbracket G\rrbracket$ [\SI{}{\joule\per\metre\cubed}] is the phase transformation (PT) Gibbs free energy jump at the phase interface between the involved phases. Phase equilibrium and resulting $\llbracket G\rrbracket$  evaluation at the interface can be achieved by considering  ortho-equilibrium (equilibrium for all components) or para-equilibrium (equilibrium only for the fast diffusional species). Para-equilibrium is often adopted in FF modeling and generally $P_e$ is not taken into account. 
 
 In the context of spheroidization and coalescence of second phase objects, the kinetic equations can be summarized as \citep{poly2017introduction}:
 \begin{equation}\label{eq:kinetic equation sd}
 \mathbf{v}=-\Omega\mathbf{\nabla_{n}}\cdot \mathbf{j^{s}}\mathbf{n},
 \end{equation}
 with $\Omega$ the atomic volume and $\mathbf{j^{s}}$ the surface atom flux,
 \begin{equation}\label{eq:fluxes}
\mathbf{ j^{s}}=-\frac{\delta_s D_s\gamma_s }{kT}\nabla_n\kappa, 
 \end{equation}
 and so,
  \begin{equation}\label{eq:kinetic equation sd2}
 \mathbf{v}=\frac{\Omega\delta_s D_s\gamma_s }{kT}\nabla_{\mathbf{n}}^{2}\kappa\mathbf{n},\text{ in case of isotropic } \delta_s D_s\gamma_s\text{ field},
 \end{equation}
 with $\delta_s D_s$ the surface diffusion multiplied by the interface thickness, $k$ the Boltzmann constant and $T$ the absolute temperature.
Interestingly, classical misunderstanding in the literature often rises from reading an observable metallurgical phenomenon without considering the underlying physical mechanisms. In other words, the preceding equations quantitatively describe many observable phenomena whose observation may suggest that they have little in common. A few examples to illustrate this point are given here: Ostwald ripening phenomenon is the expression of the long time evolution of the phase equilibrium mechanism defined by Eqs.\ref{eq:global kinetic equation} and \ref{eq:kinetic equation sspt}; surface diffusion mechanism defined by Eq.\ref{eq:kinetic equation sd} and  Eq.\ref{eq:fluxes} is also of prime importance to explain sintering phenomenon \citep{Wakai2011} or grooving of GB/PB at free surface; abnormal grain growth (AGG) or critical grain growth (CGG) phenomena can be easily understood as some local heterogeneities of terms defining Eq.\ref{eq:kinetic equation ReX}; Smith-Zener pinning of GB or PB by second phase particles can be totally explained by the curvature term in Eq.\ref{eq:kinetic equation ReX}, or Eq.\ref{eq:kinetic equation sspt} respectively, and so on. A recurrent limit of existing numerical frameworks often consists to separately consider each fundamental mechanism despite their concomitant nature and their impact on each other in real materials.  Of course, in all the previous discussed mechanisms, the plasticity due to the hot deformation but also potentially induced by the transformation themselves can be finely evaluated/predicted in FE framework and taken into account in the previous migration equations through the terms $\llbracket\rho\rrbracket$.\medbreak
 
If Eqs.\ref{eq:global kinetic equation} to \ref{eq:kinetic equation sd2} represent classical description of a large amount of metallurgical phenomena at the mesoscopic scale during metal forming (i.e. correspond to the prime importance driving pressures), they are nonetheless approximations of lower scale mechanisms. Thus, if these kinetic equations and the description of the corresponding physical mechanisms are indeed in constant evolution and improvement at the grain interface scale \citep{Chen2020,Bhattacharya2021,Florez2022}, they constitute at the polycrystalline scale and in the metal forming state-of-the-art a thermodynamically/kinematically justified and validated homogenized physical framework. Moreover, it can be easily parameterized through homogenization of dynamic molecular simulation results concerning the description of the interface properties.

\section{Level-set function, description of polycrystalline microstructure  and meshing adaptation}\label{sec:LS}

The LS method was firstly introduced by Dervieux and Thomasset \citep{dervieux2006} under the terminology of "pseudo-density function" in 1979 before to be extensively formalized and disseminated by the works of Osher and Sethian \citep{Osher1988} as a numerical tool to trace the spatial and temporal evolution of interfaces, enhanced later for curvature flow problems with multiple junctions \citep{Merriman1994, Zhao1996} and applied to recrystallization and grain growth in \citep{Bernacki2008,Loge2008,Bernacki2011}. The principle for modeling polycrystals is to deal with a front-capturing description of grains $G$ through LS functions $\psi$ in the space $\Omega$  :

\begin{align}
  \left\{
  \begin{array}{l}
    \psi\left(\mathbf{x},t\right) = \pm d\left(\mathbf{x},\Gamma\left(t\right)\right), \quad \mathbf{x} \in \Omega, \quad \Gamma\left(t\right)=\partial G \\
    \psi\left(\mathbf{x},t \in \Omega\right) = 0 \quad \Longleftrightarrow
 \quad \mathbf{x} \in \Gamma\left(t\right),
\end{array}
\label{eqn:LS}
\right .
\end{align}

\begin{figure}[ht!]
  \centering
  \includegraphics[scale=0.5]{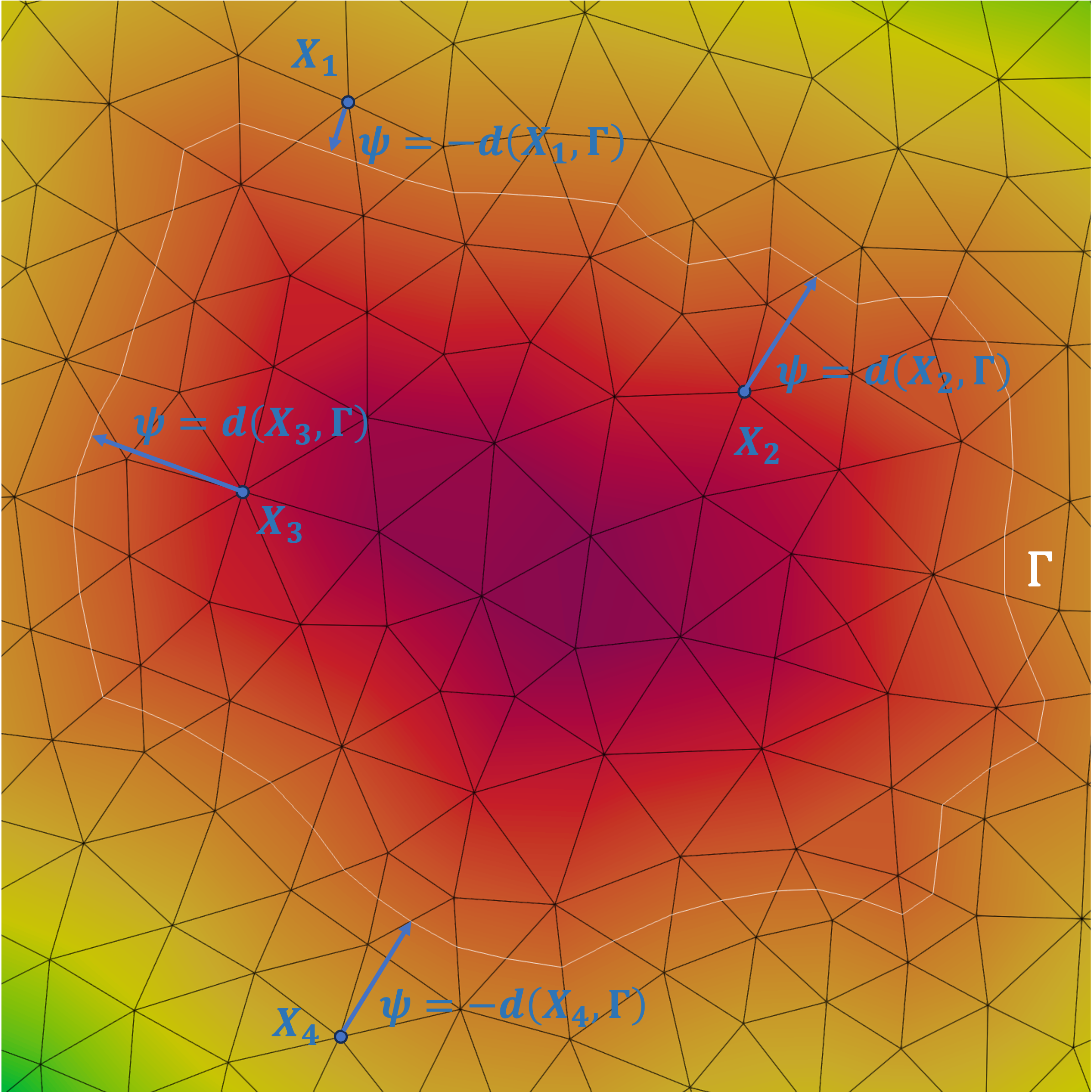}
  \caption{LS field describing a grain in an unstructured FE mesh (in black) in context of a P1 interpolation. The white interface corresponds to the 0-isovalue of the LS, i.e. the grain interface $\Gamma$. }\label{fig:LSFunction}
\end{figure}

More precisely each sub-domain $G$ (grain) in a given domain $\Omega$ (polycrystal) is classically described implicitly by computing the signed distance function $\psi\left(\mathbf{x},t\right)$ representing the distance to the sub-domain boundaries $\Gamma\left(t\right)=\partial G $ (grain boundaries). In a P1 (linear) interpolation, the function $\psi\left(\mathbf{x},t\right)$ is calculated at each node on the FE mesh and is often chosen, by convention, positive inside of the grain and negative outside as illustrated in Fig.\ref{fig:LSFunction}. This choice is also in line with the equations described in the previous section for grain interface properties (see Eq.\ref{eq:eqdtokappa}). 

Using such an implicit description simplifies the polycrystal description as it has not to be strictly correlated to a discretization of the grain/phase boundary interfaces. The experimental image can be seamlessly integrated into a finite element (FE) mesh by calculating distance functions to the grayscale values at each grain interface. Alternatively, it can be evaluated directly within a regular grid that corresponds to the original experimental data. Moreover, when only statistical data, such as average grain size or grain size distribution, is known, we must construct a digital material model based on representative microstructures  \citep{Rollett2004b}. Voronoi or Laguerre-Voronoi tessellations (LVT) are methodologies that can be utilized to generate digital polycrystals made of polyhedral grains, while maintaining adherence to the known data. The Voronoi Tessellation Method (VTM) generates random Voronoi nuclei, and defines each Voronoi cell as the space that is closer to a specific nucleus than to any other nucleus. Despite the accurate geometric correlation between Voronoi tessellations and many cellular structures, VTM has some limitations, one of which is the inability to adhere to a specific statistical volume distribution of cells. Xu and Li \citep{Xu2009} identified discrepancies between the statistical properties of grains commonly observed in an equiaxial polycrystal and the results derived from VTM. One approach to enhance traditional VTM is to use the Laguerre-Voronoi Tessellation Method (LVTM), which assigns a radius or weight to each nucleus and reflects this distribution in the Voronoi tessellation \citep{Imai1985}. It's worth noting that the main challenge of LVTM is similar to the challenge in creating polycrystal microstructures and powder representative volume elements (RVEs), which is to reflect a given statistical size of spheres with the maximum achievable density. This criterion is not only essential for powder RVEs but also for LVTM in order to minimize disparities between the assigned weight and the resulting final volume of the Voronoi cells. Although often not clearly discussed in literature, maintaining the radius size distribution for Laguerre-Voronoi sphere packing is distinct from maintaining the grain size distribution for the resulting Laguerre-Voronoi cells. Different strategies have been developed over the past 40 years \citep{Jodrey1985, He1999, Benabbou2009, Bagi1993}, collectively referred to as Sphere Packing Methods (SPM). These methods are typically categorized into two main types: sequential addition models and collective rearrangement models. Both types are employed in the generation of polycrystals \citep{Hitti2012, Ilin2016a, Ilin2016b, Quey2018, Depriester2019}.

In addition to the morphology of the grains or and the phases composition, other attributes playing on the properties of the interfaces or the driving pressures in play can be sought to be respected in an exact or statistical way.  The two quantities of interest, the GB energy $\gamma$ and GB mobility $\mu$, must then be seen, as detailed in section \ref{sec:gb}, as functions from the GB space $\mathcal{B}$ to $\mathbb{R}^{+}$. Thus, when EBSD maps are available, they can be used to define exactly the misorientation map; 3D data allowing to add a complete determination of the $\mathbf{n}_{ij}$ inclination parameter. When these information are missing, a classical strategy consists to use a random grain orientation leading to a Mackenzie-like disorientation distribution function (DDF) \citep{mackenzie_second_1958}.

Theoretically, each grain of a single-phase or multi-phase polycrystal should be represented by its own LS function. To decrease computation time and memory usage, non-adjacent grains in the initial microstructure—separated by a certain number of grains $\delta$ or a minimum distance—can be grouped to form Global Level Set (GLS) functions using graph coloring techniques. However, this approach makes it impossible to distinguish between grains that share the same GLS function. Consequently, when two child grains of a GLS grow and meet, numerical coalescence occurs—meaning the grains merge to form one grain.  Various strategies can be employed to prevent or minimize numerical coalescence events: for instance, selecting a small initial separation $\delta$ or minimum distance to limit computation time while also minimizing coalescence  \citep{Fabiano2014}, performing complete optimal coloring at each time step, or considering re-coloration algorithms at each time step to manage risky configurations \citep{KrillIII2002,Elsey2009,Scholtes2015}.

In the context of polyhedral cells, FE meshing of digital microstructures is typically straightforward. It generally involves discretizing the facets of cells and then the volume within each cell \citep{Quey2011}. For real polycrystals observed through Scanning Electron Microscopy (SEM) or 3D X-ray imaging techniques \citep{Ludwig2009, Proudhon2016}, approximate Voronoi/Laguerre-Voronoi meshing is not ideal. Though a substantial amount of research has been done on meshing methods for real 3D microstructures \citep{Young2008, Zhang2005}, applying these methods to complex topologies is not an easy task \citep{Rollett2004b, Brahme2006, sinchuk_x-ray_2022}. The main difficulty lies in dealing with multiple junctions, where balancing respect for experimental data and achieving a high-quality mesh can be complex.

Once a mesh has been generated, modeling large plastic strains and subsequent microstructure evolutions within the FE framework is another significant challenge. Many researchers opt to avoid front-tracking algorithms, where grain boundaries are explicitly meshed, and instead utilize implicit interface methods such as LS and MPF. While results using explicit interface methods are either limited to minor deformations or rely on full reconstruction of the computational mesh at each time step \citep{Hallberg2013}, implicit interface methods allow for a wider range of metallurgical phenomena with large deformations \citep{Bernacki2009,Scholtes2015,Scholtes2016,Maire2017}.

However, this strategy typically demands fine FE meshes at grain interfaces to achieve acceptable accuracy in relation to the driving pressures under consideration. Though global isotropic mesh refinement or high-order interpolation of LS functions can be used to attain desired accuracy in the interface description, these methods substantially increase computational resources. Therefore, adaptive local isotropic or anisotropic remeshing is generally preferred. There are various methods for generating locally adapted meshes to zero-isovalue of LS, but typically, metric field and topological meshers based on local mesh topology optimizations are used \citep{Gruau2005,Coupez2000,Shakoor2015a}.

Regarding metric calculations, the most common approach is to use a posteriori error analysis to obtain an optimal mesh for a given number of nodes \citep{Almeida2000, Coupez2011, Mesri2016}. This approach can be extended for situations where a significant number of LS functions must be considered. For instance, when describing strictly disjoint objects with LS functions, one simple solution is to adapt the mesh based on a posteriori error estimator to $\psi_{max}(\mathbf{x},t)=\max\limits_{i=1,...,N_{LS}} \psi_i(\mathbf{x},t)$ with $N_{LS}$ being the number of LS functions used. However, this approach, used in \citep{Scholtes2015, Maire2016}, is not straightforward when the objects considered are not strictly disjoint, which is the case in polycrystalline microstructures, as the  $\psi_{max}$ function is then not derivable at the grain boundary network. 

\begin{figure}[ht!]
  \centering
  \begin{subfigure}{0.49\textwidth}
    \centering
    \includegraphics[scale=0.27]{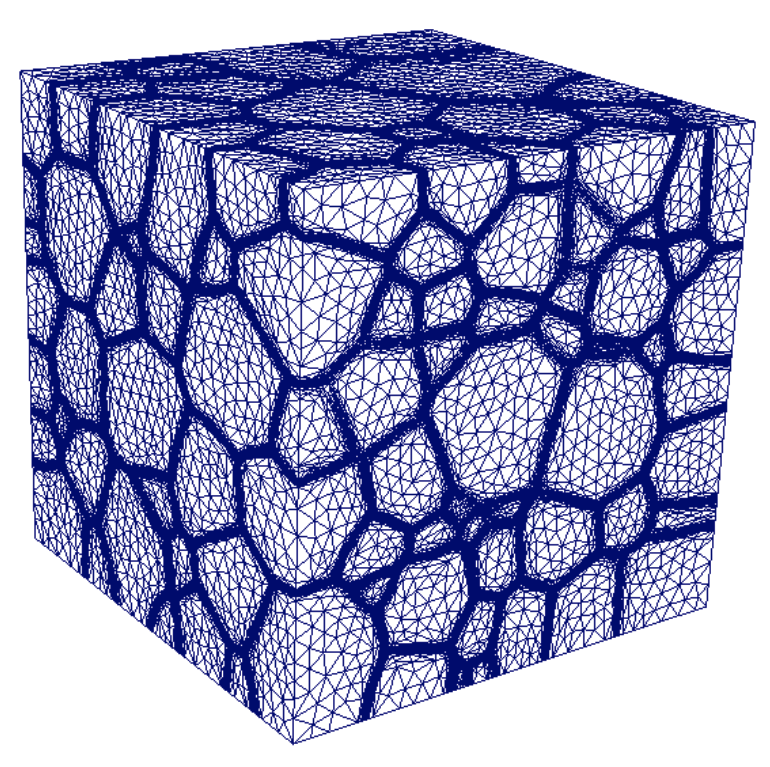}
    \caption{}
    \label{fig:Mesh3D}
  \end{subfigure}
  \begin{subfigure}{0.49\textwidth}
    \centering
    \includegraphics[scale=0.35]{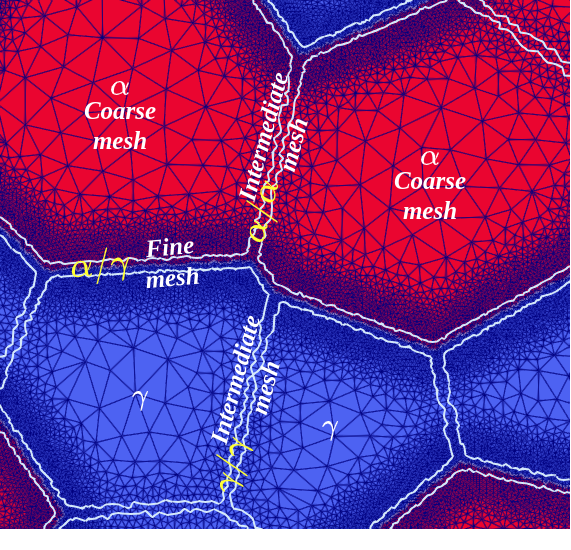}
    \caption{}
    \label{fig:MeshMultiPhase}
  \end{subfigure}
   \begin{subfigure}{1\textwidth}
    \centering
    \includegraphics[scale=0.4]{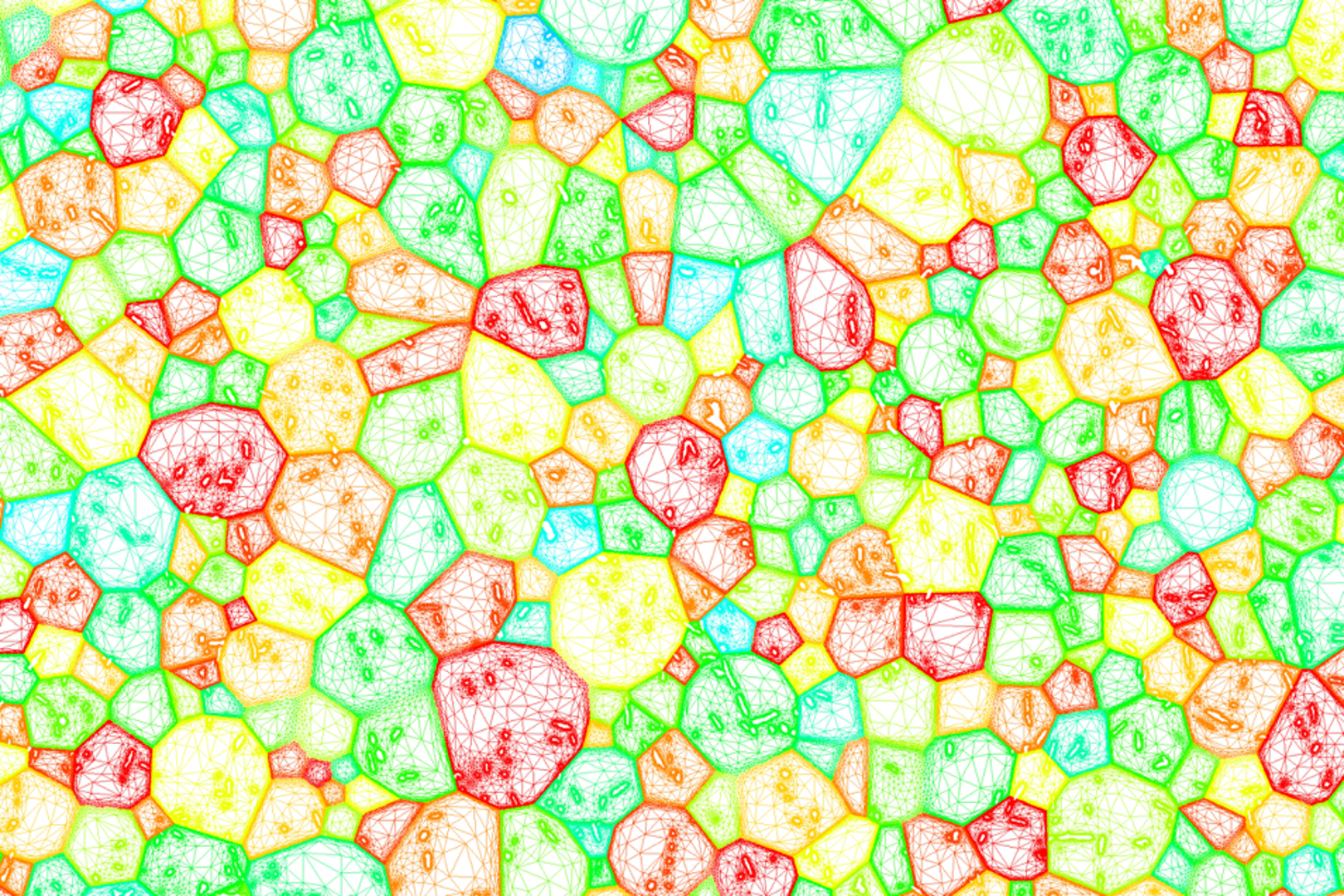}
    \caption{}
    \label{fig:Mesh2D}
  \end{subfigure}
  \caption{(a) 3D equiaxed microstructure example where where local anisotropic metric is considered for the FE mesh generation thanks to a posteriori approach \citep{Scholtes2015}. The mesh is adapted to the grain interfaces (generated statistically with a Voronoï tessellation), (b) A 2D two-phase cases where an isotropic strategy is used with a mesh refinement in grain and phase interfaces but with an increased precision in the phase interface \citep{Chandrappa2023}, and (c) 2D Inconel 718 microstructure example where local anisotropic metric is considered for the FE mesh generation thanks to a geometric approach \citep{Bernacki2009}. The mesh is adapted to the grain interfaces (generated statistically with a Laguerre-Voronoï approach) and to $\delta$ phase interfaces immersed thanks to a SEM image. The color code corresponds to the GLS functions. }
  \label{fig:Mesh}
\end{figure}

Alternatively, automatic geometric methods can be employed for creating locally refined isotropic or anisotropic meshes tailored to polycrystals. This is based on the normal and/or mean curvature of the grain interfaces  \citep{Hitti2012,Bernacki2009}. Fig.\ref{fig:Mesh3D} shows a 3D case for 304L stainless steel where an a posteriori metric is adopted \citep{Scholtes2015}. Fig.\ref{fig:MeshMultiPhase} exhibits a 2D two-phase case where an isotropic strategy is applied. In this case, the mesh refinement occurs in both grain and phase interfaces, but precision is increased at the phase interface \citep{Chandrappa2023}. Fig.\ref{fig:Mesh2D}  presents a 2D example where an anisotropic metric is considered using a geometric approach \citep{Bernacki2009} for an Inconel 718 microstructure. The mesh is adapted to the grain interfaces (statistically generated using a Laguerre-Voronoi approach) and to second phase particle (SPP) interfaces, which were immersed after thresholding a SEM image.

\section{Classical isotropic framework for LS modeling of GG and ReX}

In the LS method, the evolution of $\psi(\mathbf{x},t)$, submitted to a velocity field $\mathbf{v}(\mathbf{x},t)$ is then given by the following convective partial differential equation \citep{dervieux2006, Osher1988}:
\begin{equation}\label{eq:Transport}
\left \{
\begin{aligned}
    \partial_t \psi\left(\mathbf{x},t\right) + \mathbf{v}\left(\mathbf{x},t\right) \cdot \nabla \psi\left(\mathbf{x},t\right)=0\\
    \psi\left(\mathbf{x},t=0\right)=\psi^0\left(\mathbf{x}\right)
\end{aligned}
\right.
\end{equation}
with $\psi^0\left(\mathbf{x}\right)$ the initial description of the LS function. 

\subsection{Modeling of isotropic grain growth}

In context of deterministic full-field approaches and neglecting torque terms \citep{Fausty2018}, the velocity field can be defined using Eq.\ref{eq:global kinetic equation} and when GG is involved, the net pressure is classically defined using Eq.\ref{eq:herring6}.  The isotropy hypothesis remains here, for the LS simulations, to consider $\gamma$ as constant and $\mu$ as only dependent of the temperature through an Arrhenius law $\mu\left(T\right)=\mu_0\exp\left(-Q/RT\right)$ with $\mu_0$ a pre-exponential constant parameter, $R$ the gas constant and $T$ the absolute temperature. 
$\kappa$ and $\mathbf{n}$ can be defined naturally by taking advantages of the possibilities offered by the LS framework. Indeed, considering that GLS functions remain distance functions all along the simulation (i.e. that $\lVert \nabla \psi \rVert = 1$), they can be defined as detailed in Eq.\ref{eq:eqdtokappa} by:

\begin{equation}\label{eq:NormalGB}
\mathbf{n} = -\nabla \psi,\quad \kappa = - \Delta \psi. \\
\end{equation}

Finally, by using coloring/recoloring algorithms \citep{Scholtes2015} and the described metric properties of the LS functions, one can solve after substituting Eqs.\ref{eq:NormalGB} into Eq.\ref{eq:herring6} and Eq.\ref{eq:herring6} into Eq.\ref{eq:global kinetic equation}, a set of $N_{GLS}$ diffusive equations as detailed by Eq.\ref{eq:Transport1}, with $N_{GLS}\ll N_G$ and $i \in \{1,2,...,N_{GLS}\}$. The numerical strategy consisting in limiting the number of involved LS functions is then generally crucial in terms of numerical cost and memory aspect.

\begin{equation}\label{eq:Transport1}
\left \{
\begin{aligned}
    \partial_t \psi _i\left(\mathbf{x},t\right) - \mu\gamma \Delta \psi _i\left(\mathbf{x},t\right)=0,\\
    \psi _i\left(\mathbf{x},t=0\right)=\psi _i^0\left(\mathbf{x}\right)
\end{aligned}
\right.
\end{equation}

The field $\psi _i^0\left(\mathbf{x}\right)$ is generally defined as the signed distance function to the union of the grains initially present in the $i$th GLS function, i.e. $$
\psi _i^0\left(\mathbf{x}\right)=\max_{j\ |\ G_j\in GLS_i}\left(d_j\left(\mathbf{x}\right)\right)\text{, if } \mathbf{x}\in\bigcup_{j\ |\ G_j\in GLS_i}G_j \text{ and the opposite otherwise}.
$$
Generally, a particular numerical treatment must be imposed to avoid kinematic incompatibilities after the resolution of Eq.\ref{eq:Transport1}. Indeed, voids can appear at multiple junctions and must be treated. A classical solution consists in correcting the GLS functions as follows \citep{Merriman1994}:

\begin{equation}
\label{eq:vacuum}
\tilde{\psi} _i\left(\mathbf{x},t\right) = \frac{1}{2}\left(\psi _i\left(\mathbf{x},t\right) - \max_{j\neq i} \psi _j\left(\mathbf{x},t\right)\right),\ \ 1\leq i\leq N_{GLS} .
\end{equation} 

\begin{figure}[ht!]
  \centering
  \includegraphics[scale=0.4]{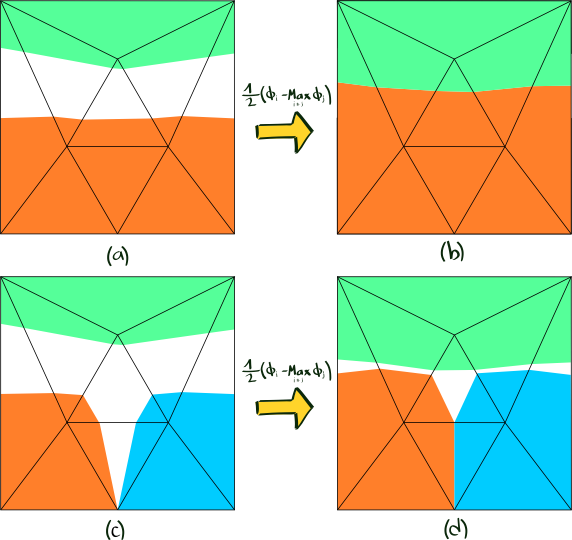}
  \caption{Global treatment to eliminate non-physical vacuum regions on a FE discretization: two colored LS (a) with a vacuum region in between them, (b) result after applying Eq. (9). Three colored LS : (c) with  a vacuum region (d) result after applying Eq.\ref{eq:vacuum}. Figure from \citep{Florez2020}.}\label{fig:vacuum}
\end{figure}

The effect of Eq.\ref{eq:vacuum} is schematized in Fig.\ref{fig:vacuum} for a $P1$ interpolation. An alternative could also consist \citep{Zhao1996} in closing void regions thanks to an energy minimization principle enforced by a Lagrange multiplier related to a constraint added to Eq.\ref{eq:Transport}.

A drawback of the LS formulation lies in the fact that after the resolution of Eqs.\ref{eq:Transport1} and \ref{eq:vacuum}, the GLS are no longer distance functions $\lVert \nabla \tilde{\psi} \rVert \neq 1$. This is particularly problematic when a specific remeshing technique depending on the distance property is used at the interface as described in the previous section. In addition, the diffusive formulation proposed in Eq.\ref{eq:Transport1} requires a distance function at least in a thin layer around the interface in order to model properly the curvature driven mechanism. Finally, the conditioning of the transport problem also depends on the regularity of the LS function \citep{Bernacki2009}. For these reasons, the GLS functions need to be reinitialized (or redistancing). Restoring the metric property at the instant $t$ is equivalent to solving the following eikonal equation for each GLS function, i.e. $\ \forall i \in \{1,2,...,N_{GLS}\}$:

\begin{equation}
\begin{cases}
\lVert \nabla \psi_i \rVert = 1, \ \forall \mathbf{x}\in\Omega \\
\psi_i\left(\mathbf{x},t\right) = \tilde{\psi}_i\left(\mathbf{x},t\right) = 0,\ \forall \mathbf{x} \in\tilde{\Gamma}_i\left(t\right)
\end{cases}	\Longleftrightarrow  \psi_i\left(\mathbf{x},t\right) =\ Redist\left(\tilde{\psi}_i\left(\mathbf{x},t\right)\right),
\label{eq:eikonal}
\end{equation}

Different approaches exist to solve this equation including the well-known Fast Marching Method introduced by Sethian \citep{Sethian1996} which propagates a front from the interface and ensures directly a $L2$ gradient norm equals to unity. Though this approach has been extended to unstructured meshes \citep{Kimmel1998}, its implementation becomes complicated when it comes to consider anisotropic triangulations \citep{Sethian2003} and parallel efficiency is poor.
In \citep{Sussman1994}, a Hamilton–Jacobi (H–J) formulation equivalent to Eq.\ref{eq:eikonal} was proposed in order to correct iteratively the LS values around the interface by solving a partial differential equation (PDE). This method is massively used in the LS modeling of ReX and GG \citep{Agnoli2014a,Fabiano2014,Bernacki2011,Hallberg2019}. It requires the definition of a purely numerical parameter known as a fictitious time step for reinitialization and the ratio between the desired reinitialized thickness and this parameter gives the number of required increments.
Coupled convection-reinitialization (CR) methods emerged wherein the LS function is automatically reinitialized during the resolution of the transport equation \citep{Coupez2007,Bernacki2008,Bernacki2009}. Their main advantage lies in the fact that only one solver is needed for the simulation instead of two for the classical H–J technique. The signed distance function can also be replaced by any smooth function which satisfies the metric property, at least in a thin layer around the interface. 
Finally, a natural way to reinitialize GLS functions consists in using a brute force algorithm to perform a complete reconstruction of the distance function. This technique works in two steps: discretize the interface (0-isovalue of the LS function) into a collection of simple elements and, for every integration point, compute the distance to all elements of the collection and store the smallest value which becomes the updated value of the distance function. Though it guarantees optimal accuracy, this Direct Reinitialization (DR) technique is generally reported as being greedy in terms of computational requirements \citep{Sussman1994,Elias2007,Jones2006}. It is nevertheless worth mentioning that these works generally address only regular grids or hierarchical meshes \citep{Fortmeier2011}. Few years ago, a direct fast and accurate approach usable in unstructured FE mesh has been proposed \citep{Shakoor2015b}. This method takes advantage of a space-partitioning technique using $k-d$ tree and an efficient bounding box strategy enabling to maximize the numerical efficiency for parallel computations. Discussions concerning the residual errors inherent to this approach are also proposed in \citep{Florez2020}. 

Interestingly, a similar flow rules than the previously described one can be followed when dealing with the modeling of pure grain growth in regular grids where signed distance functions to interfaces are convolved with Gaussian kernels to generate a variety of geometric motions, including multi-phase motion by mean curvature \citep{Elsey2009,Elsey2011,Esedoglu2010}. First developments concerning this approach was proposed by Esedoglu, Elsey and coworkers \citep{Elsey2009, Esedoglu2010}. To be more precise, let $d^k_i\left(x\right)$ be the signed distance function at times $kdt$ of the $i$th grain, where $dt$ is the chosen time step. The solution of the curvature flow equation (Eq.\ref{eq:Transport1} with a dimensionless reduced mobility equal to one) for the following time step with $d^k_i$ an initial condition for grain $G_i$ can be obtained as a solution of the heat equation:
\begin{equation}\label{eq:convolution1}
\tilde{d}^{k+1}_i\left(\mathbf{x}\right)=\left(G\star d^{k}_i\right)\left(\mathbf{x}\right),\text{ with } G\left(\mathbf{x}\right)=\frac{1}{\left(4\pi dt\right)^{dim/2}}e^{-\lVert \mathbf{x}\rVert^{2}/4dt},\text{ and }dim\text{ the space dimension },
\end{equation}
\begin{equation}\label{eq:convolution2}
\tilde{\tilde{d}}^{k+1}_i\left(\mathbf{x}\right)=\frac{1}{2}\left(\tilde{d}^{k+1}_i\left(\mathbf{x}\right) - \max_{j\neq i} \tilde{d}^{k+1}_j\left(\mathbf{x}\right)\right),\ \ 1\leq i\leq N_{G}
\end{equation}
\begin{equation}\label{eq:convolution3}
d^{k+1}_i\left(\mathbf{x}\right) = Redist\left(\tilde{\tilde{d}}^{k+1}_i\left(\mathbf{x}\right)\right),\ \ 1\leq i\leq N_{G}.
\end{equation}
The direct resolution of Eq.\ref{eq:convolution1} obviously allows for unbeatable computation times compared to the finite element or finite difference resolution of the isotropic grain growth problem in 2D or 3D.  Therefore, much more massive calculations can be proposed \citep{Elsey2011}. In this case, memory management can become a problem, which explains also the possible use of coloration/recoloration in order to minimize the number of LS function \citep{Elsey2009}. In this case, equations \ref{eq:convolution1} to \ref{eq:convolution3} are then applied for $i$ ranging from $1$ to $N_{GLS}$. The parallelization of existing formulations has also been considered \citep{Elsey2013,Miessen2015,miessen2017highly}. Another expression of $G$ can also be used to improved the approximation of $\kappa$ or to introduce new terms to the kinetic equation \citep{Esedoglu2010}. By the nature of these approaches, based on convolution products on regular grids for solving the heat equation, they have been little extended to the modeling of other mechanisms or enriched in recent years. Any exceptions are cited subsequently.

\subsection{Modeling of isotropic recrystallization}
When plastic stored energy as to be considered, the kinetic equation described in Eq.\ref{eq:kinetic equation ReX} applies and the system defined by Eq.\ref{eq:Transport} can be rewritten in Eq:\ref{eq:Transport2}:

\begin{align}\label{eq:Transport2}
\left \{
\begin{array}{l}
\partial_t \psi _i\left(\mathbf{x},t\right) - \mu\gamma \Delta \psi _i\left(\mathbf{x},t\right) + \mathbf{v}^{\llbracket\rho\rrbracket}_{i} \cdot \nabla \psi _i\left(v,t\right)=0\\
\mathbf{v}^{\llbracket\rho\rrbracket}_{i}=\mu\tau \llbracket\rho\rrbracket_{i}\mathbf{n}_{i},\quad \psi _i\left(\mathbf{x},t=0\right)=\psi _i^0\left(\mathbf{x}\right).
\end{array} 
\right.
\end{align}

By extrapolating the shape of the velocity term, $\mathbf{v}^{\llbracket\rho\rrbracket}_{i}$ of Eq.\ref{eq:Transport2}, for the interface between grain $G_i$ and $G_j$, the corresponding velocity $\mathbf{v}^{\llbracket\rho\rrbracket}_{ij}$  can be written as: 
\begin{equation}
\label{eq:energybetweengrains}
\mathbf{v}^{\llbracket\rho\rrbracket}_{ij} = \mu\tau \llbracket\rho\rrbracket_{ij}\mathbf{n}_{ij}.
\end{equation} 

Classically in LS or MPF approaches, a constant stored energy is considered in each grain \citep{Bernacki2008,Bernacki2009,Loge2008,Scholtes2016,Hallberg2013,Elsey2009,Maire2017}. Then, for each interface between grain $G_i$ and $G_j$,  it is assumed that:
\begin{equation}
\label{eq:energybetweengrains2}
\llbracket E\rrbracket_{ij}=\tau\llbracket\rho\rrbracket_{ij} = e_j - e_i,
\end{equation} 
where $e_i$ and $e_j$ are the mean stored energies in the grains $G_i$ and $G_j$, respectively. These averages can directly come from either constant approximative values where only a gradient of the stored dislocations between the nuclei and the non-recrystallized grains is considered as in \citep{Bernacki2008,Bernacki2011,Elsey2009}, or simplified mechanical formulations as in \citep{Maire2017}. They can also be evaluated thanks to an existing dislocation field in the FE mesh of the calculation domain $\Omega$ as in \citep{Bernacki2009,Loge2008,Hallberg2013,Scholtes2016,Ruiz2020a,Ruiz2020b} or come from a dislocation density field measured from experimental data, immersed in the FE mesh and averaged per grain  as in \citep{Agnoli2015,Villaret2020,Grand2022}. More local approximations of the energetic field can also be considered \citep{Ilin2018,grand_modeling_2023}.

\begin{figure}[h!]
  \centering
  \begin{subfigure}{0.49\textwidth}
    \centering
    \includegraphics[scale=0.255]{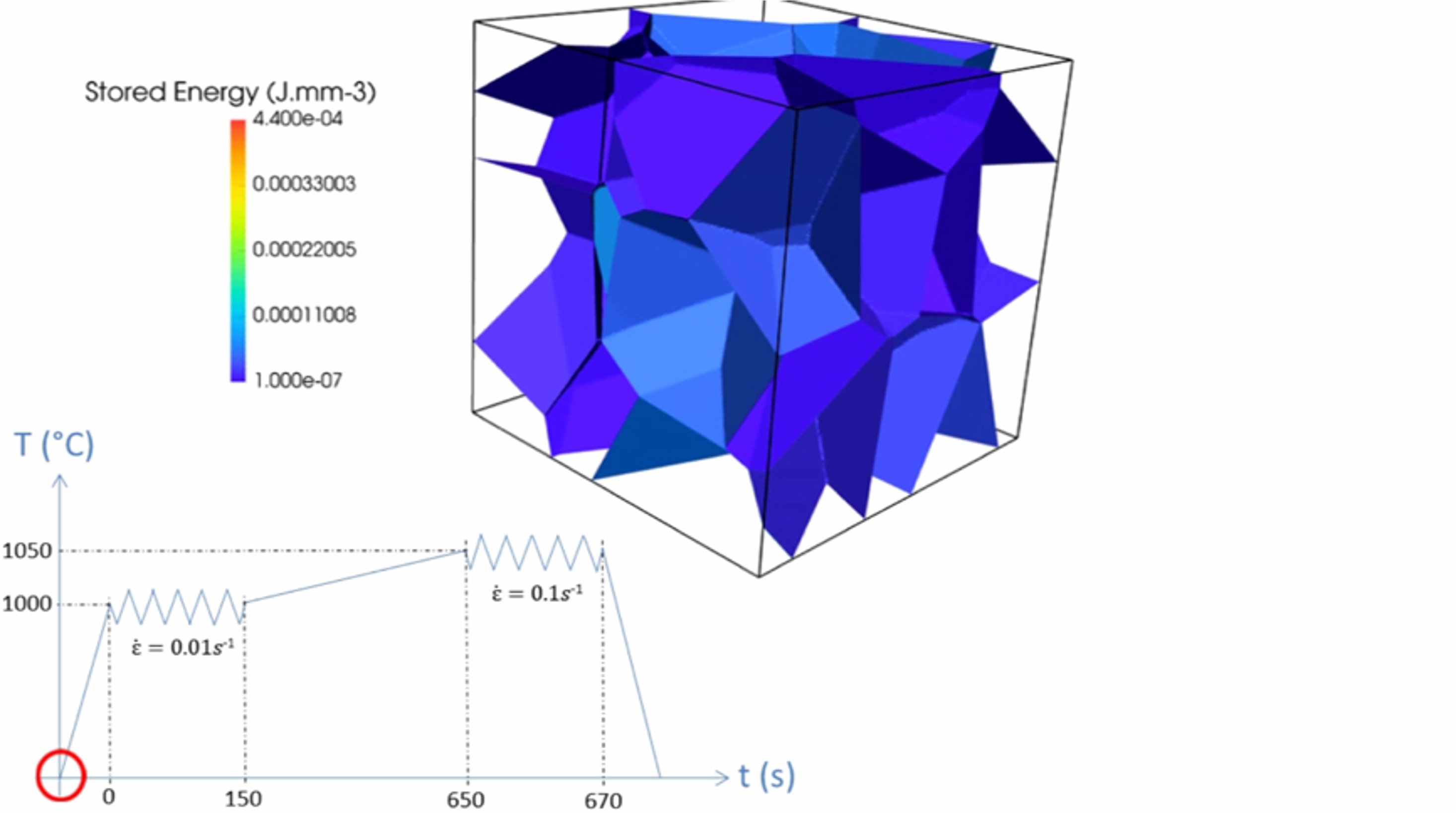}
    \caption{}
    \label{fig:DDRXa}
  \end{subfigure}
  \begin{subfigure}{0.49\textwidth}
    \centering
    \includegraphics[scale=0.255]{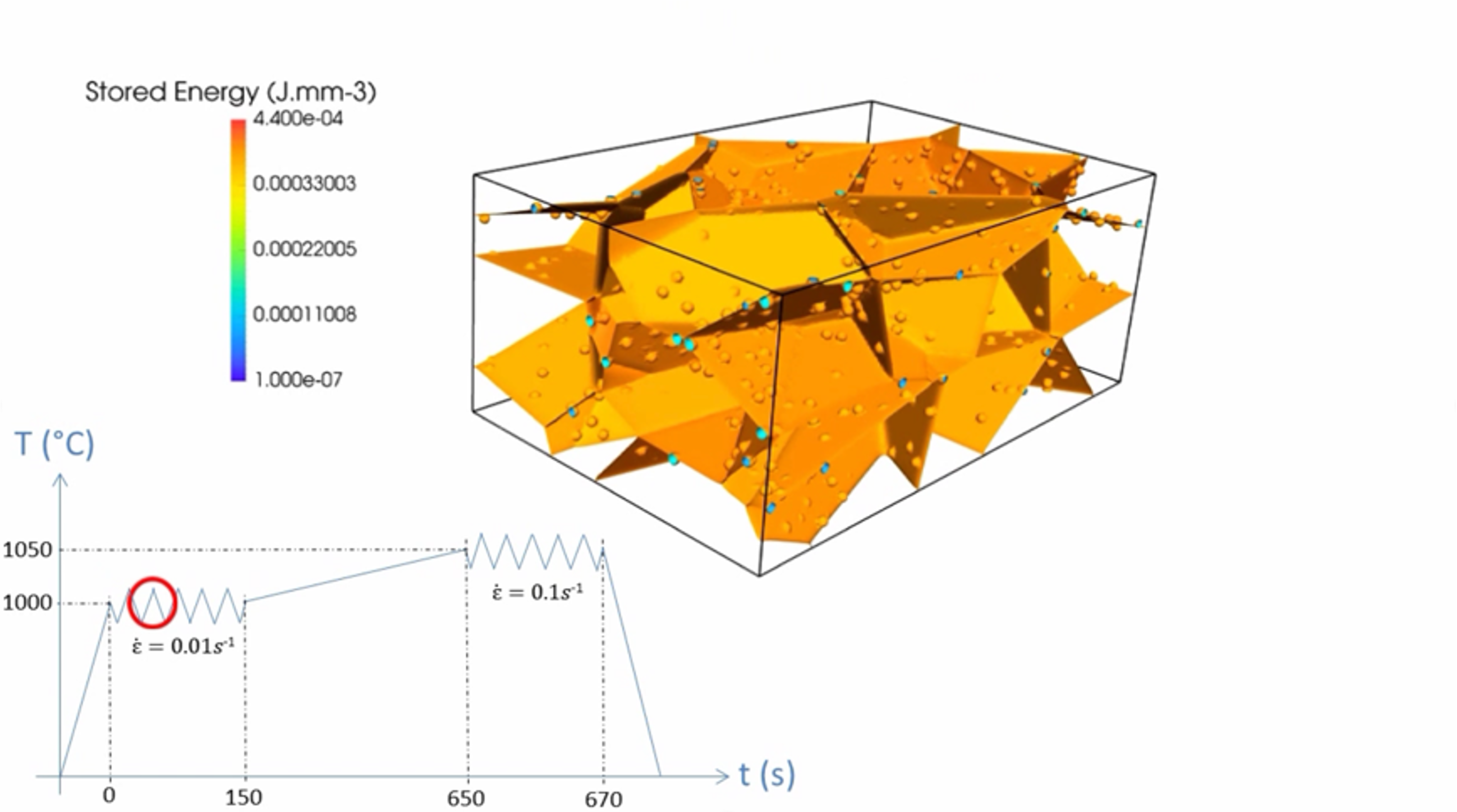}
    \caption{}
    \label{fig:DDRXb}
  \end{subfigure}
  \begin{subfigure}{0.49\textwidth}
    \centering
    \includegraphics[scale=0.187]{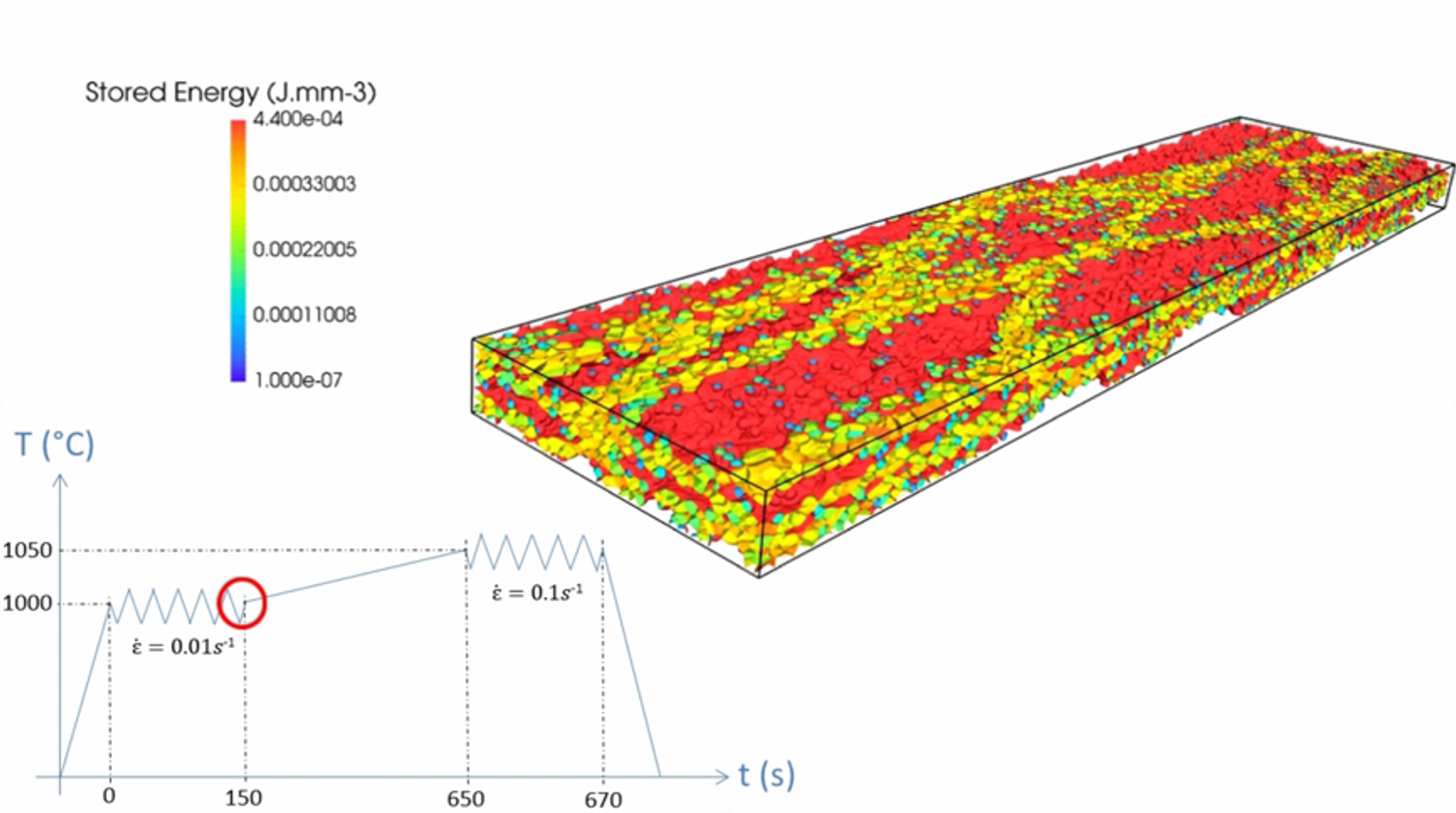}
    \caption{}
    \label{fig:DDRXc}
  \end{subfigure}
   \begin{subfigure}{0.49\textwidth}
    \centering
    \includegraphics[scale=0.187]{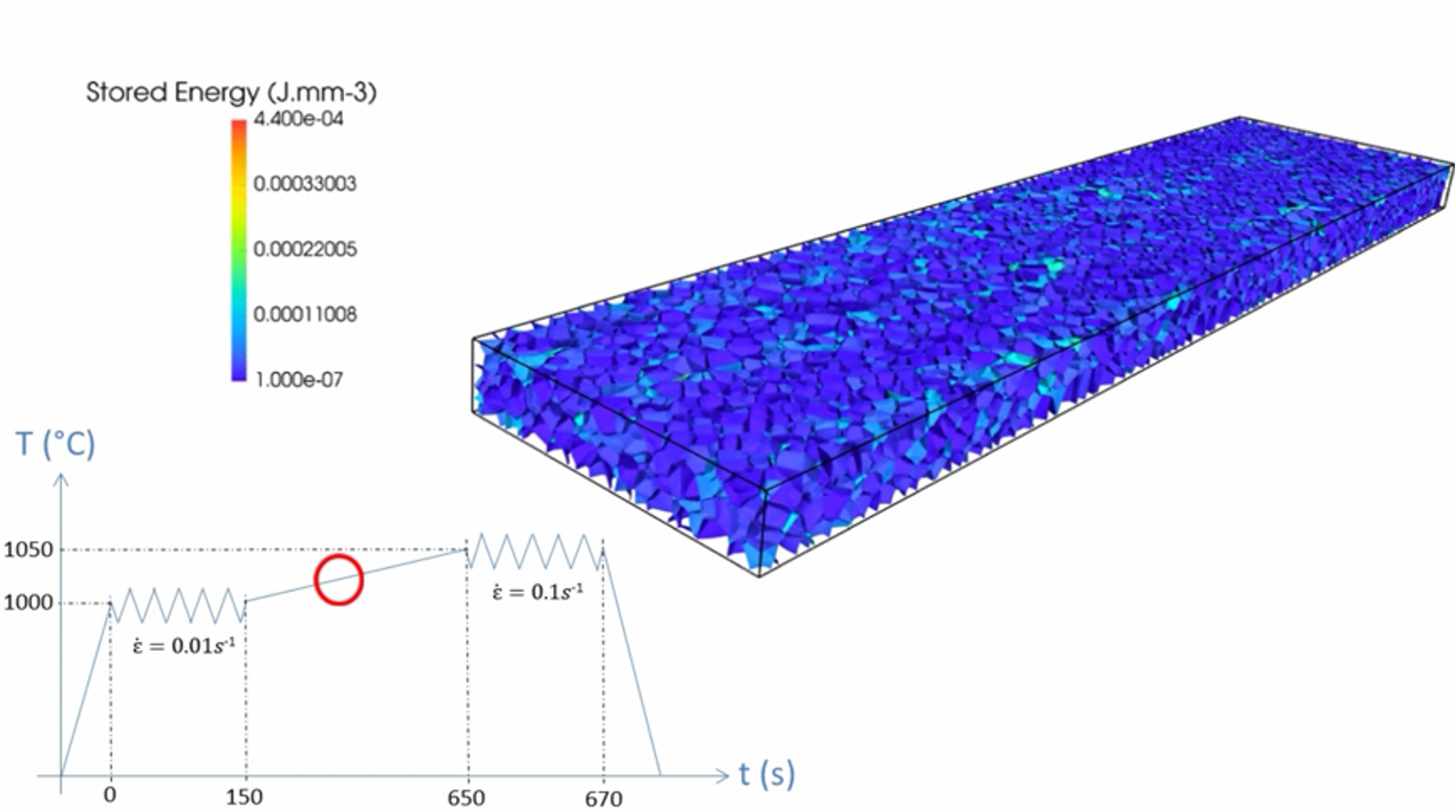}
    \caption{}
    \label{fig:DDRXd}
  \end{subfigure}
  \begin{subfigure}{0.49\textwidth}
    \centering
    \includegraphics[scale=0.187]{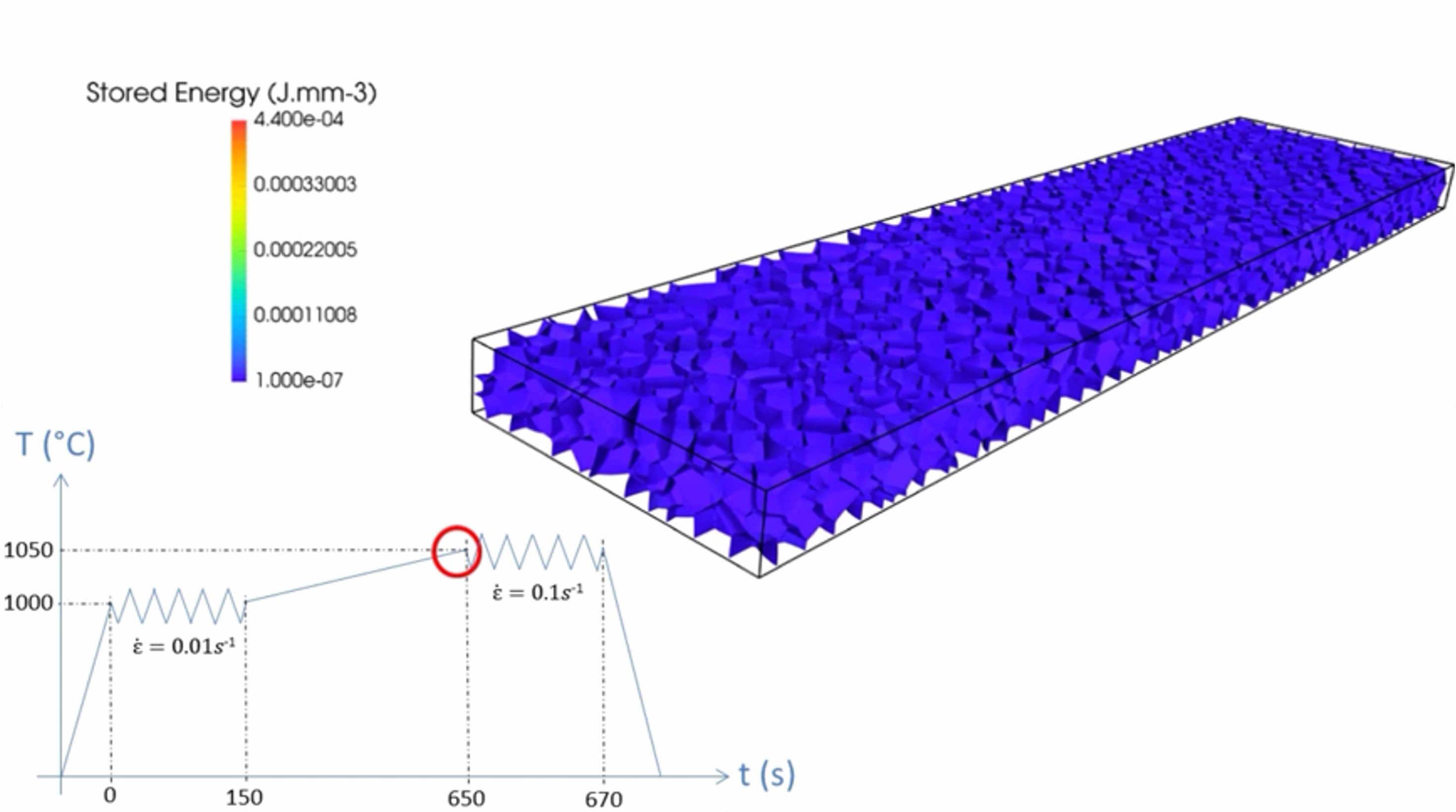}
    \caption{}
    \label{fig:DDRXe}
  \end{subfigure}
  \begin{subfigure}{0.49\textwidth}
    \centering
    \includegraphics[scale=0.187]{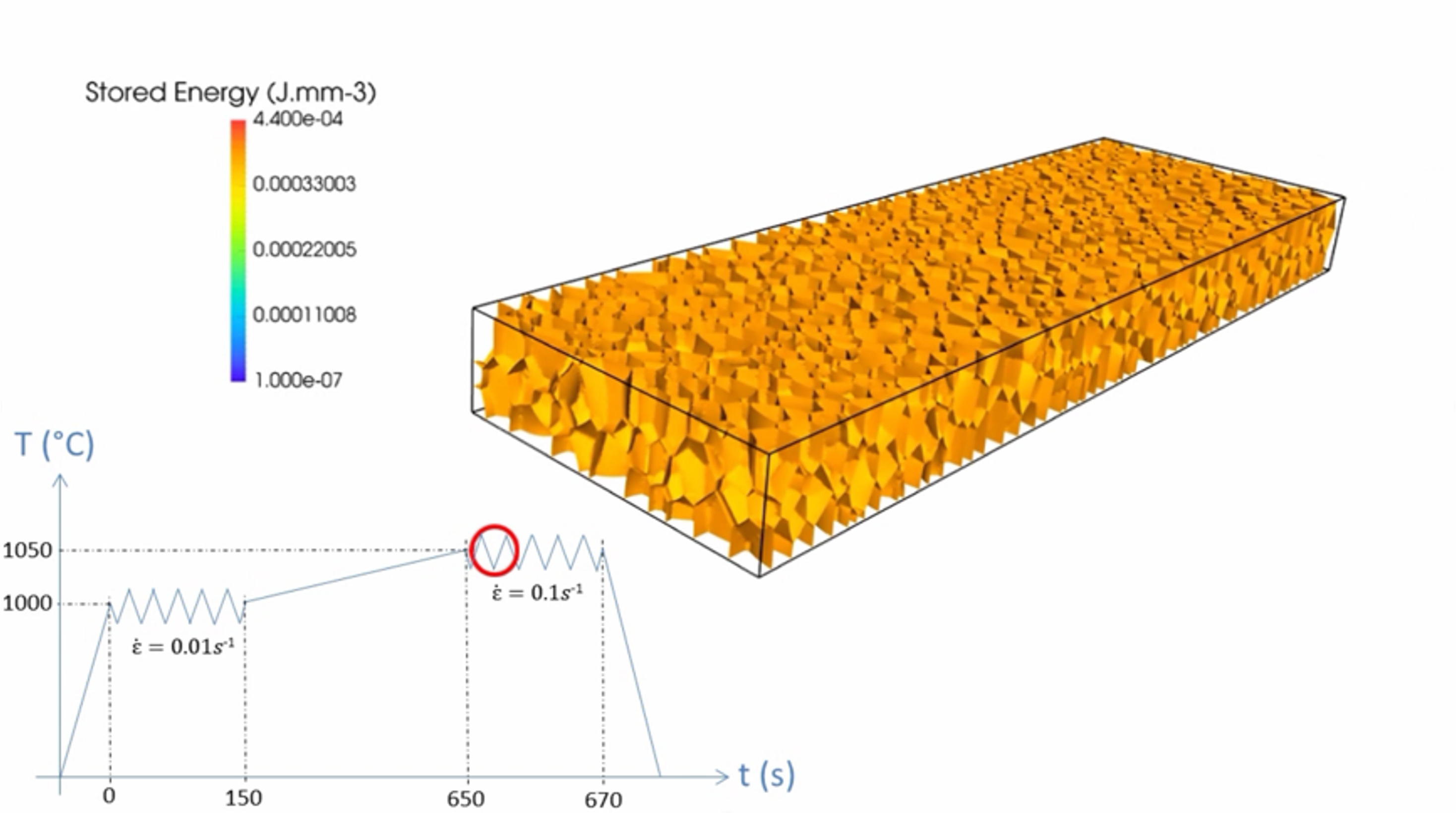}
    \caption{}
    \label{fig:DDRXf}
  \end{subfigure}
    \begin{subfigure}{0.49\textwidth}
    \centering
    \includegraphics[scale=0.187]{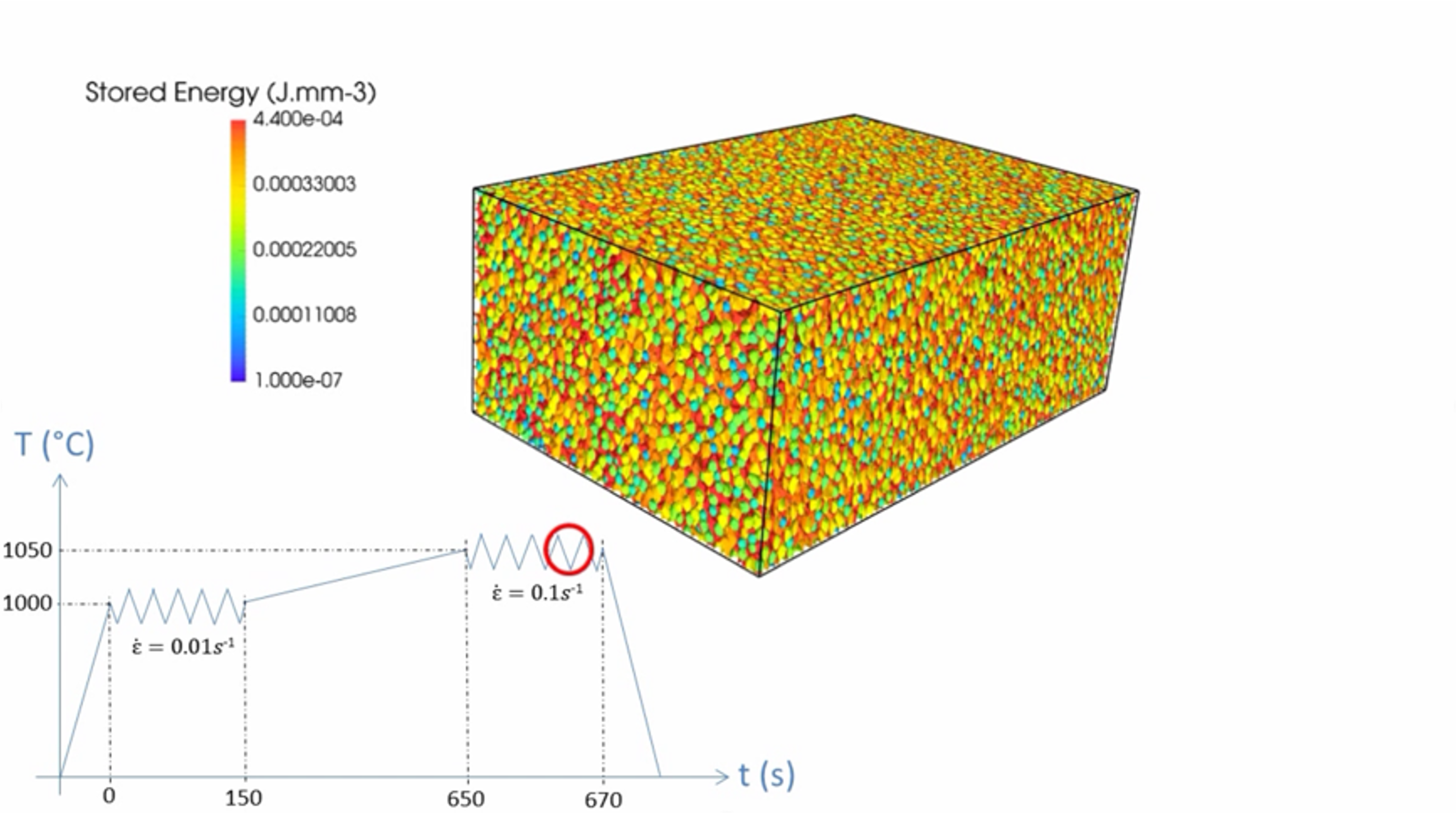}
    \caption{}
    \label{fig:DDRXg}
  \end{subfigure}
  \begin{subfigure}{0.49\textwidth}
    \centering
    \includegraphics[scale=0.187]{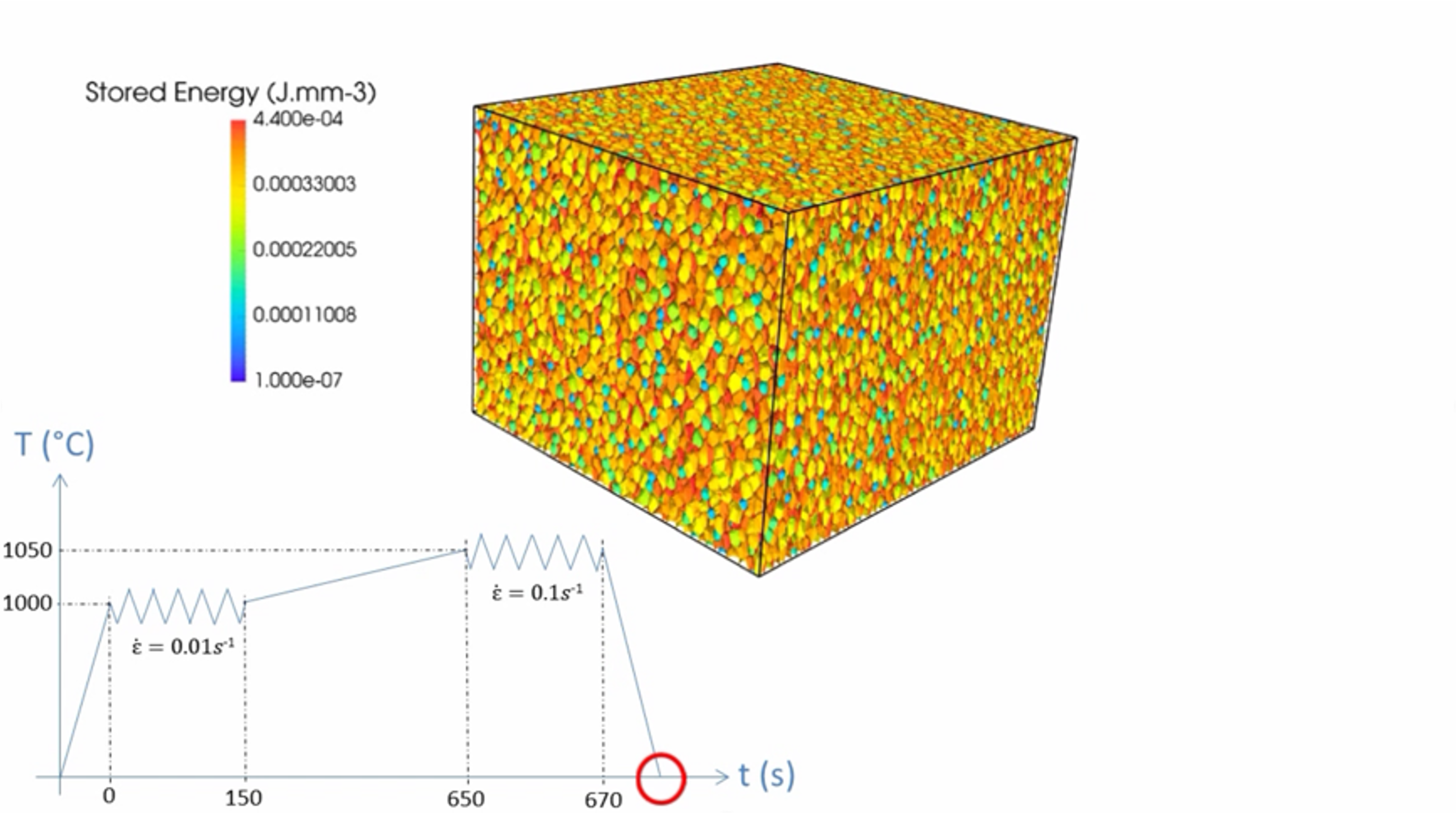}
    \caption{}
    \label{fig:DDRXh}
  \end{subfigure}
  \caption{Complex thermomechanical path for a 304L stainless steel. In each picture from (a) to (h), the red circle describes the corresponding position in the thermomechanical path (t(\SI{}{s}) in abscissa and T(\SI{}{\celsius}) in ordinate, zigzag lines symbolize deformation steps, straight lines symbolize annealing steps) and the microstructure predicted at the same time with the grain boundary network colored by the stored energy. (a) corresponds to the initial state, (b) corresponds to the beginning of DDRX with appearance of few nuclei at the existing grain interfaces, the deformation is applied along the z-direction with a strain rate $\dot{\epsilon}=$ \SI{0.01}{\per\second}, (c) corresponds to the end of the first deformation; (d) and (e) described the post-deformation evolution during an increase of temperature, MDRX, SRX and GG mechanisms occur; (f) and (g) describe a second deformation along the x-direction ($\dot{\epsilon}=$ \SI{0.1}{\per\second}) with a second DDRX evolution and finally (h) described the final state obtained after quenching. From \citep{Maire2017}. A video of this simulation is available \href{https://www.youtube.com/watch?v=CC1BxzJa2kE}{online}.}
  \label{fig:DDRX}
\end{figure}

Special attention has to be paid to the velocity field $\mathbf{v}^{\llbracket\rho\rrbracket}_{i}$   in the vicinity of multiple junctions as emphasized in \citep{Bernacki2008}. In fact, rather than dealing with $\mathbf{v}^{\llbracket\rho\rrbracket}_{i}$ per grain as described in Eq.\ref{eq:Transport2} and considering the contributions of each neighbor as in Eq.\ref{eq:energybetweengrains}, a global common velocity can be built in the calculation domain and used for each convection-diffusion system. Generally, the following formulation is adopted:

\begin{equation}
    \mathbf{v}^{\llbracket\rho\rrbracket}\left(\mathbf{x},t\right)=\sum_{i=1}^{N_{GLS}}\sum_{\substack{j=1 \\ j \neq i}}^{N_{GLS}}\chi_{G_i}\left(\mathbf{x},t\right)\mu_{ij}\exp{\left(-\beta|\psi_j\left(\mathbf{x},t\right)|\right)}\llbracket E \rrbracket_{ij}\left(-\mathbf{n}_j\right),
    \label{eq:vstoredBernacki}
\end{equation}
where $\chi_{G_i}$ is the characteristic function of the grain $G_i$, $\mu_{ij}$ is the interface mobility between the neighboring grains $i$ and $j$ (equal to $\mu$ in isotropic context), the exponential term is a continuous decreasing function varying from $1$ to $0$ on either side of the interface and has the function of smoothing the velocity field across the interface, $\beta$ is a positive parameter that controls the degree of smoothness. 3D results obtained this way are summarized in Fig.\ref{fig:DDRX} for a complex thermomechanical path applied onto 304L stainless steel \citep{Maire2017}. The microstructure is initially generated by a Laguerre-Voronoï strategy as detailed in \citep{Hitti2012}. The system of equations \ref{eq:Transport2} is solved by considering Eqs \ref{eq:vacuum}, \ref{eq:energybetweengrains2}, \ref{eq:vstoredBernacki}, an optimized graph recoloration technique \citep{Scholtes2015}, a direct reinitialization technique \citep{Shakoor2015b} and the simplified mechanical framework detailed in \citep{Maire2017} where Eq.\ref{eq:YLJKM} is considered (with $\xi=0$) and associated with a critical stored energy law, a nucleation rate law and a critical nucleus radius law to defined the apparition of recrystallized grains.\\
Concerning the method based on Gaussian kernels (Eqs.\ref{eq:convolution1} to \ref{eq:convolution3}), it can not be used in context of discontinuous dynamic recrystallization (DDRX) where domain deformation has to be considered but was extended to take into account a constant stored energy per grain in static recrystallization evolution context \citep{Elsey2009}.\\
It must be highlighted that this numerical framework is well adapted for configurations where the anisotropy of GB properties is not of prime importance i.e. where isotropic grain growth and DDRX can be representative of the considered material. This is often the case for low stacking fault energy materials. When texture aspects and subgrain definition, organization and evolution can be of prime importance for microstructure evolution prediction like for high stacking fault energy materials, the discussed framework can become insufficient. In this context, anisotropy of GB interface energy and mobility must be introduced into the LS formalism. This aspect is discussed in the following section.

\section{Anisotropy of GB properties and CDRX modeling}\label{sec:aniso}

\subsection{LS formulations in context of anisotropic GB properties}

The study of GB energy and mobility has garnered significant attention since its initial observation of anisotropy by Smith \citep{Smith1948} and Kohara \citep{Kohara1958}. Existing research has explored two primary modeling approaches: the isotropic approach, where constant values are employed for the GB energy $\gamma$ and temperature-dependent mobility $\mu\left(T\right)$, as introduced in earlier sections \citep{Anderson1984,Gao1996,Lazar2011,Bernacki2011,Garcke1999}, and the heterogeneous approach, which proposes energy and mobility values for each grain boundary \citep{Rollett1989,Hwang1998,Upmanyu2002,Fausty2018,Zollner2019,Miyoshi2016,Chang2019, Miyoshi2019,Miessen2015,Fausty2020,Holm2001}. Heterogeneous models aim to replicate complex microstructures, such as twin boundaries, by accounting for individual grain orientations and the disorientation angle between grains \citep{Miyoshi2017,Fausty2020}. However, the effect of misorientation axis and GB inclination is often overlooked. To address this limitation, anisotropic models have been developed, encompassing the dependence of GB properties on misorientation and inclination \citep{Kazaryan2002,Fausty2021,Hallberg2019,Salama2020}. It is essential to clarify the distinction between 3-parameter (heterogeneous) and 5-parameter (anisotropic) full-field formulations, as the literature often ambiguously categorizes heterogeneous GB properties as anisotropic. Additionally, confusion arises between the concepts of respecting an anisotropic GB dynamics and/or respecting Herring's equation at equilibrium  \citep{herring1999}, with or without considering the torque terms.

The scarcity of GB property data has led to the prevalence of studies utilizing heterogeneous GB properties. Early measurements of GB properties, primarily GB reduced mobility, were conducted on bicrystals \citep{Viswanathan1973, Demianczuk1975, Maksimova1988,Gottstein1992,Winning2002,Ivanov2006} , resulting in the well-known Sigmoidal model \citep{Rollett2017}. However, advancements in experimental and computational technologies have enabled 3D techniques using X-ray \citep{Zhang2017,Zhang2020,Jensen2020,Fang2021}  or molecular dynamics \citep{Janssens2006,Olmsted2009a,Olmsted2009b} to study GB and recrystallization. At the mesoscopic scale, limited studies have been conducted in 2D using anisotropic GB properties derived from mathematical models \citep{Kazaryan2002,Fausty2021} or by fitting data from molecular dynamics \citep{Hallberg2019}. Nevertheless, these 2D models overlook a portion of the 3D space, as the GB inclination is measured in the sample plane, and GB properties are simplified.

Addressing the study of GB in 3D frequently involves employing heterogeneous GB properties based on mathematical descriptors \citep{Fjeldberg2010,Chang2019,Song2020,Miyoshi2021} or databases of GB energy values  \citep{Kim2011,Kim2014}. Two key questions frequently arise in the current state of the art: Can GB properties be accurately described in 2D using classical Read-Shockley \citep{Read1950} and Sigmoidal \citep{Rollett2017} models? Is the effect of anisotropy stronger in 3D? Answering the latter requires conducting 3D simulations instead of 2D to achieve a more comprehensive description of GB properties in the 5D GB space.

In the current state of the art, four main formulations using a FE-LS approach are notable. The first is the isotropic formulation mentioned earlier, which has been successfully applied to model various annealing phenomena, including GG, recrystallization, and GG in the presence of second phase particles \citep{Bernacki2008,Bernacki2009,Bernacki2011,Scholtes2015,Maire2017}. While this approach demonstrates good agreement with experimental data in predicting mean grain size and grain size distribution (GSD), it may struggle to replicate complex grain morphologies, specialized grain boundaries, and textures. The second formulation extends the isotropic approach by incorporating heterogeneous GB energy and mobility values through axis-angle parameterization, and even inclination for the energy \citep{Hallberg2014,Jin2015}.

\begin{align}
    \partial_t \psi - \mu\left(\pmb{\mathbb{M}},T\right) \gamma\left(\pmb{\mathbb{M}},\mathbf{n}\right) \Delta \psi = 0.
    \label{eq:transportClassicHet}
\end{align}
With this formulation, predominant in the LS literature, it is expected to obtain more physical grain shapes. Indeed, some GBs can evolve faster thanks to higher grain boundary mobility values, and triple junctions may have different stable dihedral angles thanks to different GB energy values enabling to respect Herring's equation. This strategy classically used in full-field formulations (not only in LS ones) can lead to confusion when it is named as ``heterogeneous''. Indeed, \textit{stricto sensu}, the heterogeneity shape of $\mu$ and $\gamma$ lead to additional terms in the driving pressure of the kinetic equation (Eq.~\ref{eq:kinetic equation ReX}) as illustrated through Eq.\ref{eq:herring4}  but also in the weak formulation derived to solve the GB motion which are not taken into account in a such strategy. In the following, the term ``heterogeneous'' will be used to distinguish this formulation from the purely isotropic model. Such strategies were also proposed to deal with heterogeneous description of $\gamma$ in method based on Gaussian kernels (Eqs.\ref{eq:convolution1} to \ref{eq:convolution3}), one can cite \citep{Elsey2013,Miessen2015,nino2023}.

A such discussion is proposed in \citep{Fausty2018} where an additional term capturing the local heterogeneity of the multiple junctions is added to the velocity equation such that: 
\begin{align}
    \mathbf{v} = \mu\left(\theta,T\right) \left( \nabla \gamma\left(\theta\right) \cdot \nabla \psi - \gamma\left(\theta\right) \Delta \psi \right) \nabla \psi.
    \label{eqn:VelFieldFull}
\end{align}
Inserting this term into the transport equation (Eq. \ref{eq:Transport}) leads to the, hereafter called, ``Heterogeneous with Gradient'' formulation \citep{Fausty2018}: 
\begin{align}
    \partial_t \psi + \mu\left(\theta,T\right) \nabla \gamma\left(\theta\right) \cdot \nabla \psi - \mu\left(\theta,T\right) \gamma\left(\theta\right) \Delta \psi = 0.
    \label{eqn:transportFull}
\end{align}
The introduction of the term $\nabla \gamma\left(\theta\right) \cdot \mathbf{n}$ only acts at multiple junctions because these are the only places where this term does not vanish. This formulation is equivalent to the Isotropic one if no heterogeneity is added. The formulation proposed in \citep{Hallberg2019} is very similar. This third formulation was proposed for triple junctions in \citep{Fausty2018} and extended to model GG using heterogeneous GB energy in \citep{Fausty2020} and both heterogeneous GB energy and mobility in \citep{Murgas2021}. 

The last one is an anisotropic formulation which was initially developed using thermodynamics and differential geometry in \citep{Fausty2021} and was improved in \citep{Murgas2021} in order to consider heterogeneous GB mobility. Both the GB normal and misorientation are taken into account and an intrinsic torque term is present:
\begin{align}
\mathbf{v} = -\mu\left(\pmb{\mathbb{M}},T\right)\overbrace{\left(-\left(\underbrace{\gamma \pmb{\mathbb{Id}} + \nabla_{\mathbf{n}}\nabla_{\mathbf{n}}\gamma}_{\pmb{\bbGamma}\left(\mathbf{n}\right)}\right):\pmb{\mathbb{K}} + \nabla_{\mathbf{n}}\gamma\left(\pmb{\mathbb{M}},\mathbf{n}\right)\cdot\nabla\psi\right)}^{P_c}\nabla\psi.
    \label{eq:VelFieldAniso5LS}
\end{align}
Remarkably, the term $P_c$ is here totally equivalent to the one introduced in Eq.\ref{eq:herring4}. As detailed in Eq.\ref{eq:herring4}, the term $ \nabla_{\mathbf{n}}\gamma\left(\pmb{\mathbb{M}},\mathbf{n}\right)\cdot\nabla\psi$ in Eq.~\ref{eq:VelFieldAniso5LS} should be null in the grain interfaces. However, the front-capturing nature of the LS approach which consists to solve Eq.\ref{eq:Transport2} at the GB network and in its vicinity, requires to consider this term which could be non-null around the interfaces near multiple junctions. 

In the current state of the art, the utilization of these various formulations leads to the observation that predicting grain growth at the polycrystal scale can be ambiguous, contingent upon the targeted attributes and available data. When moderate anisotropy is involved, the Isotropic formulation can effectively reproduce the evolution of mean grain size and grain size distribution. However, in scenarios with texture configurations, the Anisotropic formulation demonstrates superior performance in predicting grain morphology, DDF (disorientation distribution function), and the evolution of interfacial energy, while still maintaining reasonable computational efficiency compared to the isotropic approach. Furthermore, it has been demonstrated that 3D simulations should be considered. This is not only crucial for improving the representativeness of the analyzed polycrystals but also to accurately describe the $\gamma$ dependence on inclination. The existing 2D models and data currently limit the practical use of inclination. 

In addition to this, the integration of torque effects and the GB stiffness tensor in simulations and analysis is an essential aspect that needs to be addressed. It is worth noting that this conclusion is consistent across the majority of existing works in the state of the art involving anisotropic 2D grain growth simulations and 3D simulations, where the inclination dependence, torque terms, or both are not accounted for. To the author's knowledge, only one paper dedicated to the extension of the classical isotropic kernel using Gaussian kernels has been focused on incorporating the $\gamma$  dependence on inclination, and it was validated in configurations without multiple junctions \citep{elsey2018threshold}.\\

\begin{figure}[h!]
  \centering
  \begin{subfigure}[c]{1.0\textwidth}
    \centering
    \includegraphics[scale=0.4]{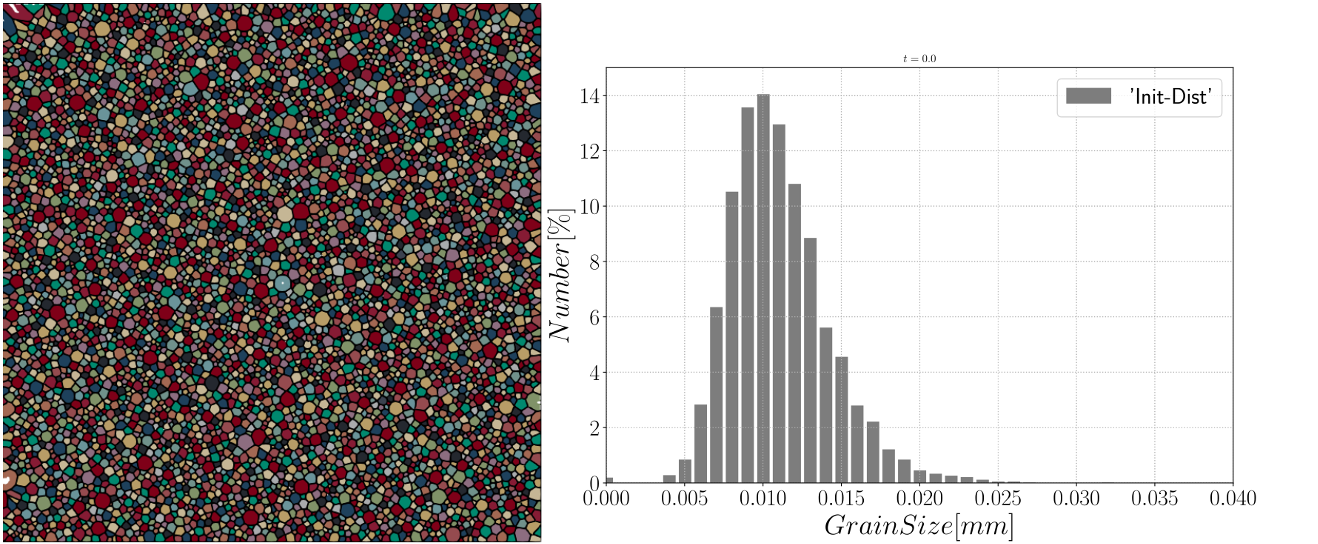}
  \end{subfigure}
  \begin{subfigure}[c]{0.49\textwidth}
    \centering
    \includegraphics[scale=0.25]{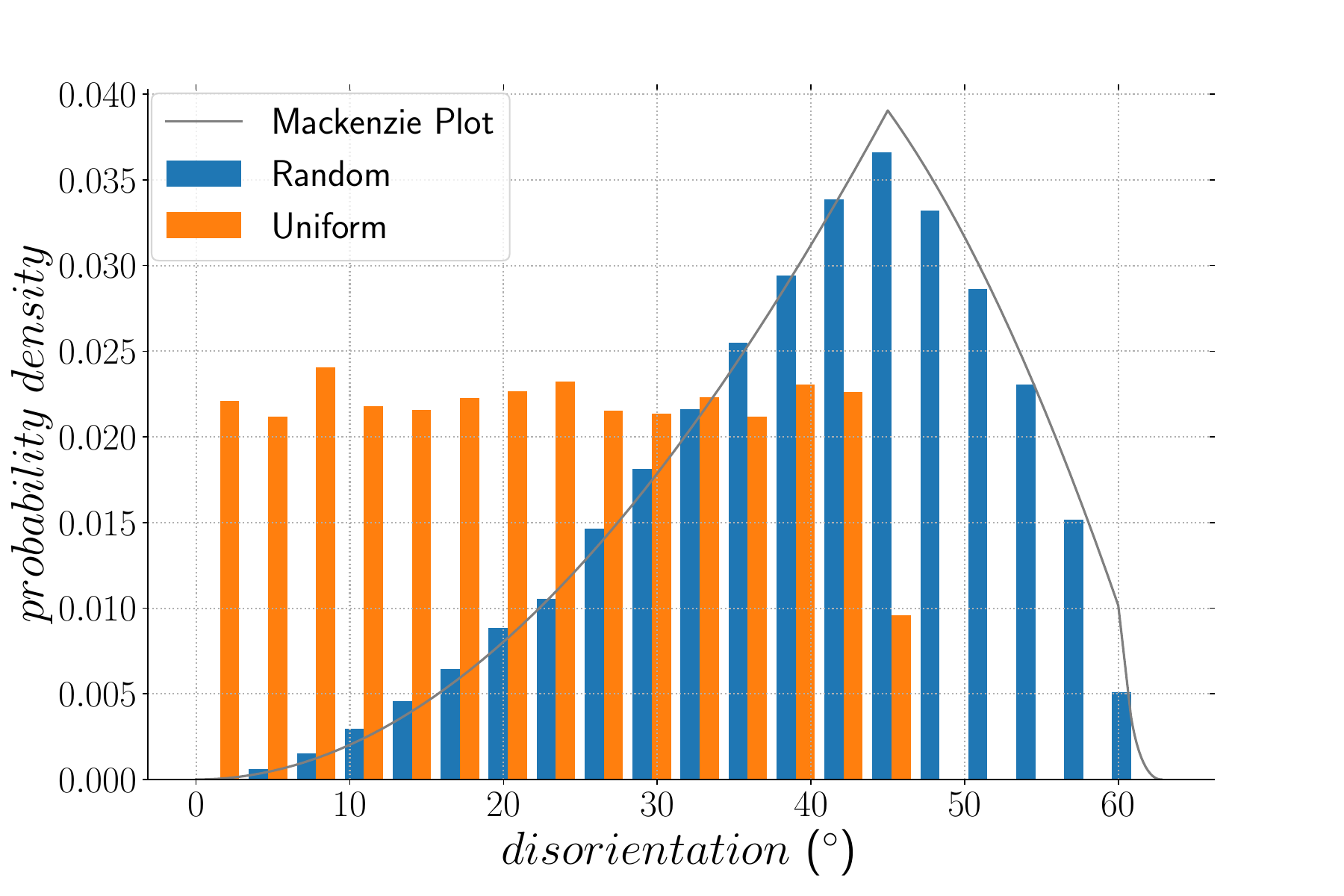}
  \end{subfigure}
  \begin{subfigure}[c]{0.49\textwidth}
    \centering
    \includegraphics[scale=0.25]{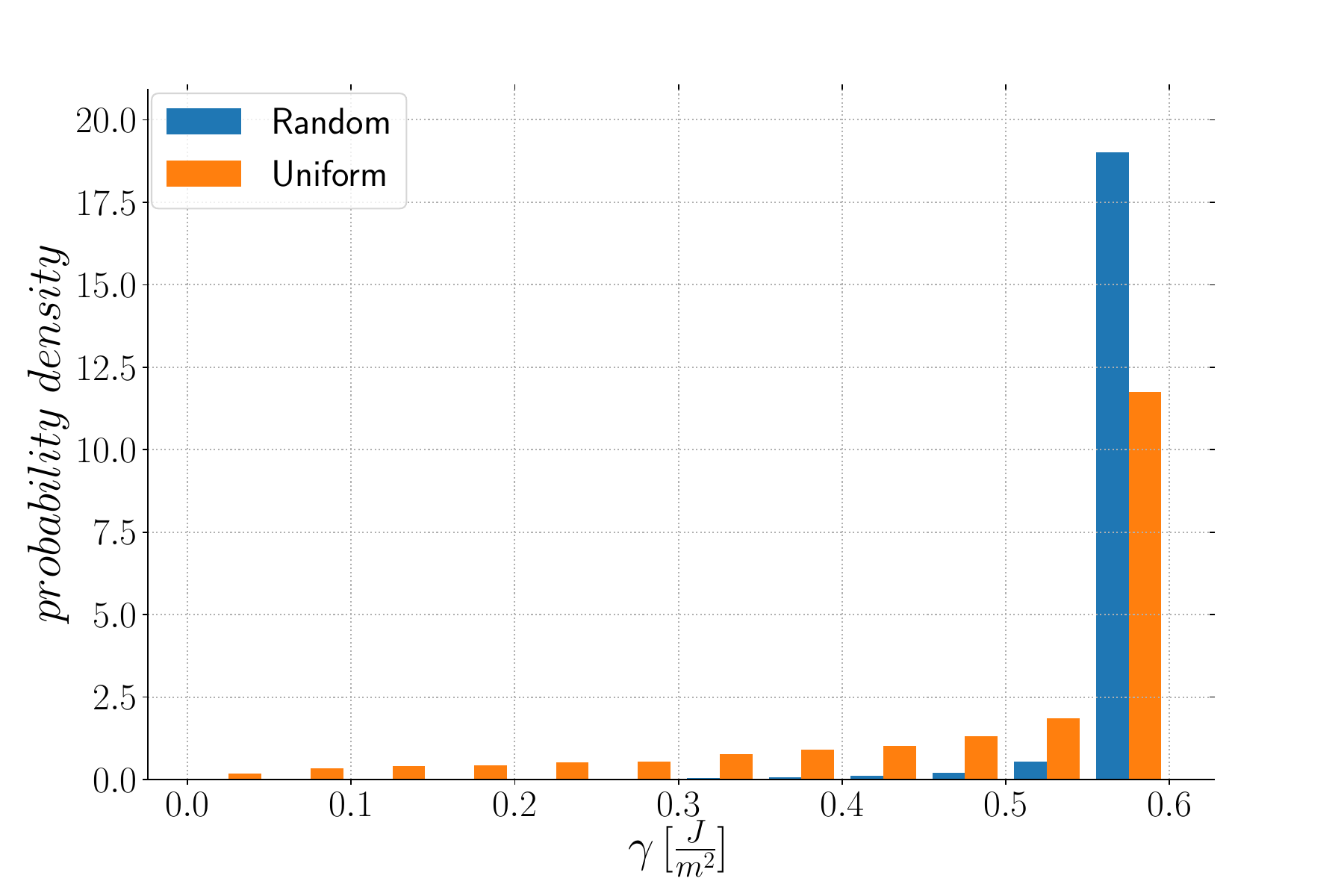}
  \end{subfigure}
   \begin{subfigure}{0.495\textwidth}
    \centering
    \includegraphics[scale=0.25]{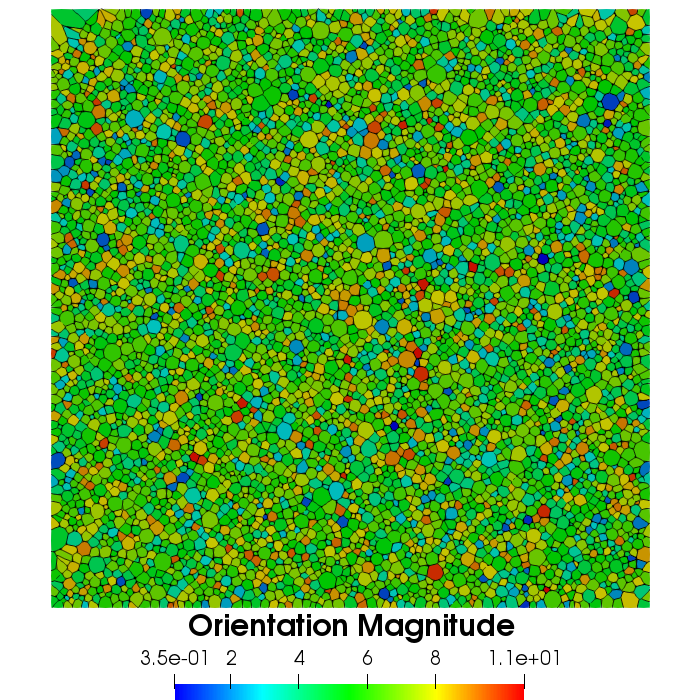}
  \end{subfigure}
  \begin{subfigure}{0.495\textwidth}
    \centering
    \includegraphics[scale=0.25]{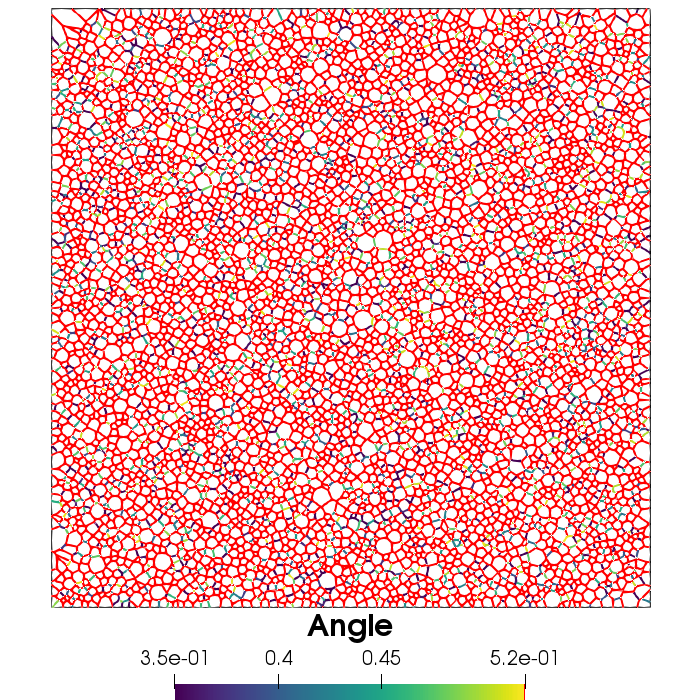}
  \end{subfigure}
  \caption{Characteristics of the initial microstructures (5000 grains) for the anisotropic simulations with from top to bottom and left to right: coloration field of the grain, initial grain radius distribution, initial considered disorientation distributions, initial considered grain boundary energy distributions, orientation magnitude field for the random configuration and disorientation angle field at grain interfaces for the random configuration. From \citep{Murgas2021}.}
  \label{fig:Aniso1}
\end{figure}

The Fig.\ref{fig:Aniso1} illustrates the characteristics of a 2D initial example discussed by Murgas et al. \citep{Murgas2021}: it consists of a square domain with length $L=1.6mm$ and 5000 grains generated using a Laguerre-Voronoi tessellation \citep{Hitti2012}. 

A misorientation dependent GB energy defined with a Read-Shockley (RS) function \citep{Read1950} is considered:
\begin{align}
  \left\{
  \begin{array}{l}
    \gamma\left(\theta\right) = \gamma_{max} \frac{\theta}{\theta_{max}} \left( 1 - ln \left( \frac{\theta}{\theta_{max}} \right) \right), \ \theta < \theta_{max} \\
    \gamma_{max}, \ \theta \ge \theta_{max}
\end{array}
\label{eq:GammaH}
\right .
\end{align}

where  $\gamma_{max}$ is the maximal GB energy. $\theta_{max}=$\SI{15}{\degree} is the disorientation defining the transition from a low angle grain boundary (LAGB) to a high angle grain boundary (HAGB). $\mu=$\SI{1.379}{\raiseto{4}\milli\meter\per\joule\per\second}, and $\gamma_{max}=$ \SI{6e-7}{\joule\per\square\milli\meter}, which are typical for a stainless steel \citep{Fabiano2014}, are considered. Two ways are used to exacerbate the GB heterogeneities. In a first one (called Random), the Euler angles defining the crystallographic orientations are generated randomly, leading to a Mackenzie-like disorientation distribution function \citep{Mackenzie1958}. In a second one (called Uniform), one Euler angle is generated randomly with a uniform distribution function and the two others are constants. As a result, the final disorientation distribution is more uniform. Both considered orientation distribution and resulting grain boundary energy distribution (GBED) are described in Fig.\ref{fig:Aniso1} (middle) and the resulting orientation field using the vector magnitude $O_G = \sqrt{\varphi_1^2 + \phi^2 + \varphi_2^2 }$ where $\left(\varphi_1, \phi,\varphi_2\right)$ are the three Euler angles is described in Fig.\ref{fig:Aniso1} (bottom) for the Random configuration.

\begin{figure}[h!]
  \centering
  \begin{subfigure}{0.48\textwidth}
    \centering
    \includegraphics[scale=0.22]{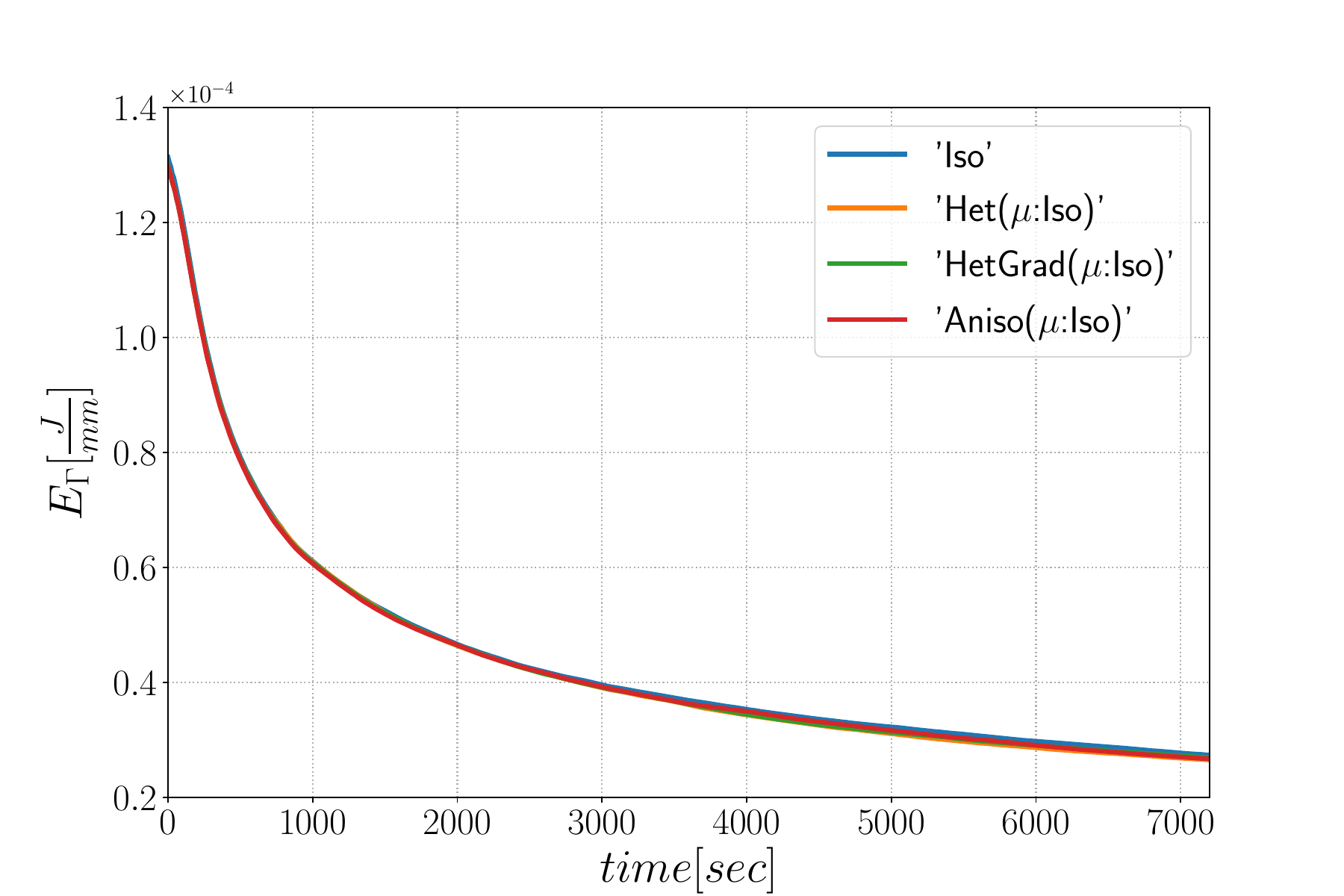} 
    \caption{$E_{\Gamma}=f\left(t\right)$ - Random Configuration}  
  \end{subfigure} 
  \begin{subfigure}{0.48\textwidth}
    \centering
    \includegraphics[scale=0.22]{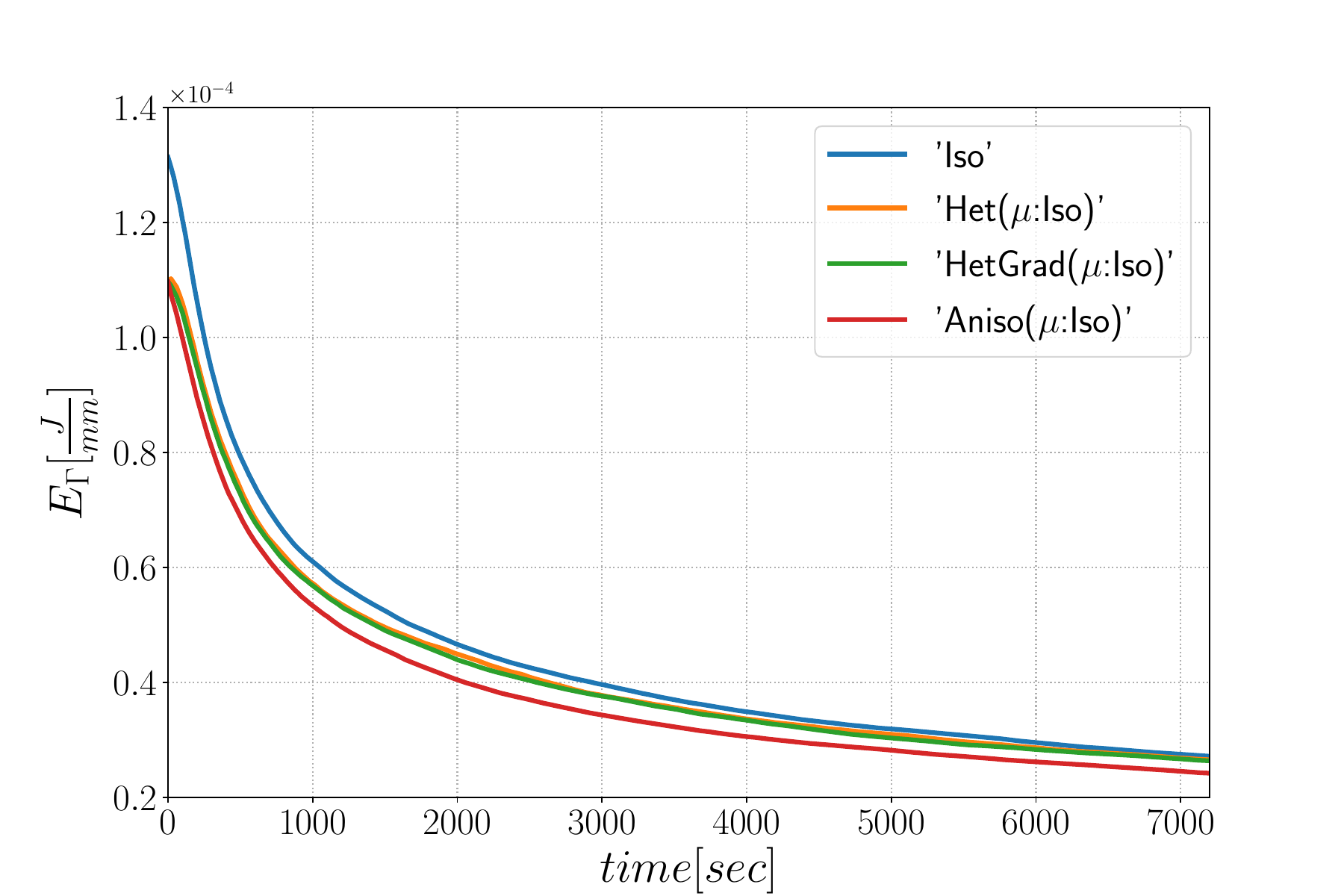} 
    \caption{$E_{\Gamma}=f\left(t\right)$ - Uniform Configuration} 
  \end{subfigure} 
  \begin{subfigure}{0.48\textwidth}
    \centering
    \includegraphics[scale=0.22]{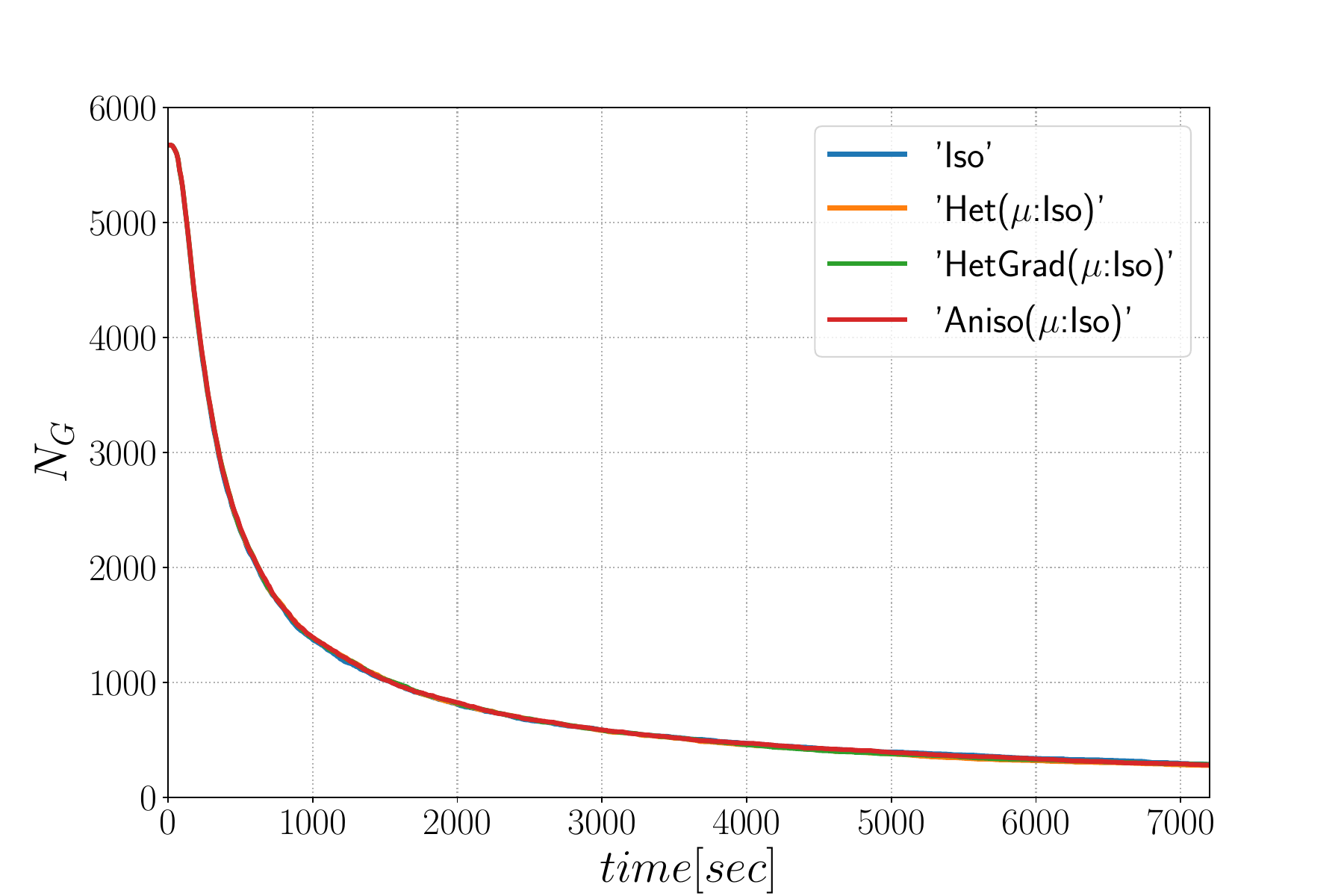}
    \caption{$N_g=f\left(t\right)$ - Random Configuration}
  \end{subfigure} 
  \begin{subfigure}{0.48\textwidth}
    \centering
    \includegraphics[scale=0.22]{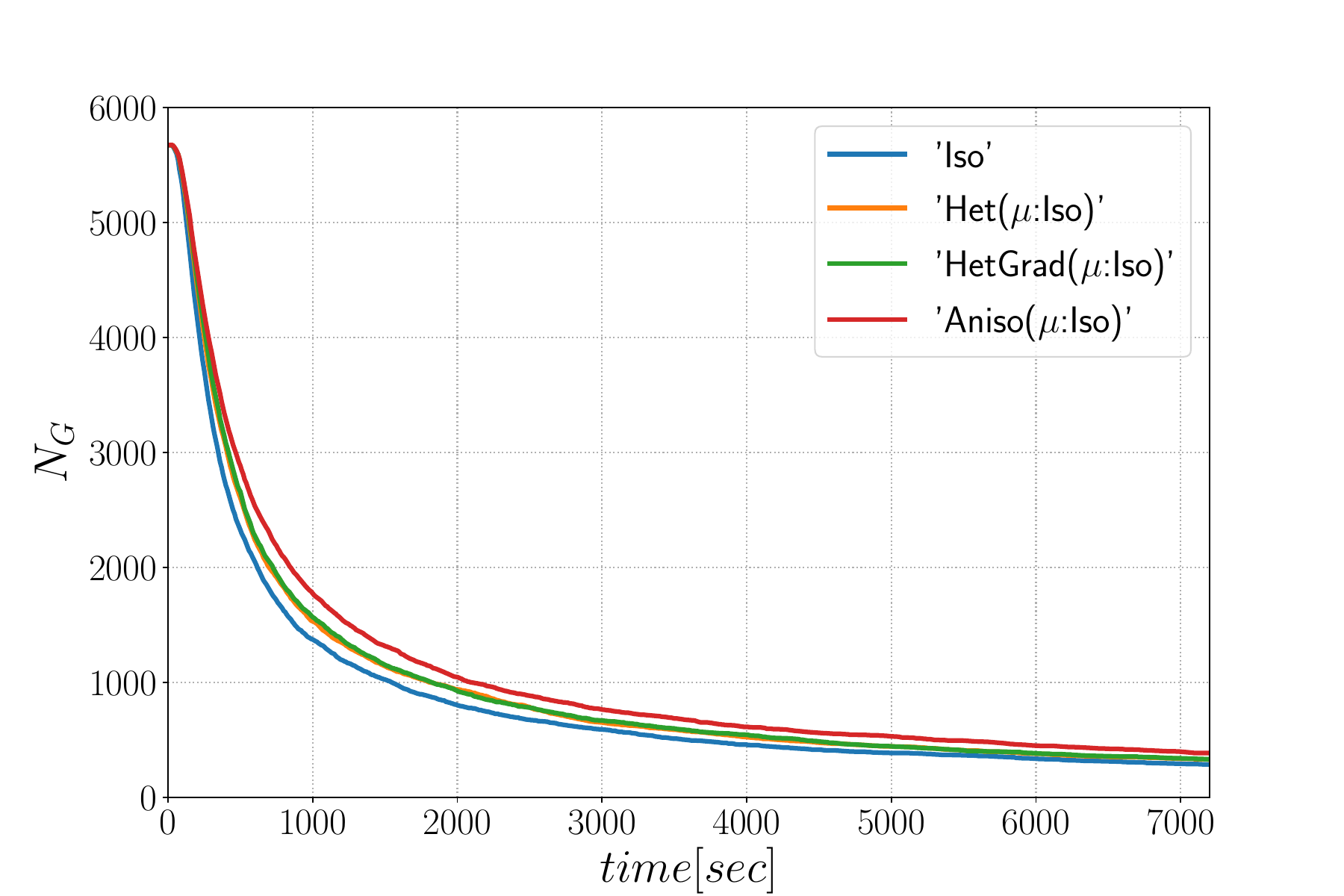}
    \caption{$N_g=f\left(t\right)$ - Uniform Configuration}
  \end{subfigure} 
  \begin{subfigure}{0.48\textwidth}
    \centering
    \includegraphics[scale=0.22]{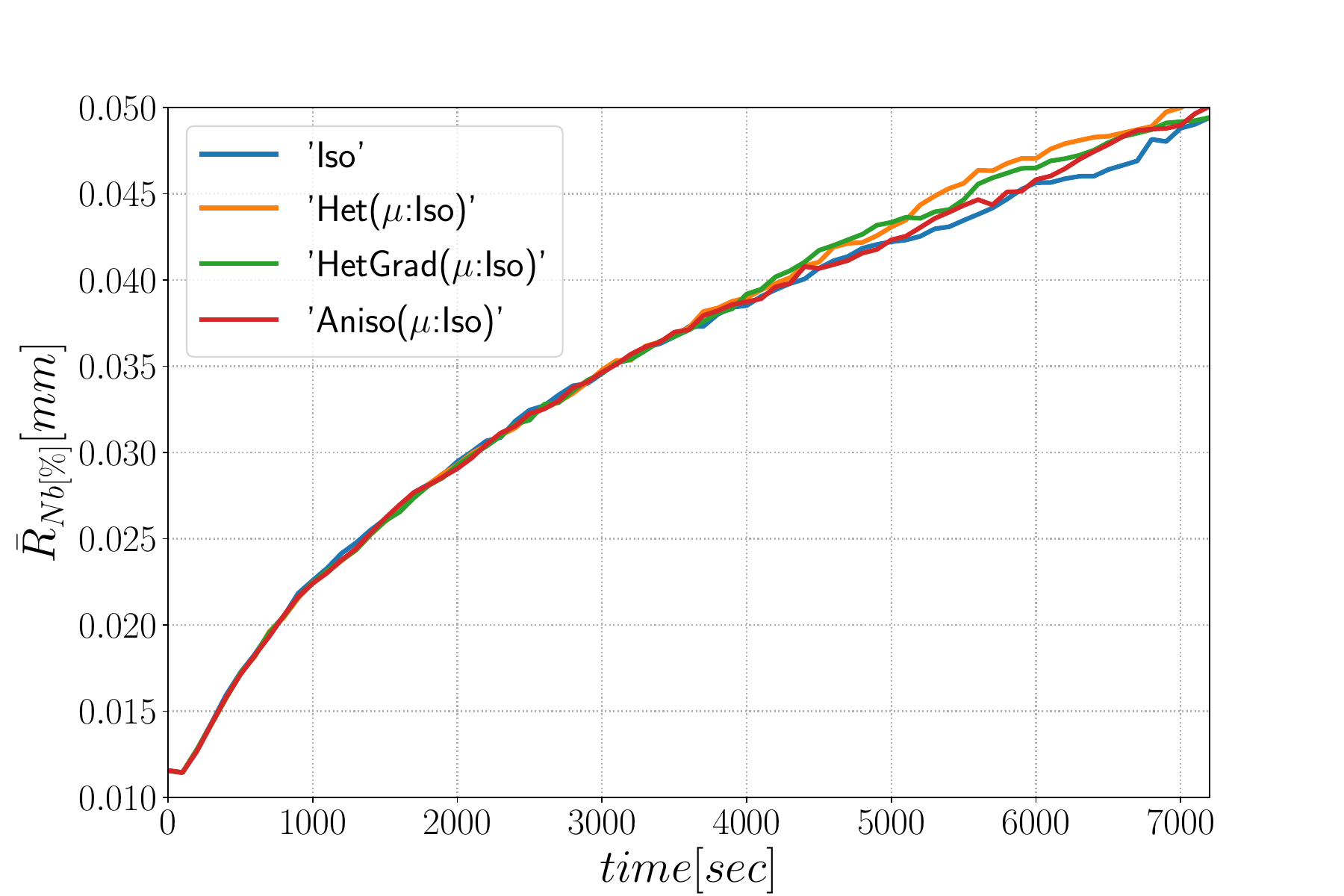}
    \caption{$\bar{R}_{Nb[\%]}=f\left(t\right)$ - Random Configuration}
  \end{subfigure}
   \begin{subfigure}{0.48\textwidth}
    \centering
    \includegraphics[scale=0.22]{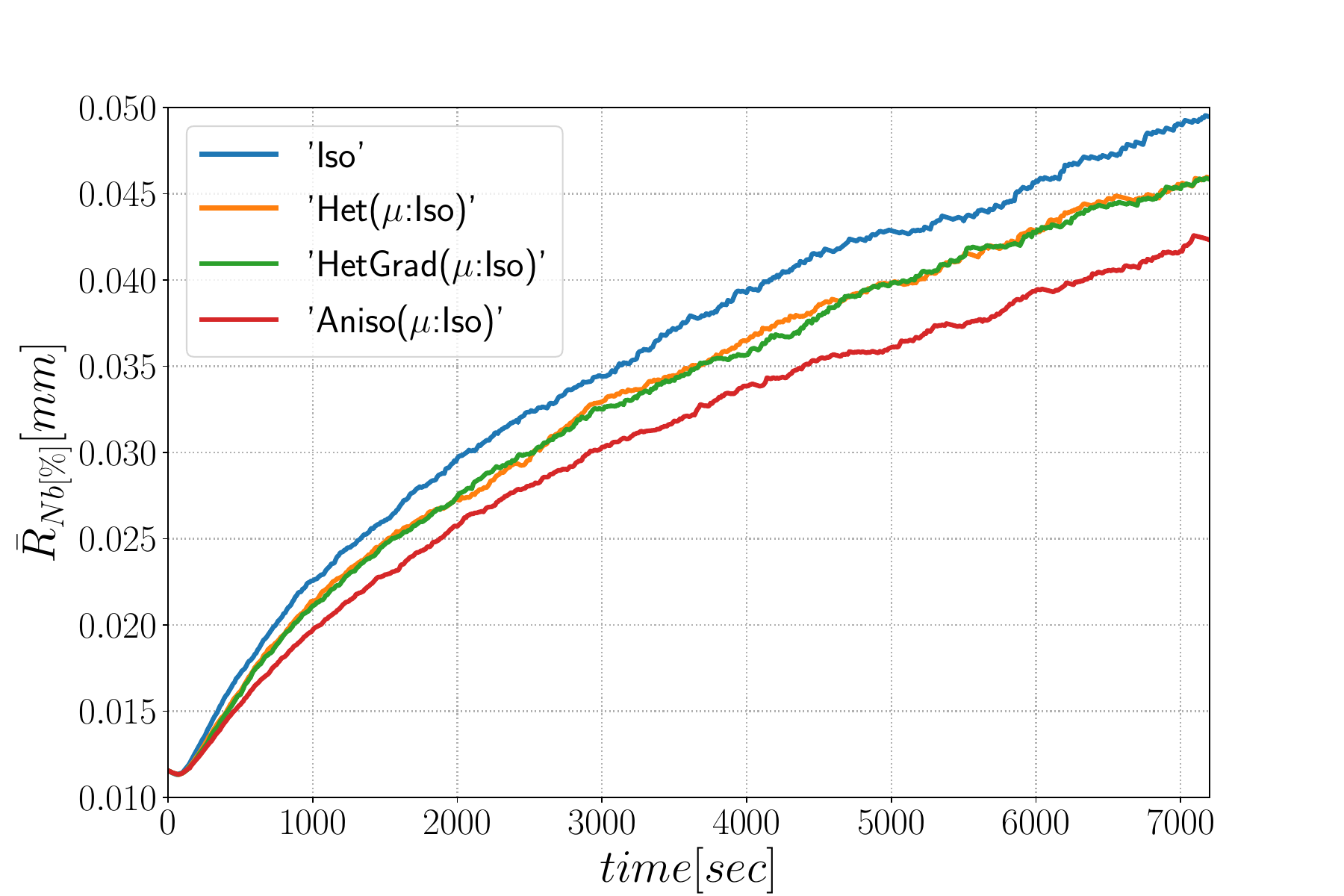}
    \caption{$\bar{R}_{Nb[\%]}=f\left(t\right)$ - Uniform Configuration}
  \end{subfigure}
  \begin{subfigure}{0.48\textwidth}
    \centering
    \includegraphics[scale=0.22]{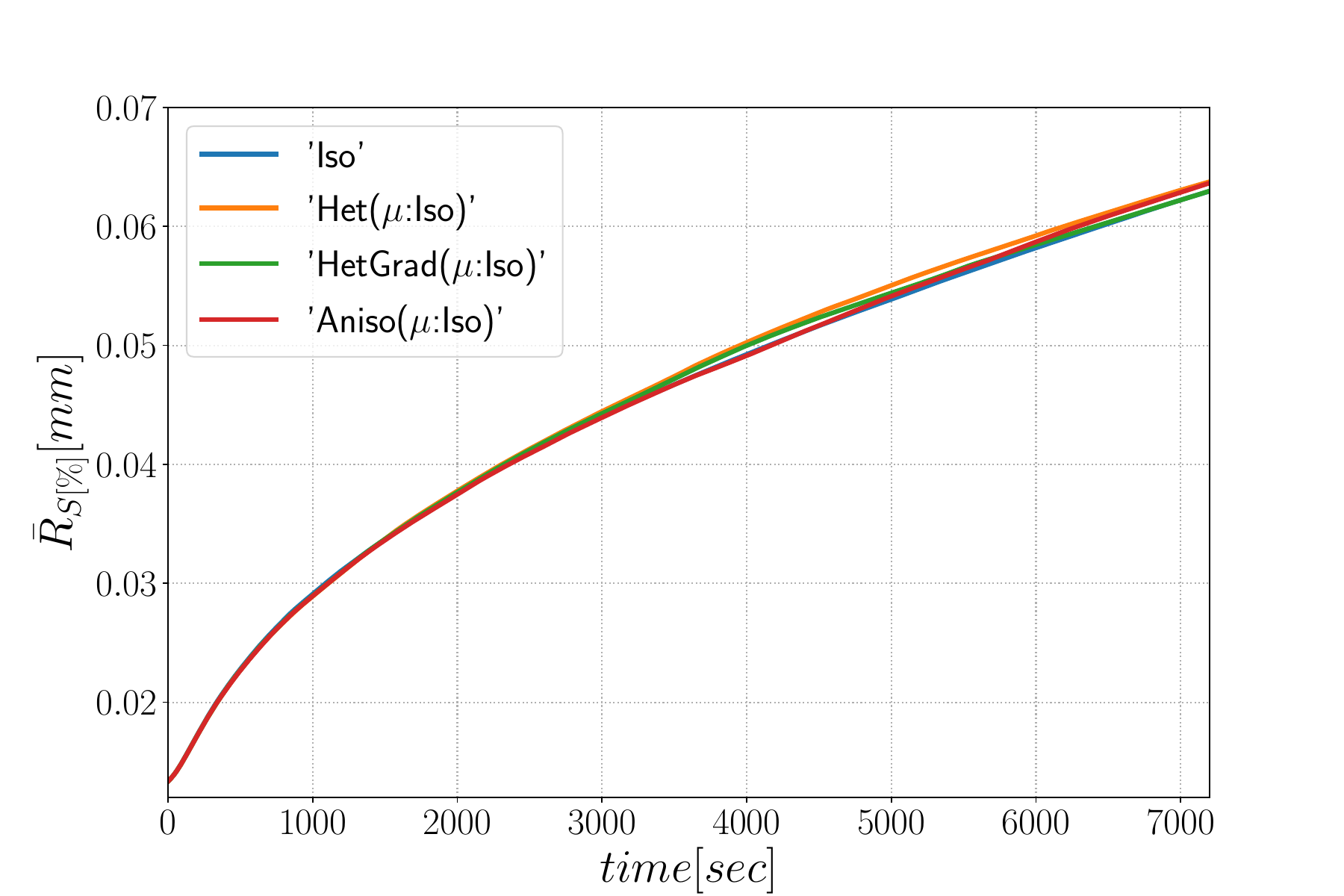}
    \caption{$\bar{R}_{S[\%]}=f\left(t\right)$ - Random Configuration}
  \end{subfigure}
   \begin{subfigure}{0.48\textwidth}
    \centering
    \includegraphics[scale=0.22]{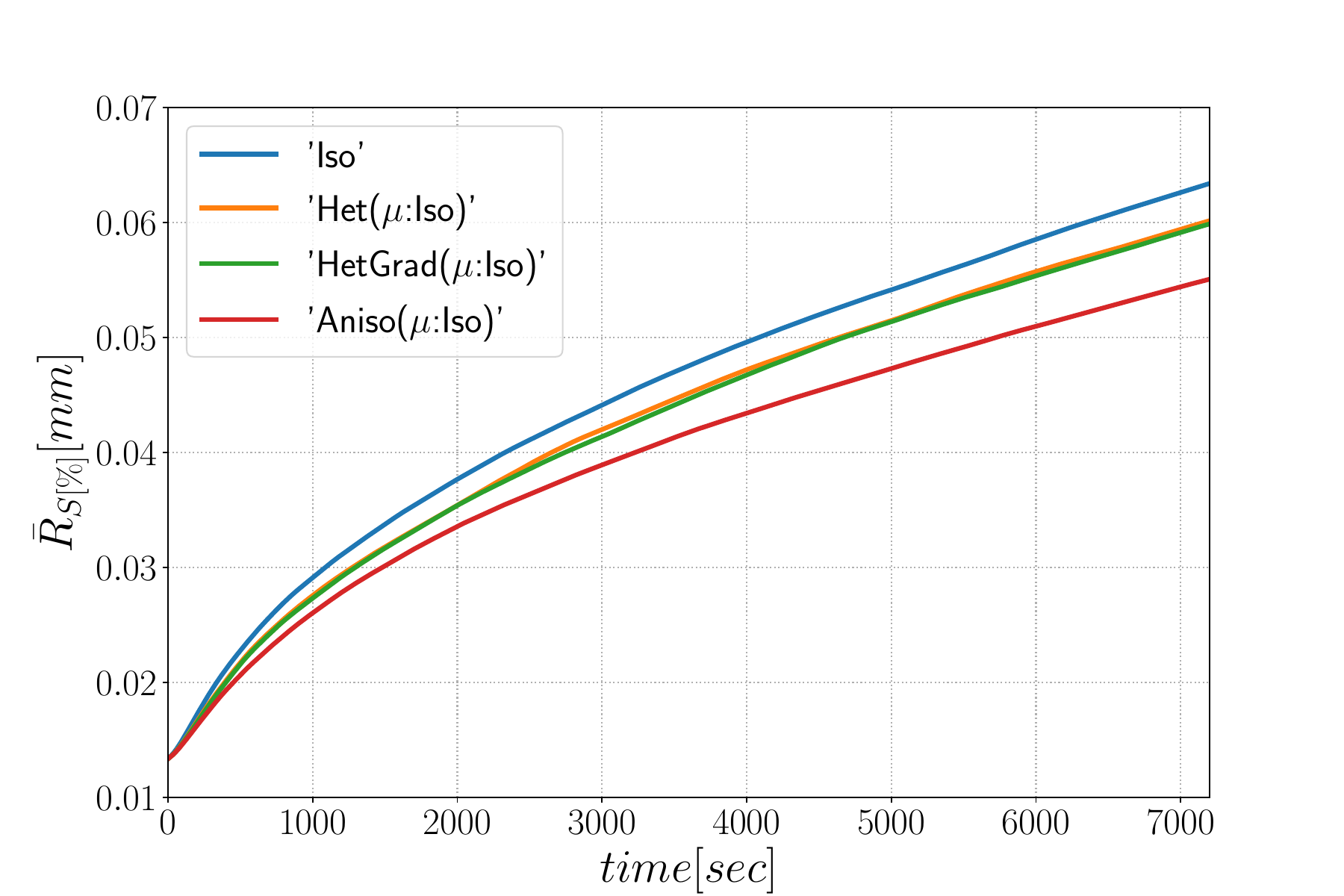}
    \caption{$\bar{R}_{S[\%]}=f\left(t\right)$ - Uniform Configuration}
  \end{subfigure}
  \caption{Time evolution for the different formulations for the Random configuration in the left side: (a) the total GB energy, (c) the number of grains, (e) the mean radius weighted in number and (g) the mean radius weighted in surface. Same information for the Uniform configuration are described in the right side. From \citep{Murgas2021}.}\label{fig:Aniso4}
\end{figure}

The time evolution are summarized in Fig.~\ref{fig:Aniso4}. First, it is noticeable that all the formulations have a similar evolution concerning the total grain boundary energy $E_{\Gamma}$, the number of grains $N_G$ and the mean grain size weighted in number $\bar{R}_{Nb[\%]}$ or in surface $\bar{R}_{S[\%]}$ when the Random configuration is considered whereas more stronger GB heterogeneities has a significant impact on the different attributes of the microstructure when the different formulations are adopted. This aspect corroborates the fact to avoid an isotropic formulation when a textured configuration exists in the material.

Another interesting aspect to deal with GB heterogeneities is be able to consider materials where the substructure (LAGBs) are of prime importance in ReX mechanisms as illustrated in the following subsection.

\subsection{CDRX modeling}

As already highlighted, the LS-FE approach was largely considered in context of DDRX for low stacking fault energy materials by associated Eq.\ref{eq:YLJKM} to a critical stored energy law, a nucleation rate law and a critical nucleus radius law. For high stacking fault energy materials, the progressive formation and evolution of subgrains must be taken into account. Dislocations could rearrange themselves to form new LAGB or accumulate into preexisting LAGB. This last phenomenon is responsible for a progressive increase of LAGB misorientation \citep{Rollett2017, Huang2016}. During recrystallization, recrystallized grains form slowly and continuously during deformation. Indeed, in this CDRX context, grain formation is induced by the progressive reorganization of dislocations into cells or subgrains and the gradual increase of misorientation angle between those subgrain. Recently, a first LS-FE framework was proposed by Grand et al. \citep{grand_simulation_2022, Grand2022, Gaillac2022,grand_modeling_2023} in order to deal with this mechanism. In these works, laws introduced by the Gourdet-Montheillet model were implemented \citep{Gourdet2003}, and while keeping the global dislocation density evolution law defined by Eq.\ref{eq:YLJKM}, several mechanisms of evolution of dislocations are taken into account :
\begin{itemize}
    \item rearrangement into LAGB that bound new subgrains. Subgrain formation is described through following equation \citep{Gourdet2003}:
    \begin{eqnarray}
    dS^+ = \dfrac{\alpha b K_2 \rho d\varepsilon}{\eta \theta_0},
    \label{eq:SurfaceNewSubgrains}       
    \end{eqnarray}
    where $dS^+$ is the surface of LAGB created. $\alpha = 1- \exp{\left( \dfrac{D}{D_0} \right) ^m}$ is a coefficient describing the fraction of dislocations recovered to form new subgrains. $D$ is the grain diameter, $D_0$ is a grain reference diameter and $m$ is a fixed coefficient. $\eta$ is the number of sets of dislocations and $\theta_0$ the disorientation of newly formed subgrains.
    \item Stacking into preexisting LAGB which is modeled according to the following equation \citep{Gourdet2003}:
    \begin{eqnarray}
    d\theta = \dfrac{b}{2 \eta} \left(1-\alpha\right) D K_2 \rho d\varepsilon.
    \label{eq:ProgressiveMisorientation}
    \end{eqnarray}
    \item Absorption during HAGB migration. This is naturally captured by affecting to the areas swept by moving boundaries a low dislocation density as described earlier.
\end{itemize}

\begin{figure}[h!]
  \centering
  \includegraphics[scale=0.1]{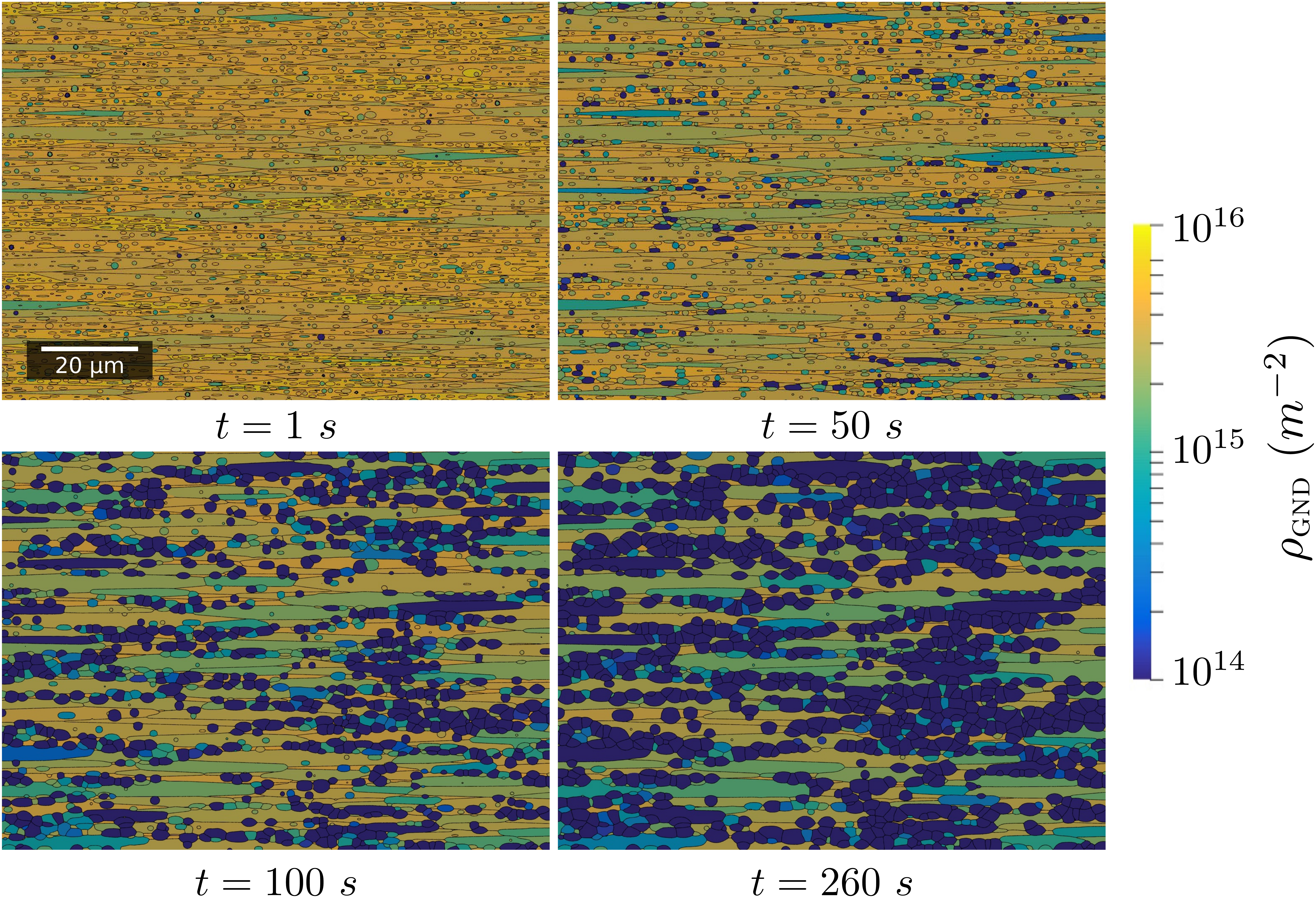}
  \caption{Illustration of CDRX modeling for Zy-4 thanks to a LS-FE approach from \citep{grand_simulation_2022}. The time scale corresponds to the time after the end of the deformation. A video is available \href{https://www.youtube.com/shorts/qLYKOb_4AOA}{online}.}
  \label{fig:CDRX}
\end{figure}

At each time step, dislocation density of each grain is updated using equation Eq.\ref{eq:YLJKM} which impacts the computation of the velocity term related to stored energy differences. Then, the length of subgrain interfaces formed into each grain is computed using Eq.\ref{eq:SurfaceNewSubgrains}. Subgrains can be added grain by grain based on the value of this grain property or globally, after having summed the length of subgrain interfaces for all grains. Subgrain orientation is initialized by applying a small misorientation to the parent grain orientation. The misorientation angle is selected to respect a distribution measured experimentally whereas the misorientation axis satisfies a uniform distribution. The misorientation axis attributed to a subgrain at its formation is kept constant. Then, during time increments, the misorientation increase described by equation \ref{eq:ProgressiveMisorientation} is realized by rotating of $d\theta$ around this axis. \\
An illustration of this framework in context of CDRX modeling for Zircaloy-4 (Zy-4) was proposed by Grand et al. and is depicted in Fig.\ref{fig:CDRX} after the deformation (the time scale corresponds to the time after the end of the deformation). The initial microstructure includes approximately 300 grains. The initial number of grains is taken low since it will increase substantially during deformation. Material parameters have been estimated based on the experimental results obtained conducting a thermomechanical testing campaign associated with extensive EBSD characterization. Thermomechanical conditions corresponding to these simulations are the following: $T=$\SI{650}{\celsius}; $\dot{\varepsilon}=$\SI{1.0}{\per\second}; $\varepsilon=$ 1.35. The number of subgrains that are formed at each deformation increment is computed individually per grain/subgrain. GB energy is described by the RS equation Eq.\ref{eq:GammaH} and GB mobility is assumed isotropic.

\section{Static/evolutive second phase particles}
Smith-Zener pinning phenomenon, where precipitates act as obstacles to the displacement of the grain boundaries and may hinder recrystallization and grain growth, was  
first discussed by Smith \citep{Smith1948} and then detailed by Zener one year after \citep{Zener1949}. Under certain conditions, Second Phase Particles (SPP) can strongly pin the microstructure, leading eventually to a limiting mean grain size (MGS). Since these first developments to equate this phenomenon, many variants have been developed in order to dispel some of the initial hypotheses (see \citep{Manohar1998} for a review). This phenomenon is widely used by metallurgists to control the grain size during the forming process of many alloys, including superalloys. Predictive tools are then needed to model accurately this phenomenon and thus optimize the final grain size and in-use properties of the materials. Classical laws predicting the limiting MGS \citep{Manohar1998} , noted $\overline{R}_{\infty}$ have the following form:

\begin{equation}\label{eq:SZP}
\overline{R}_{\infty}=K\dfrac{\overline{r}}{f^{m}},   
\end{equation} 

where $K$ and $m$ are fitted parameters which can be assumed constant \citep{Manohar1998} or dependent of the material and/or the characteristics of the SPP population \citep{Manohar1998, Agnoli2014a}. Since thirty years, numerous full field modeling of the Smith-Zener phenomenon have been proposed, including MC/CA framework \citep{Srolovitz1984, Anderson1989,Hassold1990,Gao1997,Kad1997,Phaneesh2012}, front tracking or vertex \citep{Weygand1999, Couturier2003}, MPF \citep{Chang2009, Tonks2015, Moelans2006, Chang2014} and LS \citep{Agnoli2012, Agnoli2014a, Agnoli2015, Scholtes2016b, Villaret2020, Alvarado2021a, Alvarado2021b}.  

In the LS framework, the concept of incorporating inert SPP within an FE framework was initially proposed for conducting 2D Grain Growth (GG) simulations \citep{Agnoli2012} and 2D static recrystallization simulations  \citep{Agnoli2014a} in Inconel 718. SPPs are integrated into the FE mesh using statistical descriptions or experimental data, and the local curvature of the grain boundaries in contact with SPPs is constrained. This approach allows for the consideration of SPPs without assuming their size or morphology, and it accommodates isotropic and anisotropic particle/grain interface energies, whether they are incoherent or coherent interfaces. The effect of particle dragging is naturally captured by modifying the local curvature when the grain boundary encounters the particles, eliminating the need for explicit assumptions about the dragging pressure exerted by the particles.

\begin{figure}
	\centering
	\includegraphics[width=0.6\textwidth]{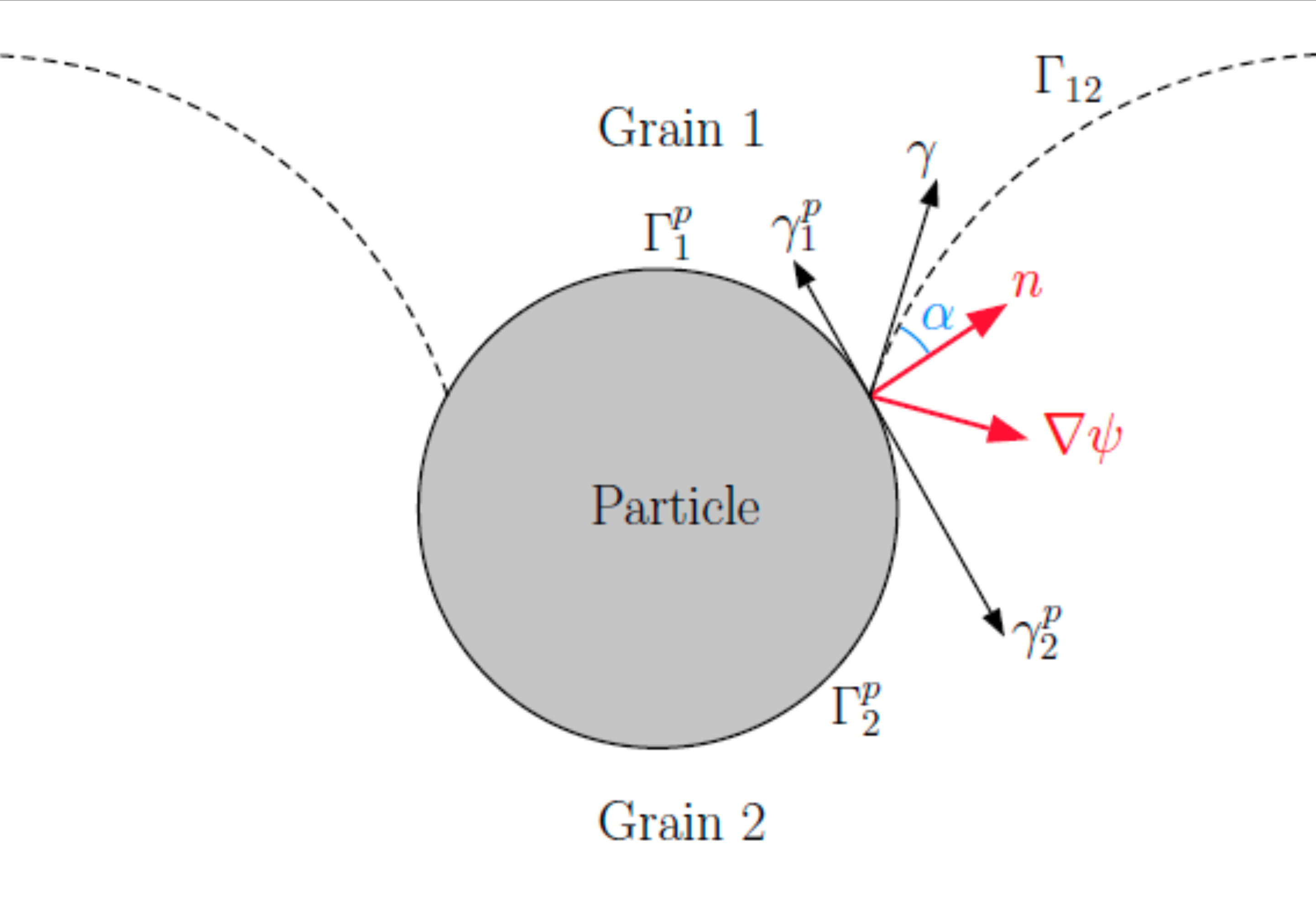}
	\caption{\label{fig:SZP} 2D illustration of the interaction between a particle and a GB (dashed lines between Grain 1 and Grain 2). $\mathbf{n}$ corresponds to the normal to the particle interface, $\nabla\psi$ to the normal to the GB and $\alpha$ to the angle established by the balance of surface tensions. From \citep{Alvarado2021a,settefrati_prediction_2018}.}
\end{figure}

Moreover, the Smith-Zener pinning effect induced by the presence of particles is naturally accounted for by imposing relevant boundary conditions at the interfaces between GBs and SPPs. Specifically, the influence of SPPs on microstructure evolution is considered by applying a Neumann-type limit condition on the GLS at the surface of the precipitates following the Young-Herring surface tension equilibrium:

\begin{equation}\label{eq:limit_condi}
 \nabla \psi \cdot \mathbf{n} = sin(\alpha)=\dfrac{\gamma_{p}^{2} - \gamma_{p}^{1}}{\gamma},
\end{equation}

where, as illustrated in Fig.\ref{fig:SZP},  $\mathbf{n}$ is here the external unitary normal vector to the precipitate, $\alpha$ is the angle established by the balance of surface tensions at contact positions between the SPP and the grain boundary. So, when the particle is assumed incoherent with the matrix, $\gamma_{p}^{1}\simeq\gamma_{p}^{2}$, which leads to $\alpha\simeq 0$, a  null Neumann boundary condition is applied at the precipitate/grain boundary interface through the respect of Eq.(\ref{eq:limit_condi}).

\begin{figure}[h!]
  \centering
 \begin{subfigure}{0.48\textwidth}
    \centering
    \includegraphics[scale=0.1]{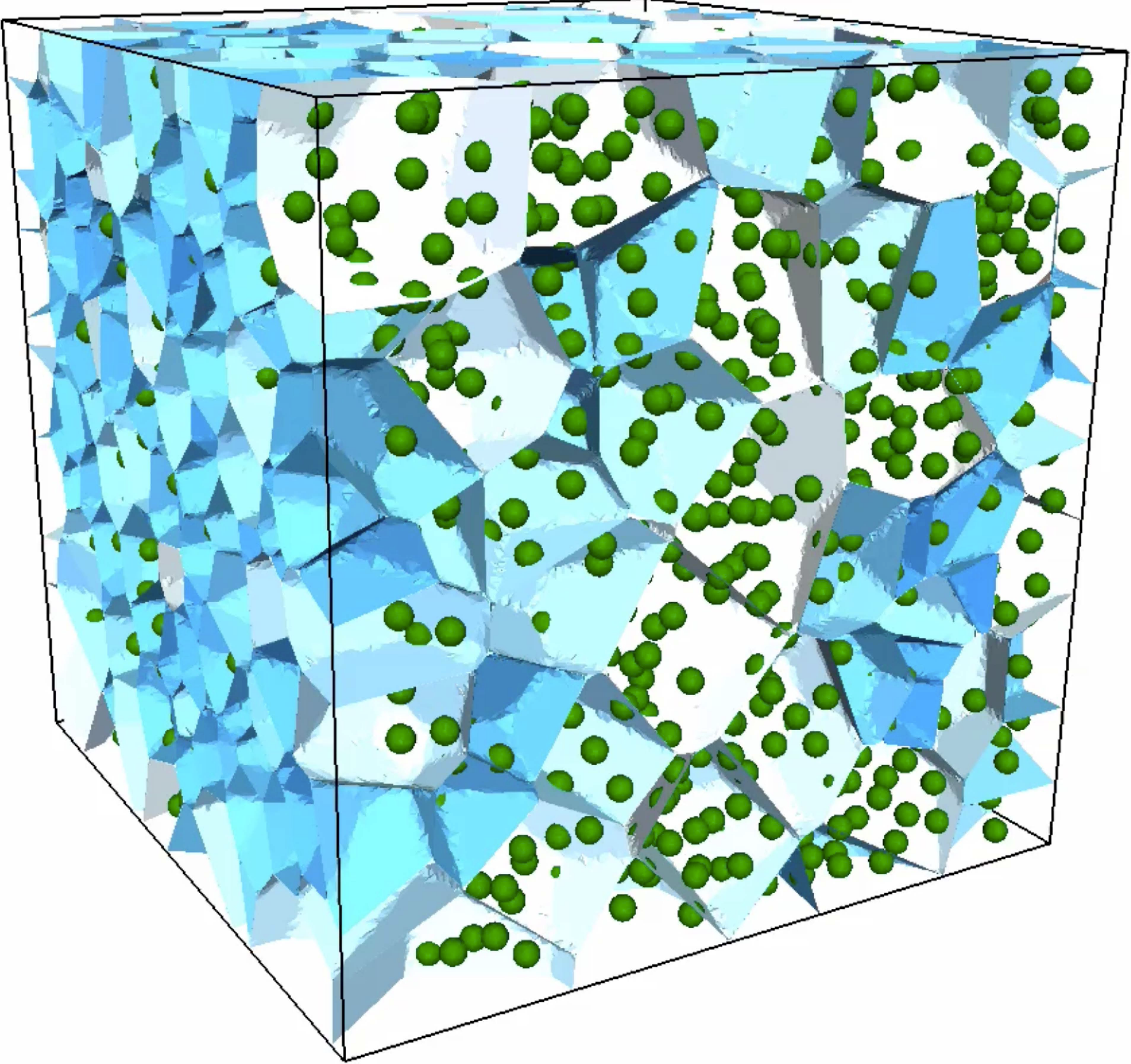} 
 
  \end{subfigure} 
  \begin{subfigure}{0.48\textwidth}
    \centering
    \includegraphics[scale=0.1]{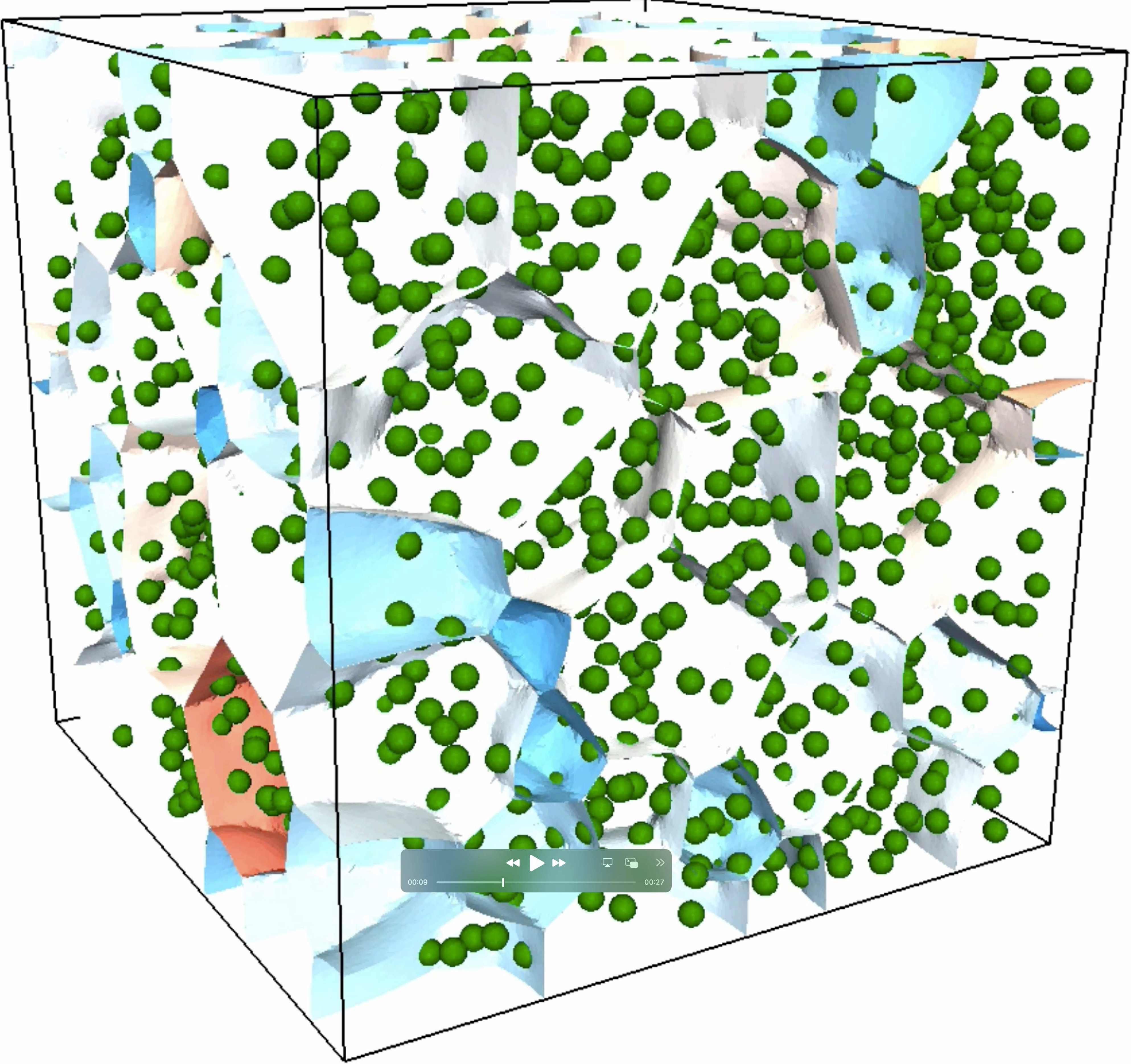}
 
  \end{subfigure} 
 
  \begin{subfigure}{0.48\textwidth}
    \centering
    \includegraphics[scale=0.1]{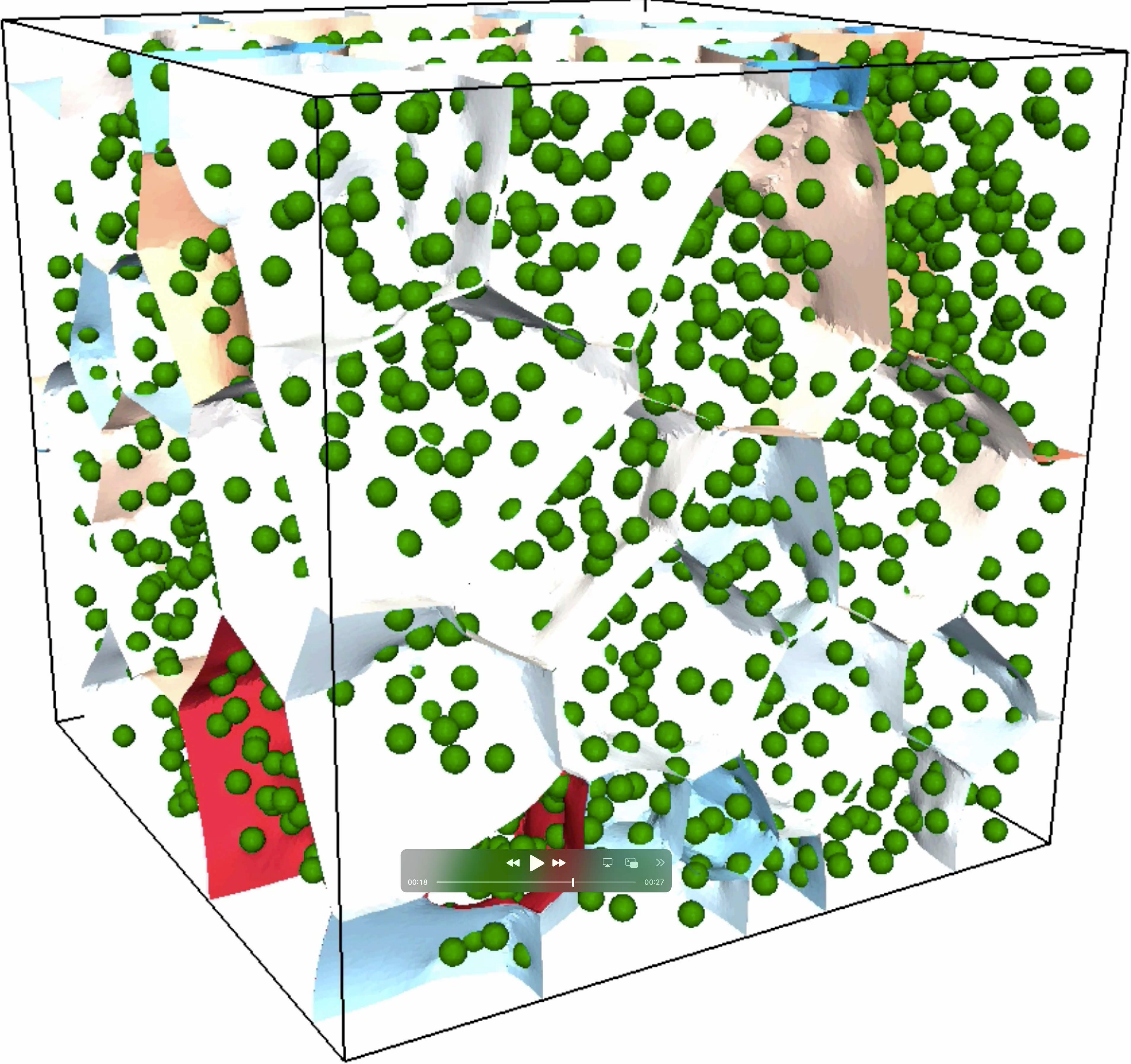}
 
  \end{subfigure}
  \begin{subfigure}{0.48\textwidth}
    \centering
    \includegraphics[scale=0.1]{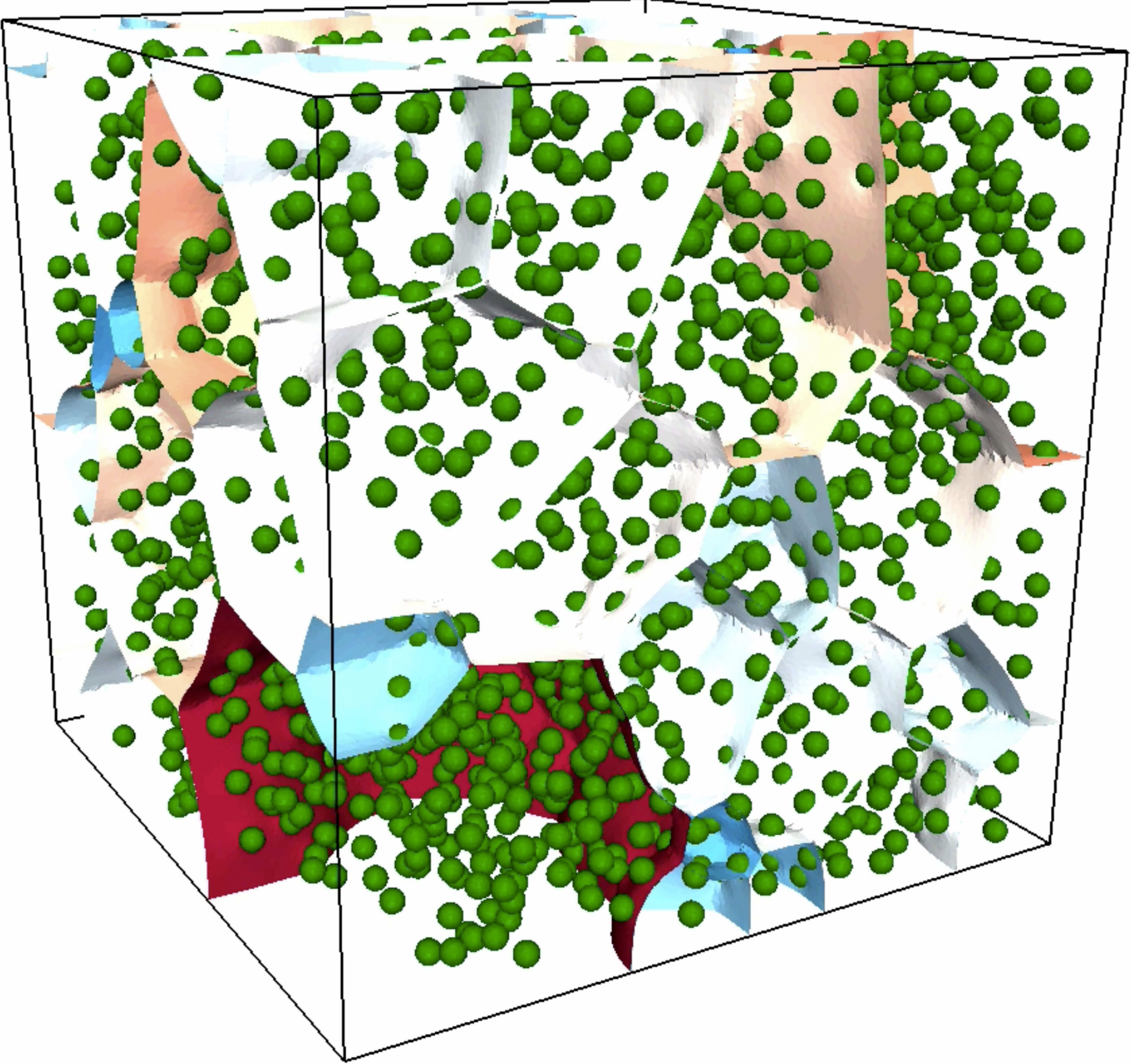}
 
  \end{subfigure}
   \begin{subfigure}{0.96\textwidth}
    \centering
    \includegraphics[scale=0.1]{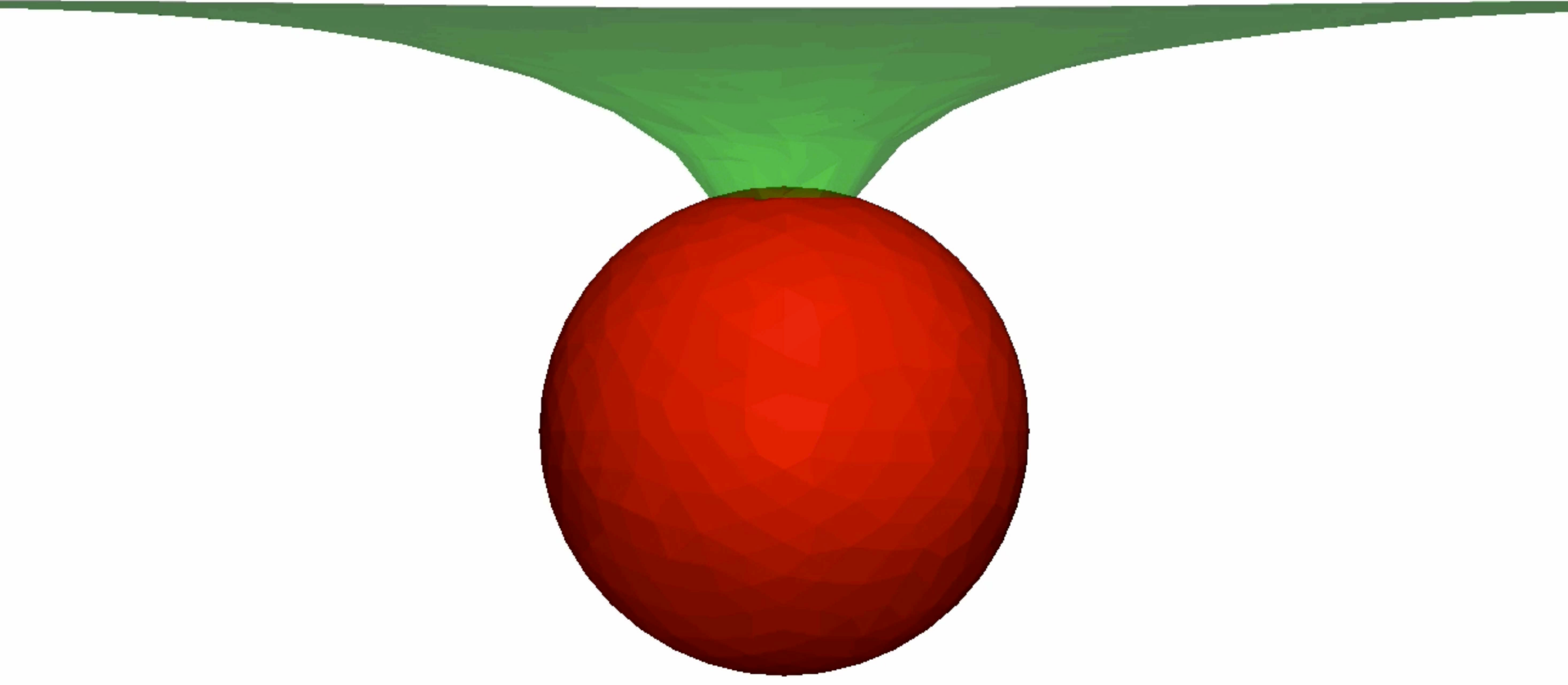}

  \end{subfigure}
  \caption{3D GG LS simulations for Inconel 718 with an idealized spherical population of second phase particles (f=3\%). SPP are described in green, the grain boundary network is described with a color code corresponding to the grain size until a stable configuration. From top to bottom and left to right: time evolution during the thermal treatment from the initial configuration to the stable grain boundary network. A video is available \href{https://youtu.be/JxL0vJEOp5A}{online}. The last image corresponds to a zoom of a SPP (in red) interacting with one GB (in green) just before unpinning. A video is available \href{https://youtu.be/xiHdIkpZsGU}{online}.}\label{fig:SZP2}
\end{figure}

This approach was used to support the idea that the phenomenon reported as abnormal grain growth in Inconel 718 could be explained by the growth of lower energy grains in a pinned microstructure and as such be a particular case of static recrystallization. Thanks to these simulations, the sensitivity of this phenomenon to the initial stored energy distribution could be studied \citep{Agnoli2015}. Optimization of parameters $K$ and $m$ of Eq.\ref{eq:SZP}  thanks to a full field simulations campaign and first 3D LS simulations were proposed in \citep{Scholtes2016b} and \citep{Scholtes2016}, respectively. Comparisons with MC simulations and experimental data have been made for ODS steels \citep{Villaret2020}. Fig.\ref{fig:SZP2} illustrates a 3D GG LS simulations for Inconel 718 realized by Scholtes et al. \citep{Scholtes2016b} with an idealized spherical population of SPP ($f$=3\%), until reaching a final stable microstructure.

However, this initial LS approach  for describing SPPs exhibits several limitations:

\begin{itemize}
	\item Simulating material deformations is not straightforward when attempting to consider SPP behavior, as SPPs are not volumetrically described in the considered FE mesh,
	\item the aforementioned issue may become critical in the context of DDRX. It is widely recognized that the interfaces of SPPs serve as conducive sites for new grain formation. Therefore, a detailed representation of the mechanical fields at the SPP interfaces is essential,
		\item the simulation time can be significantly extended due to remeshing operations around particles and grain boundaries,  mainly in 3D,
	\item the evolution of SPPs due to diffusive mechanisms—such as precipitation/dissolution, Ostwald Ripening, agglomeration, and spheroidization—cannot be accounted for, as SPPs are considered static in the current LS formulations.
\end{itemize}
For these reasons, a new LS approach to model ReX and GG mechanisms in presence of meshed SPPs and able to reproduce evolving particles was proposed by Alvarado et al. \citep{Alvarado2021a,Alvarado2021b}.
In this LS formulation, the description of SPPs is made by a new LS function, $\psi_{SPP}$, over the domain calculation without considering holes in the FE mesh. The GLS fields describing the grains are initially modified with simple topological operations (following by a reinitialization step) to introduce the presence of the SPPs, without modifying the particle interface:

\begin{equation}\label{majphi}
    \forall i\in [\![ 1,N_{GLS}]\!]\quad \hat\psi_i\left(\mathbf{x},t=0\right)=min\left(\psi_i\left(\mathbf{x},t=0\right),-\psi_{SPP}\left(\mathbf{x},t=0\right)\right).
\end{equation}
This operation is followed by a reinitialization procedure, $\psi_i\left(\mathbf{x},t=0\right) =\ Redist\left(\hat{\psi}_i\left(\mathbf{x},t=0\right)\right)\forall i \in \{1,2,...,N_{GLS}\}$, as the resulting GLS functions are not (when the intersection is not empty) a distance function, even if GLS and $\psi_{SPP}$ are.
Of course, the function $\psi_{SPP}\left(\mathbf{x},t=0\right)$ can be easily estimated as the distance function to the union of simple objects (as circles for spherical particles) but also obtained through the FE-immersion of an experimental map \citep{Agnoli2014a, Scholtes2016}.

As already discussed (see Eq.(\ref{eq:vacuum})), the appearance of voids or overlaps especially at the multiple junctions after solving the convective-diffusive equations was first treated by \citep{Merriman1994} and implement in several cases, in 2D and 3D, using the LS method \citep{Scholtes2016, Bernacki2011, Ilin2018}. In order to respect the Young-Herring equilibrium without hollowing out the SPPs, this classical treatment is extended by taken into account $\psi_{SPP}$ in the procedure:

\begin{eqnarray}\label{eq:vacuums_edges_mod}
\hat\psi_{i}\left(\mathbf{x},t\right) = \frac{1}{2}\left(\psi_{i}\left(\mathbf{x},t\right) - \max\left(\max_{j \neq i} \left(\psi_{j}\left(\mathbf{x},t\right)\right),\psi_{SPP}\left(\mathbf{x},t\right)\right)\right),\quad \forall i \in \{1,2,...,N_{GLS}\},\quad \forall \mathbf{x}\in\Omega.\\
\psi_i\left(\mathbf{x},t\right) =\ Redist\left(\hat{\psi}_i\left(\mathbf{x},t\right)\right)\quad \forall i \in \{1,2,...,N_{GLS}\}
\end{eqnarray}

In the zones without SPPs, this methodology is equivalent to the classical numerical treatment (Eq.(\ref{eq:vacuum})) whereas when SPPs are present, it enables, by successive iterations, to impose the Young-Herring equilibrium for incoherent SPPs.

In real thermomechanical processes, which typically encompass significant temperature changes, it becomes crucial to model the evolution of SPPs. This modeling is key to quantitatively and qualitatively predicting the evolution of the microstructure, particularly with respect to grain size distribution. Hence, the ability to model the transformation of particles during GG is indispensable for managing actual industrial processes and forecasting microstructural evolution accurately.

 In this LS framework, once the initial mesh is generated, the polycrystal can be created statistically or experimentally from an EBDS map. Subsequently, the grains fields $\psi_{i}$ are adjusted to incorporate the new particles field $\psi_{SPP}$. The convective equation, Eq.(\ref{eq:Transport}), is then applied to $\psi_{SPP}$. This involves computing a velocity, $v_{spp}$, using prescribed data related to the time-dependent evolution of the particle's radius. Following this, a smoothed velocity field $v$ is computed using a Laplacian equation (see Eq.\ref{eq:velLaplace}) with Dirichlet boundary conditions  $v_{spp}$  established at the particle interfaces. Finally $v$ is used to compute the velocity field $\mathbf{v}$ oriented towards the center of each precipitate to be applied to $\psi_{SPP}$ :

\begin{equation}\label{eq:velLaplace}
\left\lbrace\begin{array}{l}
\Delta v=0 \\
v={v}_{spp}\quad at\quad \Gamma_{spp}
\end{array}\right.
\end{equation}

and 
\begin{equation}\label{vel}
\mathbf{v}=v \cdot \mathbf{n}=-v \cdot \nabla \psi_{SPP} ,
\end{equation}

with $\mathbf{n}$ the unitary inside normal vector to the SPP, $v_{spp}$ the velocity to impose to the SPPs and $\mathbf{v}$ the resulting velocity field that we really impose, through a convection equation to $\psi_{SPP}$.

\begin{figure}[h]
	\centering
	\includegraphics[width=0.6\textwidth]{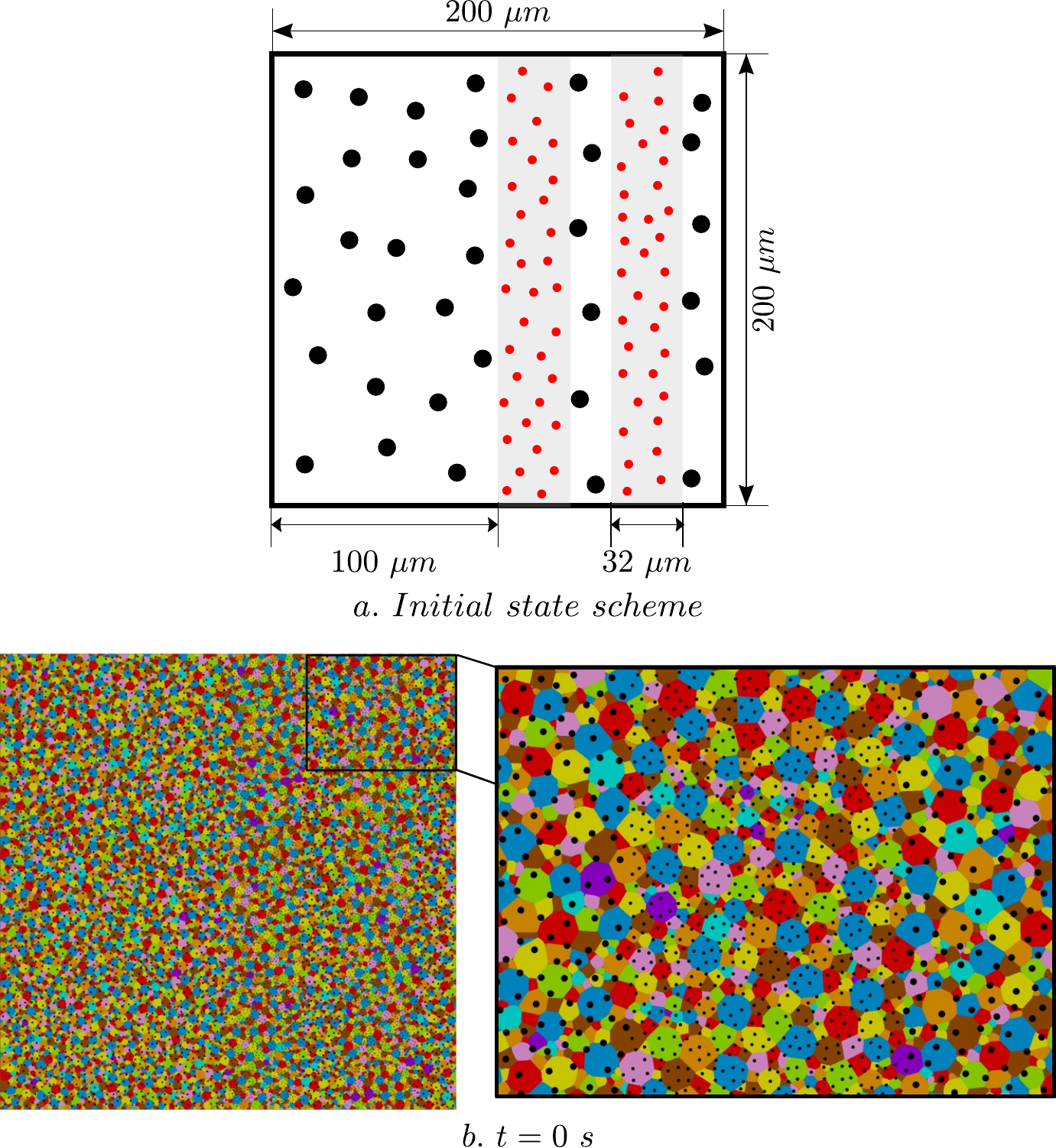}
	\caption{ Schematic and initial microstructure for a heterogeneous SPPs dispersion in a LS framework (SPPs are in black). From \citep{Alvarado2021a}.}\label{fig:ini_state_bandes}
\end{figure}

\begin{figure}[h!]
  \centering
 \begin{subfigure}{0.49\textwidth}
    \centering
    \includegraphics[scale=0.08]{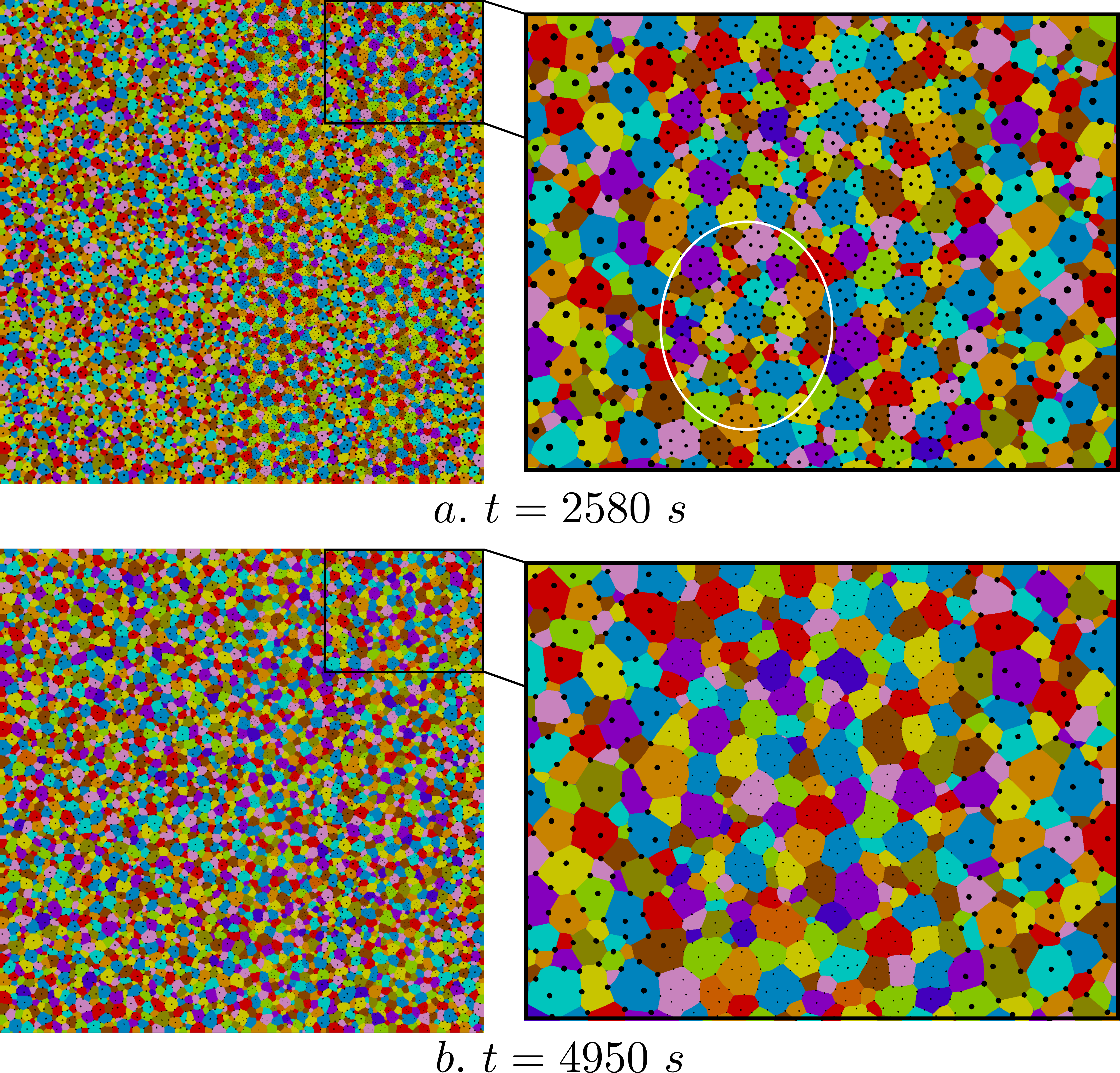} 
  \end{subfigure} 
  \begin{subfigure}{0.49\textwidth}
    \centering
    \includegraphics[scale=0.08]{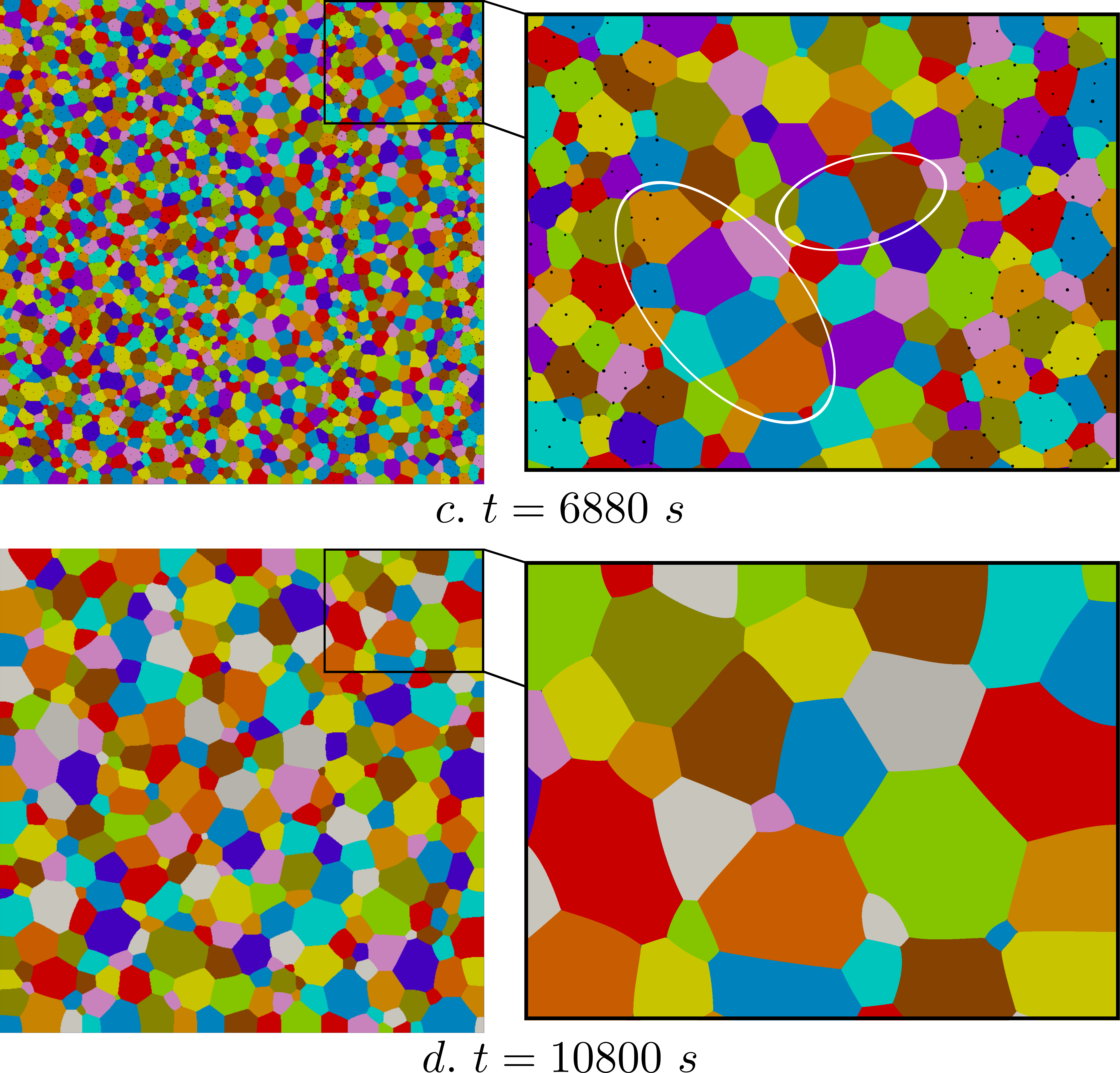}
  \end{subfigure} 
  \caption{Microstructure evolution at different times for a heterogeneous SPPs distribution (SPPs are in black). From \citep{Alvarado2021a}. A video is available \href{https://youtu.be/lRO_5RDsVIU}{online}.}\label{fig:dissolution_bandes}
\end{figure}

This methodology is illustrated in Figures \ref{fig:ini_state_bandes} and \ref{fig:dissolution_bandes} where a simulation domain of 200x200 $\mu m$ with an initial number of grains around 50000 is considered in context of $AD730^{TM}$ nickel base superalloy. The mean grain radius (in number) $\bar{R}=5$ $\mu m$ and a spherical particles population with a initial surface fraction $f_{spp}=6\%$ divided in large SPPs with $f_{spp}^1=4\%$ of radius $r_{spp}^1=2$ $\mu m$ and small SPPs with $f_{spp}^2=2\%$ of radius $r_{spp}^2=1$ $\mu m$, distributed in two bands in the domain as illustrated in Figure \ref{fig:ini_state_bandes}.a.

Material properties and particle velocity $v_{spp}$ for the chosen annealing temperature profile were obtained through experimental investigations \citep{Alvarado2021b}. The small particles are dissolved before the large particles, thus the bands regions present a classical GG mechanism sooner than the entire domain.

Fig. \ref{fig:dissolution_bandes} from \citep{Alvarado2021a} illustrates the evolution of the microstructure at different stages of the simulation (\SI{2580}{s}, \SI{4950}{s}, \SI{6880}{s} and \SI{10800}{s}) showing the particle and GG evolution during annealing.

For the initial isothermal treatment where the particles do not evolve, the grain size are smaller in the zones composed of small SPPs. This can be easily explained by a bigger resulting pinning pressure in this zone than in the zone with large SPPs (see Figure \ref{fig:dissolution_bandes}.a). When the temperature increases, the particles begin to dissolve and the grains evolve, especially at the regions composed of small particles where the small grains of this regions begin to grow (see Figure \ref{fig:dissolution_bandes}.b). Once the small SPPs are completely dissolved, the grains can grow freely and as represented in Figure \ref{fig:dissolution_bandes}.c (white ellipses) some grains can grow more than others, thus a heterogeneous grain evolution begins to take place, where some grains are likely to grow more and more leading potentially to abnormal grain growth.

The thermal treatment ends with a maintain of temperature at \SI{1120}{\celsius} which is superior to the solvus temperature, so no particles remain and all the domain converge towards a classical GG mechanism as illustrated in Fig.\ref{fig:dissolution_bandes}.d.

 This recent LS formalism opens also the possibility to simulate material deformations more naturally by taken into account the mechanical behavior of SPPs. Validation and comparisons of this proposed LS-FE numerical framework comparatively to previous formulations \citep{Agnoli2014a, Scholtes2016} and experimental data were proposed in \citep{Alvarado2021a} and \citep{Alvarado2021b}, respectively. A natural perspectives of these works will be to consider a more physcis-based modeling for SPP evolution, likely utilizing the tools described in the following section.

\section{Modeling of diffusive solid-state phase transformation}

Until recently, the LS approach in the context of polycrystalline microstructure was primarily used for modeling recrystallization and related phenomena in single-phase materials. However, solid-state phase transformation (SSPT) entails crystallographic changes in the parent phase, which occur through rearrangement of the lattice structure to form a different, more stable product phase, all while remaining in the same solid state. SSPT can either be displacive or diffusive. Displacive transformation \citep{James1986}  is characterized by spontaneous, coherent, and coordinated atomic movement over relatively short distances. On the other hand, diffusive transformation \citep{Gamsjager2002} involves a more gradual reorganization of the lattice through short and long-range atomic diffusion.

Two fundamental mechanisms drive diffusive phase transformation (PT): (i) the diffusion of solutes across phase interfaces and within the grain bulk, resulting in changes in chemical composition, and (ii) interface migration leading to lattice rearrangement or structural changes. PT plays a critical role in generating a range of materials with diverse microstructural characteristics during thermomechanical treatments.

In the context of diffusive SSPT, phase field methods (PFM) have gained popularity and extensive usage due to their thermodynamic consistency and their ability to model complex morphological changes. The pioneering work of Wheeler et al. \citep{Wheeler1992} and Steinbach et al. \citep{Steinbach1996, Tiaden1998} on solidification using PFM laid some of the mathematical groundwork for phase field modeling in multi-component, multi-phase systems involving solute diffusion. Despite this, the majority of existing numerical frameworks are primarily devoted to modeling recrystallization and grain growth in single-phase materials or to modeling phase transformations without considering recrystallization and associated phenomena. In light of this, a novel LS approach has recently been proposed to consider both aspects simultaneously \cite{Chandrappa2023}.\medbreak

Modeling diffusive SSPT at the mesoscopic scale typically necessitates two governing equations: a diffusion equation that regulates the distribution of solute atoms  across different phases, and another equation that manages the migration of the resulting interface network. Traditionally, the diffusion equation can be resolved within a  LS framework. However, the presence of material discontinuities at the phase interfaces requires explicit consideration of interface jump conditions when solving the diffusion equation in the LS framework. This requires the explicit localization of the interface at every moment to numerically manage the necessary jump conditions.

To circumvent this complex step, Chandrappa et al. \citep{Chandrappa2022a,Chandrappa2023} proposed considering a diffuse interface hypothesis across the phase interfaces during the resolution of the diffusion equation. In simpler terms, while the multi-phase grain interface network is migrated using an LS description, a global diffusion equation is considered based on a diffusive interface assumption for the phase interfaces. This approach allows for the resolution of a single diffusion equation throughout the entire computational domain, eliminating the need for any interface jump conditions. \\

The transition between a distance function to a diffuse interface description can be established, thanks to a hyperbolic tangent relation : 
\begin{equation}
\phi\left(\mathbf{x},t\right) = \frac{1}{2}tanh\left(\frac{3\psi\left(\mathbf{x},t\right)}{\eta}\right)+\frac{1}{2},
\label{tanhf}
\end{equation}where $\eta$ is a diffuse interface thickness parameter. In the following, we shall refer this function ($\phi$) yielding the diffuse interface as the phase-field function. In the following, the particular case of austenite decomposition to ferrite by carbon diffusion is illustrated even if the framework can be generalized.
the total carbon concentration field ($C$) can be expressed as a continuous variable as the solute flux:
\begin{equation}
C = \phi C_\alpha + (1-\phi)C_\gamma,\quad \bm{J} = \phi \bm{J}_{\alpha} + (1-\phi)\bm{J}_{\gamma}.
\label{cmix}
\end{equation}

The diffuse phase interface is assumed to be composed of a mixture of the two phases. A constant concentration ratio between the phases is enforced. This stipulation ensures that the redistribution of solute atoms at the interface aligns with a partitioning ratio (denoted as $k$) equal to that at the equilibrium::

\begin{equation}
k = \frac{C_\alpha}{C_\gamma} =\frac{C_\alpha^{eq}}{C_\gamma^{eq}}
\label{keq},
\end{equation}where $C_\alpha^{eq}$ and $C_\gamma^{eq}$ are the equilibrium concentrations of $\alpha$ and $\gamma$ phases respectively at temperature $T$.

Following Fick's laws of diffusion, the diffusion equation for carbon partitioning can be expressed as:
\begin{equation*}
\partial_t C = -\nabla\cdot\bm{J} = -\nabla\cdot\left[\phi \bm{J}_{\alpha} + (1-\phi)\bm{J}_{\gamma}\right],
\end{equation*}
with, 
\begin{equation*}
\bm{J}_{\alpha} = -D_\alpha^C \nabla C_\alpha; \quad  \bm{J}_{\gamma}=-D_\gamma^C \nabla C_\gamma.
\end{equation*}
We then obtain,
\begin{equation}
\partial_t C = \nabla\cdot\left[\phi D^C_\alpha \nabla C_\alpha + (1-\phi)D^C_\gamma \nabla C_\gamma \right],
\label{diffmixeqn}
\end{equation}where $D^C_\alpha$ and $D^C_\gamma$ represent the diffusivity of the carbon element in ferrite and austenite phases respectively.

Invoking eqs.\eqref{cmix} and \eqref{keq} in eq.\eqref{diffmixeqn}, a modified carbon diffusion equation \citep{Tiaden1998} is obtained:
\begin{equation}
\partial_t C = \nabla\cdot\left(D^*(\phi)\left[\nabla C -\frac{C(k-1)}{1+\phi(k-1)}\nabla\phi \right]\right),
\label{diffmixform}
\end{equation}where $D^*(\phi)$ is called "mixed diffusivity" and is defined as,
\begin{equation*}
D^*(\phi) = \frac{D_\gamma^C + \phi(kD^C_\alpha - D^C_\gamma)}{1+\phi(k-1)}.
\label{mixeddiffus}
\end{equation*}

With further simplifications, the above eq.\eqref{diffmixform} can be transformed into a Convective-Diffusive-Reactive (CDR) form as follows:

\begin{equation*}
\partial_t C = \nabla\cdot\left[D^*(\phi)\nabla C - C\bm{A}(\phi)\right] 
\end{equation*}

\begin{equation}
\partial_t C +  \left(\bm{A}-\nabla D^*\right)\cdot\nabla C - D^*\Delta C + RC = 0,
\label{CDReqn} 
\end{equation} where,
\begin{equation*}
\bm{A}(\phi)= \frac{D^*(\phi)(k-1)}{1+\phi(k-1)}\nabla\phi, \quad and \quad R=\nabla\cdot\bm{A}.
\end{equation*}

Interestingly, when applying the weak form of the prior equation in a FE resolution, the gradient of the mixed diffusivity term ($\nabla D^*$) vanishes. This aspect carries significant value with regard to numerical stability, especially considering the abrupt changes of this term across a phase interface.

Concerning interface migration, following Eq.\ref{eq:kinetic equation sspt}, P can be defined as,
\begin{equation}
    P=\Delta G + \llbracket E \rrbracket - \kappa \sigma.
    \label{eq:drivpress}
\end{equation}

The $\Delta G$ component acts only across the phase interfaces while vanishing across the grain interfaces of similar phases. Also, the sense and value of interface mobility, and interface energy could be different depending on the type of interface (i.e., $\alpha/\gamma$ phase interface, $\alpha/\alpha$ grain interface, and the $\gamma/\gamma$ grain interface).
Eq.\ref{eq:Transport2} can be then generalized in the following form:
\begin{equation}
\begin{gathered}
\partial_t \psi_i + \left[\bm{v_{\Delta G}}+\bm{v^{\llbracket \rho \rrbracket}}\right]_i\cdot\nabla\psi_i - \left[\sum_{l\in \mathcal{S}}\chi_{l}\mu_{l}\sigma_{l}\right]\Delta\psi_i = 0 \qquad \forall i \in \{1,2,...,N_{GLS}\},
\label{lsCDeqn}
\end{gathered}
\end{equation}where $\mathcal{S}=\{\alpha\gamma, \alpha\alpha, \gamma\gamma\}$, $\bm{v_{\Delta G}}=\chi_{\alpha \gamma}\mu_{\alpha \gamma}\Delta G_{\alpha \gamma} \bm{n}$, and $\bm{v^{\llbracket \rho \rrbracket}}=\left[\sum_{l\in \mathcal{S}}\chi_{l}\mu_{l}\llbracket E \rrbracket_{l} \right]\bm{n}$.

As previously introduced, these equations must be followed by a multiple-junction treatment (Eq.\ref{eq:vacuum}) and a reinitialization step (Eq.\ref{eq:eikonal}). Moreover, as for $\bm{v^{\llbracket \rho \rrbracket}}$  (see Eq.\ref{eq:vstoredBernacki}), $\bm{v_{\Delta G}}$ component can be built as an unique function in the FE mesh:
\begin{equation}
    \bm{v_{\Delta G}}(\mathbf{x},t)=\sum_{i=1}^{N_{GLS}}\sum_{\substack{j=1 \\ j \neq i}}^{N_{GLS}}\chi_{G_i}(\mathbf{x},t)\mu_{ij}\exp{\left(-\beta|\psi_j(\mathbf{x},t)|\right)}\chi_{\alpha\gamma}(\mathbf{x},t)\Delta G_{\alpha\gamma}\mathscr{F}_s(\mathbf{x},t)(-\bm{n}_j),
    \label{vdeltaG}
\end{equation}where $\chi_{\alpha\gamma}$, as seen earlier, helps to filter this component of velocity field only on the phase interfaces and
\begin{equation}
    \mathscr{F}_s(\mathbf{x},t)=\chi_\alpha(\mathbf{x},t)-\chi_\gamma(\mathbf{x},t)=2\chi_\alpha(\mathbf{x},t)-1,
    \label{sensefunc}
\end{equation} with $\chi_\alpha(\mathbf{x},t)$ and $\chi_\gamma(\mathbf{x},t)$ the characteristic functions of $\alpha$ and $\gamma$ phase, respectively.

\begin{figure}[h!]
	\centering
	\includegraphics[width=0.8\textwidth]{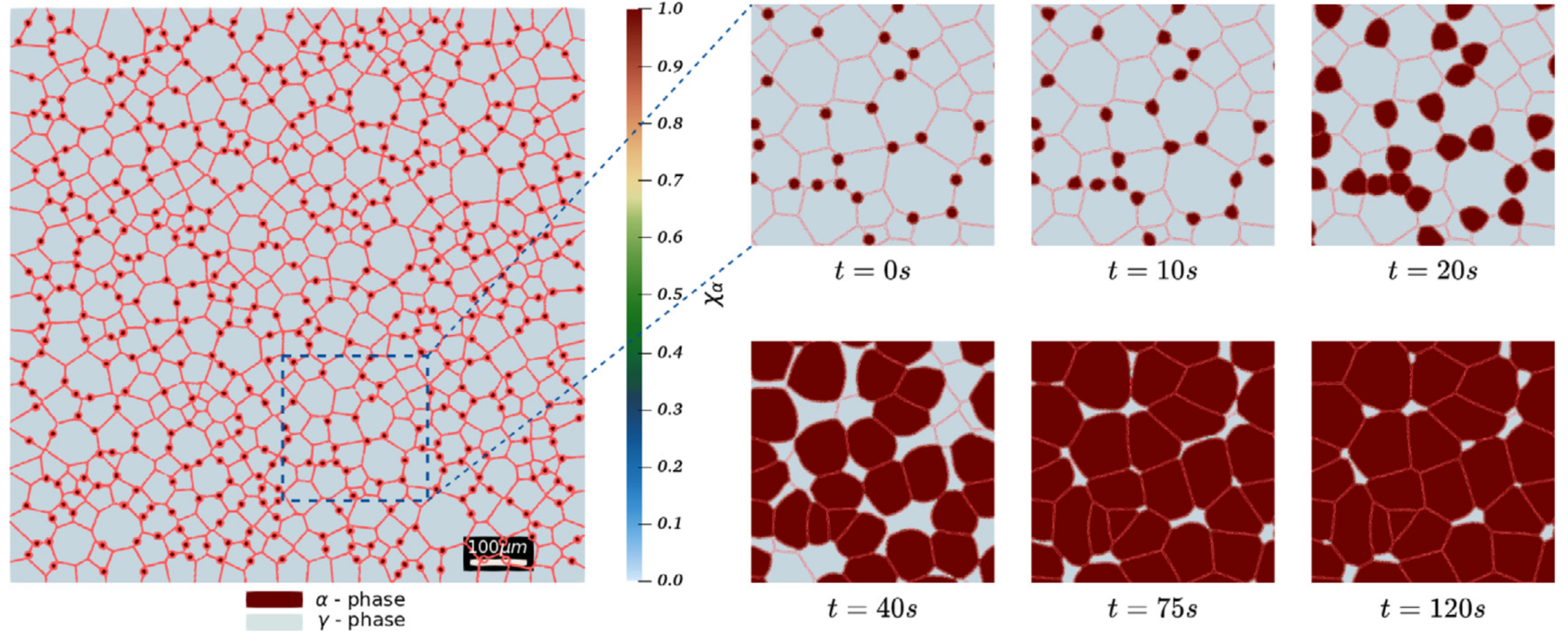}
	\caption{$\alpha$ phase nucleation and growth in a $\gamma$ phase polycrystal thanks to a LS-FE framework. From \citep{Chandrappa2023}.}\label{fig:sspt} 
\end{figure}

The last ingredient missing to completely prescribe the above kinetics is the change in Gibbs free energy between the two phases. $\Delta G_{\alpha\gamma}$ is typically dependent on the local composition of the solutes, temperature, and the pressure. In many works, the description for $\Delta G_{\alpha\gamma}$ has been
established by thermodynamic evaluations based on Calphad data \citep{Steinbach2007} or ThermoCalc software \citep{Thermocalc}. For certain sharp interface descriptive models, the diffusion in the product phase is assumed to be instantaneous and so $\Delta G_{\alpha\gamma}$ is simply assumed to be proportional to the deviation in concentration at the interface in the parent phase ($C_{\gamma, eq}$) from the equilibrium concentration in this phase ($C_{\gamma, \gamma \alpha}$) \citep{Mecozzi2011, Liu2018}. In Chandrappa et al. \citep{Chandrappa2022a,Chandrappa2023} as in the works of Mecozzi \citep{Mecozzi2007}, $\Delta G_{\alpha\gamma}$ is described based on a local linearization of the phase diagram and is basically assumed to be proportional to a small undercooling ($\Delta T = T^{eq}-T$). More details can be found in \citep{Chandrappa2023,Chandrappa2022a}. Fig.\ref{fig:sspt} (from \citep{Chandrappa2023}) illustrates the described numerical framework in a 2D case which corresponds to $\alpha$ phase nucleation and growth in a $\gamma$ phase polycrystal.

\section{Conclusion, limits, and perspectives}

As highlighted, the level-set approach has spread extensively in computational metallurgy at the polycrystalline scale and is now used to model numerous mechanisms. The known advantages of this approach, explaining its relevance, primarily lie in its 'front-capturing' nature, where complex topological events (like the emergence and disappearance of grains) can be addressed in a completely natural way. Since the description of interfaces is not tied to a set of nodes and edges of the considered mesh or regular grid, the kinetics of the interfaces can thus be decoupled from that of the medium used for domain discretization. This strong point is, of course, common to multi-phase field type approaches. Moreover, if one take advantage of the allusion to multi-phase field approaches here to delve deeper. One can affirm that the level-set approach eliminates one of the inherent difficulties of phase field methods in defining, controlling, and optimizing the thickness to be considered in phase field approaches for a similar computational cost. The kinetic approach inherent to the LS method also negates the need to debate the comparative calibration of potential different Ginzburg-Landau functionals for the same mechanism.\medbreak

Of course, if the level-set approach only had advantages, it would have spread much more in computational metallurgy compared to Monte Carlo, Cellular Automata or multi-phase field approaches, which are much more widely used in the community. This can easily be explained by the fact that each strong positive statement made in the previous paragraph can be offset by a negative argument. Let's try the exercise: While the front-capturing type description is often a sign of simplicity in topological events, it can also become a drawback when it comes to properly handling multiple junctions or when knowing their precise position and the neighboring discretization of the interfaces converging there can be indispensable for a rigorous consideration of the anisotropy of interface properties, as discussed in section \ref{sec:aniso}. This explains why this problem still remains stiff in level-set or multi-phase field approaches compared to vertex type approaches, for example. In regard to the link to the finite element mesh, it is also useful to remember that while a discretization of the interface is not necessary, the precision of the interface description and its attributes remain however tied to the fineness of the mesh around the interfaces and/or the order of interpolation used. Therefore, it cannot be asserted, as is sometimes written, that the level-set description allows one to be completely independent from any work on the embedding finite element mesh. Finally concerning the comparison with multi-phase field approach, it is probably useful to recall that the level-set framework can be seen somewhere as a particular extreme configuration of multi-phase field approach when a top hat potential is used as nicely depicted in the work of Steinbach \citep{Steinbach_2009}. The level-set approach brings precision in the description of the interface where the multi-phase field approach brings regularization of singularities at the interfaces.\medbreak

Finally, in the author opinion, by taking into account the balance of the two previous paragraphs, the main interest of the LS approach remains probably more in its versatility to take into account numerous mechanisms in a global numerical framework even in context of large deformation of the calculation domain. This likely explains its significance in modeling complex thermomechanical paths \citep{Maire2017}, and thus its unique potential for predicting microstructures in the context of hot metal forming processes \citep{de_micheli_digimu_2019}. On this matter, it seems evident to assert that level-set approaches are far from having demonstrated their full potential in the mechanisms that can be modeled. As presented in the previous section, solid-state transformations could become a new playground for these methods while taking into account the driving pressures also acting on grain boundaries. Other diffusion mechanisms are also within reach, such as surface diffusion \citep{poly2017introduction}. Finally, level-set approaches are also a powerful tool for accounting for all the mechanisms mentioned for processes related to powder metallurgy \citep{zouaghi_modelling_2012,bruchon_finite_2012,bruchon_3d_2011} where particles and their polycrystalline structures can be taken into account.\medbreak

However, its main weakness, like other full-field approaches, lies in its computational cost mainly in 3D. While these calculations are of interest in the modeling and local understanding of mechanisms, their connection with component scale calculations remains very tenuous today due to the cost of these calculations. The most advanced tools involve approaches where RVE simulations are conducted at a limited number of integration points within component-scale simulations. Thus, in R\&D, low dimension/number of RVE and indirect coupling with component scale simulations correspond to the more advanced tools. In this context, an evident perspective lies in the notable emergence of machine learning-trained surrogate models. These models, which are based entirely on artificial intelligence, serve as reduced order models derived from full-field simulations databases. Their primary advantage lies in their ability to offer predictive models at a significantly reduced computational cost \citep{montes2021accelerating,hashemi2021machine,oommen2022learning,yan2022novel}. Despite the evident appeal of these initial efforts, they are hindered by a few strong key limitations. Firstly, the representativeness of the training bases may be limited due to the inherent inaccuracies of the promoted full-field models. Secondly, the comparatively simplistic mechanisms primarily investigated thus far, such as grain growth and spinodal decomposition, do not adequately reflect the complexities found in actual metal forming conditions. Lastly, the considered microstructures lack the required intricacy, and hence may not provide an accurate representation of real materials. In context of hot metal forming, such nascent approaches are then currently unable to provide physically representative simulations at the mesoscopic scale. This constitutes, however, a very promising field in the coming years either for developing reduced order models, or even more intriguingly, to accelerate high-fidelity level-set simulations through mixed formulations (high-order time step resolution interposed with low-order ones) \citep{oommen2022learning}.\medbreak

As highlighted,  the representativity of Eq.\ref{eq:global kinetic equation} at the mesoscopic scale coupled to the usual 5D-space and models to described the grain boundary energy and mobility is more and more discussed and recent studies have resulted in contradictory conclusions on the subject. Indeed, the recent improvements of 3D GG and ReX experiments now make it possible to discuss in a much more precise way the veracity of this picture. Non-destructive in-situ experiments allow to avoid bias inherent to 2D observations and open the path to precise reverse engineering of the $\lVert\bm{v}\rVert/\kappa$ ratio. In this context, numerical reverse engineering by using anisotropic level-set formulations taken into-account the torque terms, will be of great interest to participate to this discussion.

\bibliographystyle{elsarticle-num}
\bibliography{biblio}





\end{document}